\documentclass[usenatbib,usegraphicx]{mn2e}

\usepackage{graphicx}
\usepackage{times}
\usepackage{aas_macros}
\usepackage{pdfpages}
\usepackage{grffile}
\usepackage{amsmath}
\usepackage{amsfonts}
\usepackage{booktabs}
\usepackage{bm}
\usepackage[pdftex,colorlinks=true,breaklinks=true,citecolor=blue,draft=false]{hyperref}

\date{Accepted ---. Received ---; in original form ---}
\pagerange{\pageref{sec:intro}--\pageref{lastpage}}
\pubyear{2016}
\begin{document}

% write title (with email and institute)
\title[Sparse imaging of interferometric observations]{Robust sparse image reconstruction of radio interferometric observations with PURIFY}
\author[Pratley \textit{et al.}]
{Luke Pratley$^1$\thanks{E-mail: Luke.Pratley@gmail.com}, Jason D.~McEwen$^1$, Mayeul d'Avezac$^2$, Rafael E.~Carrillo$^3$, \newauthor
Alexandru Onose$^4$, 
Yves Wiaux$^4$\\
${}^1$Mullard Space Science Laboratory (MSSL), University College London (UCL), Holmbury St Mary, Surrey RH5 6NT, UK\\
${}^2$Research Software Development Group, Research IT Services, University College London (UCL), London WC1E 6BT, UK\\
${}^3$Signal Processing Laboratory (LTS5), Ecole Polytechnique F\'ed\'erale de Lausanne (EPFL), Lausanne CH-1015, Switzerland\\
${}^4$Institute of Sensors, Signals, and Systems, Heriot-Watt University, Edinburgh EH14 4AS, UK}
\maketitle

\begin{abstract}
Next-generation radio interferometers, such as the Square Kilometre Array (SKA), will revolutionise our understanding of the universe through their unprecedented sensitivity and resolution. However, to realise these goals significant challenges in image and data processing need to be overcome. The standard methods in radio interferometry for reconstructing images, { such as CLEAN}, have served the community well over the last few decades and have survived largely because they are pragmatic.  However, they produce reconstructed \mbox{inter\-ferometric} images that are limited in quality { and scalability} for big data. In this work we apply and evaluate alternative interferometric reconstruction methods that make use of state-of-the-art sparse image reconstruction algorithms motivated by compressive sensing, which have been implemented in the PURIFY software package. In particular, we implement and apply the proximal alternating direction method of multipliers (P-ADMM) algorithm presented in a recent article. First, we assess the impact of the interpolation kernel used to perform gridding and degridding on sparse image reconstruction. We find that the Kaiser-Bessel interpolation kernel performs as well as prolate spheroidal wave functions, while providing a computational saving and an analytic form. { Second, we apply PURIFY to real interferometric observations from the Very Large Array (VLA) and the Australia Telescope Compact Array (ATCA) and find images recovered by PURIFY are higher quality than those recovered by CLEAN}. Third, we discuss { how} PURIFY { reconstructions exhibit additional} advantages over those recovered by CLEAN. The latest version of PURIFY, { with} developments presented in this work, is made publicly available.
\end{abstract}

% look up list of allowable keywords for this section
\begin{keywords}
techniques: image processing - techniques: interferometric
\end{keywords}

\section{INTRODUCTION}
\label{sec:intro}
Radio interferometry has been critical for imaging the radio universe at higher resolution and sensitivity than possible with a single radio telescope. However, radio interferometers are limited by the number of possible pairs of antennae in an array, which limits the number of possible measurements made during an observation. Consequently, image reconstruction methods are needed to reconstruct the true sky brightness distribution from the raw data acquired by the telescope, which amounts to solving an ill-posed inverse problem. Traditional methods, which are mostly variations of the H\"{o}gbom CLEAN algorithm \citep{hog74}, do not exploit modern state-of-the-art image reconstruction techniques.

Next-generation radio interferometers, such as the LOw Frequency ARray (LOFAR; \citealp{van13}), the Murchison Widefield Array (MWA; \citealp{tin13}), the Australian Square Kilometre Array Pathfinder (ASKAP; \citealp{hot14}), and the Square Kilometer Array (SKA; \citealp{dew13}), must meet the challenge of processing and imaging extremely large volumes of data. These experiments have ambitious, high-profile science goals, including detecting the Epoch of Re-ionisation (EoR) \citep{koo15}, mapping large scale structure \citep{maa15}, and investigating cosmic magnetism \citep{joh15}. If these science goals are to be realised, state of the art methods in image reconstruction are needed to process big data and to reconstruct images with high fidelity.

Compressive sensing is a robust framework for signal reconstruction. The theoretical framework of compressive sensing motivates sparse regularisation approaches for solving inverse problems, like those encountered in radio interferometry. 
The framework of compressive sensing was first applied to radio interferometry in the study of \cite{wia09a}, in the synthesis framework, where it was shown that compressive sensing approaches can produce higher quality reconstructed images than standard interferometric imaging methods. In \citet{car12} the analysis framework was considered and the sparsity averaging reweighted analysis (SARA) algorithm was developed and applied to radio interferometric imaging, demonstrating excellent performance \citep[see also][]{car12a}.  It has also been shown that the compressive sensing framework can be applied to wide-field-of-view observations \citep{mce11} and can correct for directional dependent effects, such as non-coplanar baselines \citep{wia09b,wol13}. In \citet{car14} state-of-the-art convex optimisation algorithms that scale to very large data-sets were developed to solve sparse regularisation problems, such as the SARA problem.  These algorithms were implemented in the first release of the PURIFY software package \citep{car14} for solving radio interferometric imaging problems by sparse regularisation.  Recently, new algorithms for solving these problems were developed by \cite{ono16}, including proximal alternating direction method of multipliers (P-ADMM) and primal dual algorithms, paving the way to image the large radio interferometric data-sets that will characterise the SKA era. 
Alternative compressive sensing approaches have also be applied to aperture synthesis \citep{li11a,dab15,gar15} and rotation measure synthesis \citep{li11b,sun15}. 

In this work we implement the P-ADMM algorithm developed by \citet{ono16} in the PURIFY software package, which has been entirely redesigned and re-implemented in C++, and apply it to observational data from the Very Large Array (VLA) and the Australia Telescope Compact Array (ATCA).  In addition, we discuss conceptual differences between the restored CLEAN image and the reconstructed PURIFY model.  The previous version of \mbox{PURIFY} supported only simple models of the measurement operator modelling the telescope.  PURIFY now supports a wider range of more accurate measurement operator models, including a number of different convolutional interpolation kernels (for gridding and degridding).  Moreover, we study how the choice of kernel can affect the quality of sparse image reconstruction. 

The remaining sections of the paper are structured as follows. Section \ref{sec:radio} reviews the basics of aperture synthesis and radio interferometry. Section \ref{sec:cs} discusses radio interferometric imaging in the context of compressive sensing and sparse image reconstruction. Section \ref{sec:gridding} discusses convolutional interpolation and the different kernels considered. These interpolation kernels are then tested and compared using simulations in Section \ref{sec:simulations}. Section \ref{sec:purify_useage} discusses the similarities and differences between images recovered by CLEAN and PURIFY and also considerations in applying PURIFY to real observational data.  The reconstruction of images from observations made by the VLA and ATCA are presented in Section \ref{sec:reconstructions}. Section \ref{sec:concusions} states the final conclusions.

\section{Aperture Synthesis and Radio Interferometry}
\label{sec:radio}
 The principles of aperture synthesis date back as far as the work of \citet{mcc47}. However, \citet{ryl60} first described how aperture synthesis could be used to construct a large scale radio interferometric telescope. Thus, the limit in resolution of single dish radio telescopes could be overcome by radio interferometric telescopes, improving our ability to observe and therefore understand the radio sky.

 In aperture synthesis, an array of antennae are collectively used to image the sky at higher resolution than possible with a single dish, hence synthesising a larger aperture. Each pair of antennae measures a phase and amplitude of a Fourier component of the brightness distribution across the sky. It is through the measurement of these Fourier components that the sky is effectively imaged. However, due to a limited number of antennae, not all Fourier components can be measured in an observation. An ill-posed inverse problem must be solved to reconstruct the true sky brightness distribution. How this ill-posed inverse problem is solved has a significant impact on the fidelity of the reconstructed image.

Each antenna in an array measures an incoming electric field across its field of view. The electric fields are then cross-correlated between pairs of antennae, using a correlator, in-order to calculate the visibility
\begin{equation}
	\mathcal{V}(\bm{b} = \bm{a}_2 - \bm{a}_1) = \langle \mathcal{E}(\bm{a}_1, t) \mathcal{E}^*(\bm{a}_2, t) \rangle_{\Delta t}\, ,
\end{equation}
where $\mathcal{E}$ is the electric field, 
$\bm{a}_1$ and $\bm{a}_2$ are the spatial positions of the two antenna, $t$ is time, and $ \Delta t$ is the time interval over which the expected value, denoted by $\langle \cdot \rangle$, is taken, which is longer than the time scale of the radio wave observed \citep{tho99, tho08}. The difference between the positions of the antennae $\bm{b} = \bm{a}_2-\bm{a}_1$ is called the baseline.

It is well known that a visibility contains spatial information about the brightness distribution across the sky. While there have been more general measurement equations developed for radio interferometry \citep{mce08, car09, smi11, pri15}, the van Cittert-Zernike theorem \citep{zer38} states that the visibility $\mathcal{V}$ is related to the sky brightness distribution $\mathcal{I}_{\lambda}$, at wavelength $\lambda$, by
\begin{equation}
	\label{eq:measurement_equation}
	\mathcal{V}(\bm{b}) = \int_{S^2} \mathcal{A}(\bm{\sigma})\mathcal{I}_{\lambda}(\bm{\sigma}){\rm e}^{-2\pi i \lambda \bm{b} \cdot \bm{\sigma}}\, {\rm d} \Omega \, ,
\end{equation}
where $\mathcal{A}$ is the primary beam of the telescope, $\bm{b}$ is the baseline separating the two antennae, and $\bm{\sigma}$ denotes a location on the celestial sphere $S^2$ with area element ${\rm d}\Omega$.
When the baselines in an array are co-planar and the field of view is narrow, Eq.~\ref{eq:measurement_equation} reduces to a Fourier relation:
\begin{equation}
	\label{eq:planar_measurement_equation}
	\mathcal{V}(u, v) = \int_{\mathbb{R}^2} \mathcal{A}(l, m)\mathcal{I}_{\lambda}(l, m){\rm e}^{-2\pi i(ul + vm)}\, {\rm d}l {\rm d}m\, ,
\end{equation}
where $(l, m)$ are the coordinates of the plane of the sky, centred on the pointing direction of the telescope, and $\bm{u}=(u, v)$ are the corresponding Fourier coordinates defined by the baseline: \mbox{$\bm{u}=\bm{b}/\lambda$}. In this context, a visibility measures a Fourier component of the sky brightness distribution in the plane of the sky, centred on the pointing direction of the telescope \citep{tho99, tho08}.

The Fourier transform relation of Eq.~\ref{eq:planar_measurement_equation} cannot be inverted directly to obtain an accurate estimate of $\mathcal{I}_{\lambda}(l, m)$ since $\mathcal{V}(u, v)$ cannot be measured for all Fourier coordinates. The missing samples of $\mathcal{V}(u, v)$ leave Eq.~\ref{eq:planar_measurement_equation} as an ill-posed inverse problem, which has an infinite number of possible solutions.  To recover a suitable, unique solution, regularisation is used to inject prior information regarding the underlying signal. 

The most common techniques used to solve for the true sky brightness distribution are CLEAN \citep[\textit{\textit{e.g.}}][]{hog74} and the maximum entropy method (MEM) \citep[\textit{\textit{e.g.}}][]{cor85}.
The basic CLEAN algorithm was developed in the 1970's \citep{hog74}. CLEAN implicitly imposes a sparse prior in a point source (Dirac) basis \citep{mar87}, and is essentially a matching pursuit algorithm \citep{mal93}. Variations of CLEAN have also been developed for resolved and extended structures, multi-frequency synthesis, and polarised sources \citep{cla80, sch84a, ste84, sau96a, cor08, off14, pra16}.
The MEM algorithm regularises the ill-posed radio interferometric inverse problem through an entropic prior, maximising an objective function comprised of an entropy term and a data fidelity term (in practice an additional flux constraint is typically imposed in radio interferometric applications of MEM; \citealt{cor85}).  
In practice, CLEAN often struggles to image diffuse structure, while MEM struggles to resolve point sources.  CLEAN, and its variants, are of widespread use in radio interferometric imaging today, while MEM has not experienced such widespread adoption.

\section{Compressive sensing for radio interferometric imaging}
\label{sec:cs}

In its fundamental form, compressive sensing provides a framework for recovering signals from small numbers of measurements and considers the efficient design of the signal measurement process \citep{candes:2006a,candes:2006b,donoho:2006,can08}. In radio interferometry, there is little control over the measurement process since the baseline configurations are typically limited by the interferometer (nevertheless, there may be scope for telescope optimisation; \citealt{wia09b,wol13}). The compressive sensing framework, however, motivates a robust method of reconstructing images from the visibilities measured by a telescope through sparse regularisation. Sparse regularisation exploits the fact that many natural signals---such as astronomical images---are sparse or compressible, \textit{i.e.}\ for a suitable representation (\textit{e.g.}\ wavelet basis) most of the coefficients for the ground truth image are zero or close to zero, respectively.  In this section we review sparse regularisation and how it is applied to radio interferometric imaging.

\subsection{Sparse regularisation}
\label{sec:sparse_regularization}

Consider the ill-posed inverse problem of estimating the image \mbox{$\bm{x} \in \mathbb{R}^{N}$} from measurements $\bm{y}\in \mathbb{C}^{M}$, where the measurements are acquired by the process $\bm{y} = \bm{\mathsf{\Phi}} \bm{x} + \bm{n}$, where the operator $\bm{\mathsf{\Phi}} \in \mathbb{C}^{M\times N}$ models the acquisition system and $\bm{n} \in \mathbb{C}^{M}$ represents noise.  This problem accurately models interferometric imaging, as discussed in more detail in the subsequent sections.  For now, we consider sparse regularisation approaches to solve this general problem.

Sparse regularisation techniques promote sparse solutions when solving ill-posed inverse problems.  Typically, natural signals are sparse in a suitable basis (\textit{e.g.} a Dirac, Fourier, or wavelet basis) or, more generally, in a sparsifying dictionary. The atoms (\textit{cf.} basis functions) of the dictionary \citep{rub10} can be represented by columns of the operator $\bm{\mathsf{\Psi}} \in \mathbb{C}^{N \times D}$, where $N$ is the number of pixels in the image and $D$ is the number of coefficients of the sparse representation, \textit{i.e.}\ $\bm{\alpha} \in \mathbb{C}^D$. The image can then be decomposed into its sparse representation by $\bm{x} = \bm{\mathsf{\Psi}} \bm{\alpha}$. 

A sparse solution to the inverse problem described above can be promoted by imposing a penalty on the number of non-zero coefficients of the sparse representation $\bm{\alpha}$ through the $\ell_0$-norm, where the $\ell_0$-norm $\| \bm{\alpha}\|_{\ell_0}$ is defined as the number of non-zero coefficients of $\bm{\alpha}$.  In principle, the inverse problem can then be solved by minimising the $\ell_0$-norm of the sparse coefficients, subject to a data fidelity constraint:
\begin{equation}
	\min_{\bm{\alpha} \in \mathbb{C}^D}\left\| \bm{\alpha} \right\|_{\ell_0}\quad {\rm subject}\, {\rm to} \quad \left \|\bm{y} - \bm{\mathsf{\Phi}}\bm{\mathsf{\Psi}} \bm{\alpha} \right\|_{\ell_2} \leq \epsilon \, .
\end{equation}
Given the solution to this problem, denoted $\bm{\alpha}^\star$, a recovered image can be synthesised by $\bm{x}^\star = \bm{\mathsf{\Psi}} \bm{\alpha}^\star$.  The solution to this minimisation problem is given by a model that matches the measurements, within error $\epsilon \in \mathbb{R}^+$, while being constructed from a minimal number of coefficients in the sparse representation.  However, this problem cannot be solved in a high dimensional setting because the $\ell_0$-norm is non-differentiable and the minimisation problem is non-convex: it is considered an NP hard problem. 

The closest convex relaxation of the $\ell_0$ problem is the \mbox{$\ell_1$ problem}:
\begin{equation}
	\label{eq:l1_synthesis}
	\min_{{\bm{\alpha}} \in \mathbb{C}^D}\left\| {\bm{\alpha}} \right\|_{\ell_1}\quad {\rm subject}\, {\rm to} \quad \left \|\bm{y} - \bm{\mathsf{\Phi}}\bm{\mathsf{\Psi}} {\bm{\alpha}} \right\|_{\ell_2} \leq \epsilon \, ,
\end{equation}
where the $\ell_p$-norm is defined by $\|\bm{r}\|_{\ell_p} = \left ( \sum_{i}|r_i|^p\right)^{\frac{1}{p}}$ (hence the $\ell_1$-norm is the sum of the absolute value of the components of a vector and the $\ell_2$-norm is the usual Euclidean norm). This $\ell_1$ minimisation problem also promotes sparsity and in some cases exhibits the same solution as the $\ell_0$ problem \citep{candes:2006a,donoho:2006}. Furthermore, since the $\ell_1$ minimisation problem is a convex problem it can be solved using efficient convex optimisation algorithms \citep[\textit{e.g.}][]{combettes:2011}.

The problem defined by Eq.~\ref{eq:l1_synthesis} is proposed in the standard synthesis setting, where one recovers the coefficients $\bm{\alpha}$ and synthesises the recovered image by $\bm{x}=\bm{\mathsf{\Psi}}\bm{\alpha}$. Alternatively, we can propose the problem in the analysis setting using the adjoint wavelet transform $\bm{\mathsf{\Psi}}^\dagger$:
\begin{equation}
	\label{eq:l1_analysis}
	\min_{{\bm{x}} \in \mathbb{R}^N}\big\| \bm{\mathsf{\Psi}}^\dagger {\bm{x}} \big\|_{\ell_1}\quad {\rm subject}\, {\rm to} \quad \left \|\bm{y} - \bm{\mathsf{\Phi}} {\bm{x}} \right\|_{\ell_2} \leq \epsilon\, ,
\end{equation}
where one recovers the image $\bm{x}$ directly, while still imposing sparsity in some sparse representation. { When the sparsifying operator $\bm{\mathsf{\Psi}}$ is an orthogonal basis the solutions of the synthesis and analysis problems are identical.}  However, for an overcomplete dictionary the solutions are very different and the analysis setting has been shown to perform very well in practice \citep[\textit{e.g.}][]{car12, car12a}.  Moreover, reweighted schemes to better approximate the solution of the $\ell_0$ problem by solving a sequence of $\ell_1$ problems can also be considered \citep{candes:2007,car12, car12a}. While these approaches can further improve the quality of the reconstructed image we do not consider them further here.

Additionally, sparse regularisation problems allow extra constraints to be imposed, such as a real and positive valued image, which is the case for { total intensity (Stokes I)} radio interferometric observations. { However, the positivity and real valued image constraints may be removed for polarimetric imaging, such as linear polarisation or the Stokes parameters. Complex valued linear polarisation reconstructions of ${\rm P = Q + iU}$ can also be performed in principle and will be rotationally invariant for rotations in ${\rm P}$ \citep{pra16}.} 

\subsection{Radio interferometric measurement operator}
\label{sec:measurement_operator}
In solving sparse regularisation problems, the measurement operator is required to compare how close the reconstructed model matches the measured data. How close the measurement operator matches the true measurement process will have an impact on reconstruction quality. 

In the context of radio astronomy, the measurement process is given by Eq.~\ref{eq:planar_measurement_equation}.  We assume co-planar baselines and a small field-field of view here; we do not consider direction-dependent effects in the measurement operator, although they can nevertheless be modelled in the framework presented \citep{wia09b, wol13}. In the compressive sensing setting, the measurements $\bm{y} \in \mathbb{C}^M$ denote the visibilities $y_i = \mathcal{V}(u_i, v_i)$ and the image \mbox{$\bm{x} \in \mathbb{R}^N$} denotes the sky brightness distribution $x_p = \mathcal{I}_\lambda (l_p, m_p)$ (for $i=1,\ldots,M$ and $p=1,\ldots,N$). The measurement operator $\bm{\mathsf{\Phi}} \in \mathbb{C}^{M \times N}$ specifies a discrete representation of Eq.~\ref{eq:planar_measurement_equation}.
Ideally, $\bm{\mathsf{\Phi}}$ would represent a direct Fourier transform from the $N$ pixels of the image to the $M$ non-uniformly spaced visibilities. However, this would require $\mathcal{O}\left(MN\right)$ computations. Consequently, a direct Fourier transform of the visibilities is not possible for the settings experienced in practice, where a single observation may be comprised of very large numbers of visibilities and high-resolution reconstructed images are required.

Alternatively, it is possible to approximate a direct Fourier transform. One can first interpolate the visibilities onto a regularly spaced grid, which requires order $\mathcal{O}(M)$ operations. Then, it is possible to take advantage of the Fast Fourier Transform (FFT), which requires order $\mathcal{O}\left( N \log N\right)$ operations. This approach requires considerably fewer computations than the direct Fourier transform \citep{bri99}, rendering a non-uniform Fourier transform computationally feasible for very large observational data-sets, but it is an approximation.  This approximation is the standard approach considered in radio astronomy.

\begin{figure}
		\center
		\includegraphics[width=0.5\textwidth]{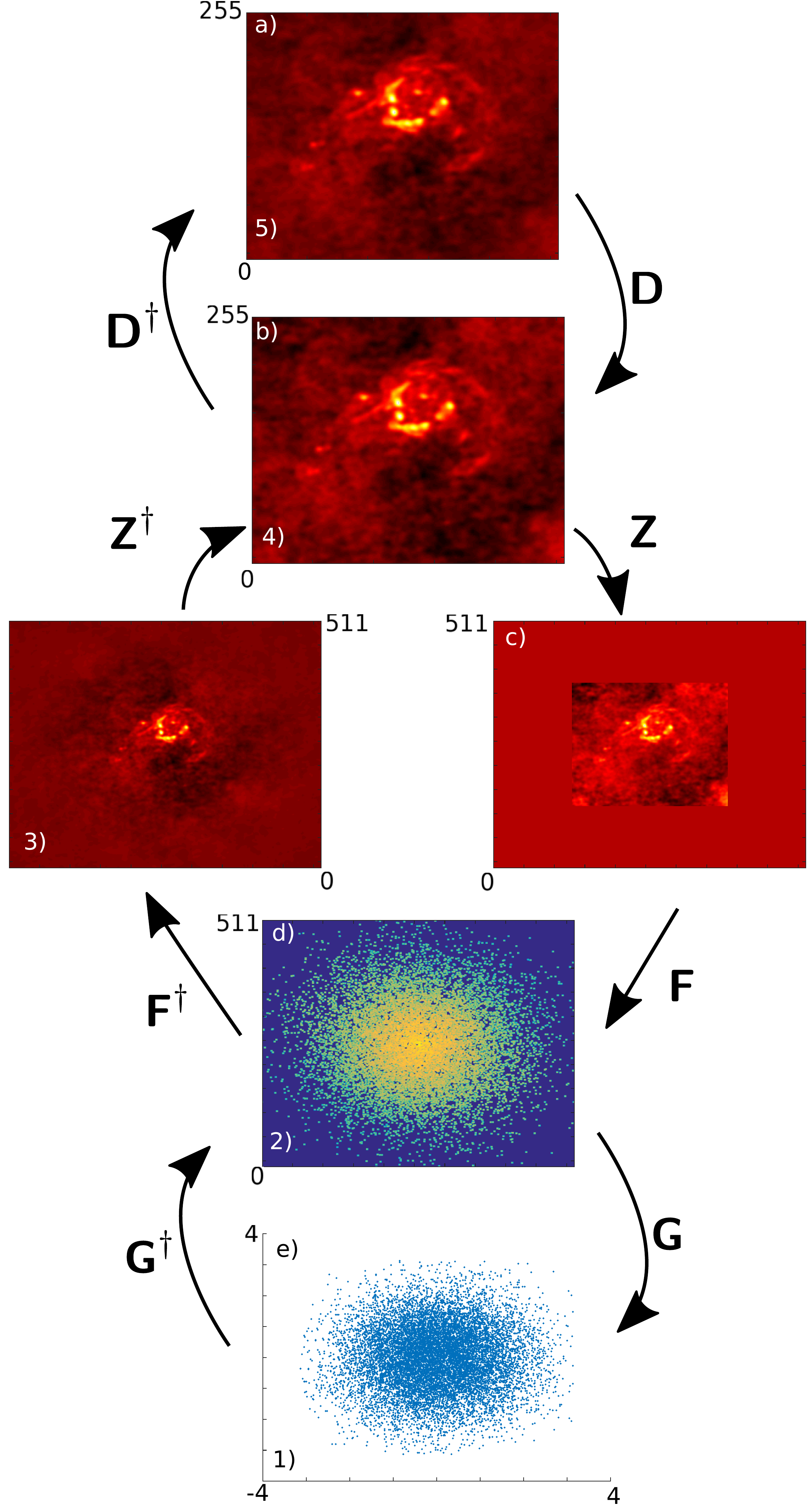}
		\caption{Representation of the application of the forward and adjoint measurement operator. The labels a) to e) represent the process of the forward measurement operator, while numbers 1) to 5) represent the process of the adjoint operator. The measurement operator consists of the following steps: a) observed image; b) image is corrected for degridding; c) image is zero-padded to twice the field of view; d) Image is Fourier transformed; e) Fourier coefficients are convolved to continuous points off of the grid. The adjoint measurement operator consists of the following steps: 1) Fourier coefficients in a continuous plane; 2) Fourier coefficients are gridded onto an oversampled grid; 3) image from the transformed Fourier coefficients; 4) image cutout; 5) image corrected for the gridding.}
		\label{fig:measurement_operator}
\end{figure}

The standard radio interferometric measurement operator $\bm{\mathsf{\Phi}}$ can be written as a series of linear operators:
\begin{equation}
\label{eq:measurement_operator}
\bm{\mathsf{\Phi}} = \bm{\mathsf{W}}\bm{\mathsf{G}}\bm{\mathsf{F}}\bm{\mathsf{Z}}\bm{\mathsf{D}}\bm{\mathsf{B}} \, ,
\end{equation}
where $\bm{\mathsf{B}} \in \mathbb{C}^{N \times N}$ is the primary beam of telescope, $\bm{\mathsf{D}} \in \mathbb{C}^{N \times N}$ is a gridding correction operator that scales the image to correct for the interpolation convolution kernel, $\bm{\mathsf{Z}} \in \mathbb{C}^{\alpha^2 N \times N}$ is a zero-padding operator that provides oversampling by factor $\alpha$ in each dimension of the Fourier domain, $\bm{\mathsf{F}} \in \mathbb{C}^{\alpha^2 N \times \alpha^2 N}$ is a fast Fourier transform operator, $\bm{\mathsf{G}} \in \mathbb{C}^{M \times \alpha^2 N}$ is a convolutional interpolation operator that uses a convolution kernel to interpolate visibilities from Fourier coefficients on a regular grid to Fourier components in the continuous Fourier plane, and $\bm{\mathsf{W}} \in \mathbb{C}^{M \times M}$ weights the measurements according to their error. Alternatively, it is possible to include the weighting $\bm{\mathsf{W}}$ by weighting the $\ell_2$-norm directly. A diagram of the process of applying the measurement operator $\bm{\mathsf{\Phi}}$ and its adjoint $\bm{\mathsf{\Phi}}^\dagger$ is shown in Figure~\ref{fig:measurement_operator}. { Since the weights are applied in the measurement operator, it is necessary to also weight the measurements, \textit{i.e.} ${\bm y} \to \bm{\mathsf{W}}{\bm y}$.}

\subsection{Radio interferometric imaging with PURIFY}
To apply compressive sensing techniques to radio interferometry, one needs to pose the sparse regularisation problems in Section~\ref{sec:sparse_regularization} and then solve them using the measurement operator of Section~\ref{sec:measurement_operator}. The software package PURIFY has been designed and written for this purpose.

The first public version of PURIFY was written in C and solved the problems described in \citet{car14}, where it was shown on simulations to produce higher fidelity reconstructed images than standard radio interferometric imaging methods. To solve $\ell_1$ minimisation problems, PURIFY calls the Sparse \mbox{OPTimisation} (SOPT) software package \citep{car12,car12a}.  This first version of PURIFY used the simultaneous-direction method of multipliers (SDMM) algorithm \citep{car14}.  Recently, new algorithms have been developed for radio interferometry imaging by \cite{ono16}, including the proximal alternating direction method of multipliers (P-ADMM) and primal dual algorithms, which have numerous advantages for the analysis of very large data-sets (see \citealt{ono16} for further discussion).

New versions of PURIFY and SOPT have been released to coincide with the current article. Both PURIFY and SOPT have been completely redesigned and rewritten in C++11 to work on Linux and Mac operating systems. The Eigen\footnote{\url{http://eigen.tuxfamily.org}} library is used for matrix and array manipulation \citep{eigen} and casacore\footnote{\url{http://casacore.github.io/casacore}} is used to read observational data in the form of measurement sets \citep{mcm07}.  SOPT is not only useful for interferometric imaging: it is a general purpose code { for solving sparse} regularisation problems and can be used to solve a variety of problems.  The first version of PURIFY was limited to measurement operators based on Gaussian kernels for convolutional gridding. The new version of PURIFY, however, supports numerous kernels, including the state-of-the-art kernels discussed in the literature \citep[\textit{e.g.}][]{fes03}, as described in Section~\ref{sec:gridding}.  Additionally, the P-ADMM algorithm of \citet{ono16} has been implemented in PURIFY and SOPT. Implementation of the primal dual algorithm of \citet{ono16} into PURIFY and SOPT is part of future work. The primal dual algorithm achieves greater flexibility, in terms of memory requirements and computational burden per iteration, by using full splitting and randomised updates. All results presented in this article are obtained with the P-ADMM algorithm, solving the analysis problem of Eq.~\ref{eq:l1_analysis}, with an additional positivity constraint { (however, it is possible to remove the positivity or reality constraints)}.  While the development of fully distributed implementations of the algorithms supported by PURIFY and SOPT is ongoing, current versions are parallelised with OpenMP, so that the gridding, degridding, and FFT calculations can be performed efficiently.  The latest versions of PURIFY\footnote{\url{http://basp-group.github.io/purify}} and SOPT\footnote{\url{http://basp-group.github.io/sopt}} are now publicly available.

\section{Convolutional gridding and degridding}
\label{sec:gridding}
The fidelity of reconstructed radio interferometric images depends not only on the technique used to solve the resulting inverse problem but also on the accuracy with which the measurement operator models the measurement process. Ideally, the measurement operator would match the measurement process exactly. However, this is not possible due to the computational time required for a direct Fourier transform. We are forced to use a measurement operator that interpolates the visibilities onto and off of a regular grid through the operator $\bm{\mathsf{G}}$, so that we may apply an FFT $\bm{\mathsf{F}}$ to regularly spaced data. Interpolation is typically performed by convolution with a suitable kernel, which then determines the convolutional degridding operator $\bm{\mathsf{G}}$. Several interpolating convolutional kernels have been suggested in the literature; we introduce a subset of these kernels in this section.  
The choice of convolution kernel affects the quality of the image, through aliasing error, and total computation time, through the support size of the kernel. Ideally, a convolution kernel will have minimal support while maximally suppressing aliasing error, allowing high quality images to be reconstructed in minimal computation time.

\subsection{Degridding}
To replicate the measurement process, Fourier coefficients need to be interpolated off of the FFT grid, \emph{i.e.} they need to be \emph{degridded}. An ideal interpolation that does not change the content of an image is the well-known (Shannon) Sinc interpolation \citep{whi15,sha49}, where a continuous band-limited image can be exactly reconstructed from the discrete Nyquist sampled signal. Sinc interpolation can also be considered in the context of interpolating the Fourier domain, which is exact for a space-limited image.  In practice, Sinc interpolation in this context can be performed by zero-padding the image domain, which up-samples the Fourier domain via Sinc interpolation.

In the context of degridding, a Sinc interpolation kernel preserves the image and frequency content of the signal when the image has a limited field of view. However, Sinc interpolation is computationally expensive because the Sinc kernel does not have finite support in harmonic space. A computationally inexpensive method, due to its small support, is to interpolate in the Fourier domain using the nearest neighbour grid point.  Nearest neighbour interpolation in the Fourier domain corresponds to convolving with a Box kernel, which corresponds to multiplying with a Sinc function in the image domain.  Since the Sinc function has infinite support in the image domain, this introduces artefacts known as aliasing error. The Sinc and nearest-neighbour approaches to interpolating visibilities represent the two extreme cases.  

We require kernels with small support in harmonic space (so they are computationally efficient) and small support in image space (to suppress aliasing error). However, the uncertainty principle means there is a fundamental limit on how localised a function can be in both harmonic space and image space. In practice, we seek a trade-off between the two extremes, so that the support of the kernel in harmonic space is not so large as to be computationally expensive, while the support in image space is also well-localised to suppress aliasing error.  

Since the interpolation is performed by a convolution, it is necessary to correct for this operation, which can be achieved by multiplication in the image domain with an appropriate window. Furthermore, interpolation accuracy can be increased by zero-padding in the image domain to up-sample the Fourier domain.
The process of degridding therefore starts by scaling the image by the diagonal operator $\bm{\mathsf{D}}$, which preemptively corrects for the interpolation kernel of $\bm{\mathsf{G}}$. This correction is calculated from the reciprocal of the inverse Fourier transform of the interpolation kernel. The image is then zero-padded using the zero-padding operator $\bm{\mathsf{Z}}$ which up-samples harmonic space. An FFT is applied to obtain an up-sampled Fourier grid using the operator $\bm{\mathsf{F}}$.  The model measurements are then interpolated off of the grid using the circular convolution operator $\bm{\mathsf{G}}$. The explicit construction of $\bm{\mathsf{G}}$ is discussed in Section~\ref{sec:kernels}.

\subsection{Gridding}
Most image reconstruction algorithms in radio astronomy require going  both backward and forward between the image and measurement domain. Typically, mapping from the measurement domain to the image domain is performed by the adjoint of the measurement operator, since the measurement operator does not have a defined inverse, given by
\begin{equation}
	\bm{\mathsf{\Phi}}^\dagger = \bm{\mathsf{B}}^\dagger\bm{\mathsf{D}}^\dagger\bm{\mathsf{Z}}^\dagger\bm{\mathsf{F}}^\dagger\bm{\mathsf{G}}^\dagger\bm{\mathsf{W}}^\dagger \, .
\end{equation}
\emph{Gridding} can be considered the reverse process of degridding. Mathematically, the gridding operator is the adjoint of the degridding operator and is performed by application of $\bm{\mathsf{G}}^\dagger$.  The full adjoint measurement operator consists of the following operations.  First the weighting $\bm{\mathsf{W}}^\dagger=\bm{\mathsf{W}}$ is applied, before the visibilities are interpolated onto an up-sampled Fourier grid using $\bm{\mathsf{G}}^\dagger$. Then an inverse FFT is performed by $\bm{\mathsf{F}}^\dagger$ to produce an image. The image is cropped to the desired field of view using $\bm{\mathsf{Z}}^\dagger$, and the convolution is corrected by $\bm{\mathsf{D}}^\dagger$. Lastly, the adjoint of the primary beam $\bm{\mathsf{B}}^\dagger$ is applied.

A consequence of interpolating the visibilities onto a grid is that the signal is now represented via a Fourier series rather than a Fourier transform. This means the imaged region has periodic boundary conditions.  In the case of a radio interferometer, the visibilities can contain information over the entire sky, and the signal may not end at the boundaries of the imaged region. In this case, the interpolation kernel is used to apodize aliasing error, where structure from outside the boundaries of the imaged region is folded back in \citep{bri99}.

\subsection{Aliasing error}

In the case where the convolution kernel does not sufficiently attenuate the image outside the imaged region, the periodicity of the image will cause features from outside the imaged region to fold into the image. Two ways to minimise aliasing error are to either image a wider field of view, so that the primary beam of the telescope naturally attenuates structures outside the field of view, or to choose a convolution kernel that attenuates the aliasing error sufficiently.

An ideal convolution kernel would set the image to zero outside the imaged field of view, which would eliminate aliasing error. This can be done with a Sinc convolution kernel, which is computationally expensive. An inexpensive kernel, like a Box kernel, is highly delocalised in the image domain, so does not suppress structure outside the imaged field of view from being folded back in.

To increase image quality and computational performance, a convolution kernel needs a minimal support in harmonic space while attenuating the image outside the field of view. Any attenuation within the imaged field of view is corrected for by $\bm{\mathsf{D}}$, calculated from the Fourier transform of the gridding kernel.

If the gridding kernel apodizes the image domain strongly within the gridded field of view, correcting by $\bm{\mathsf{D}}$ will induce numerical errors \citep{sch80}. This means that while the suppression due to the gridding kernel can reduce aliasing error, correcting for it has the potential to cause numerical error.

\subsection{Interpolation kernels}
\label{sec:kernels}
Next, we introduce the convolution kernels used in this work. The width (support) of the gridding kernel $J$ is given in units of grid cells. The oversampling ratio in each dimension is denoted by $\alpha$.

The degridding matrix is a circular convolution matrix that interpolates the measurements off of the discrete Fourier grid onto the continuous Fourier plane. The convolution can be seen as a weighted average of the nearest neighbour grid points. The interpolation kernel determines the weighting of each grid point. Weighting is maximum at the location of the measurement and typically decreases in value when the grid points are further from the measurement location.

In 1-D Fourier space, the degridding matrix $\bm{\mathsf{G}}$ is constructed from a kernel $d(u)$ by \citep{fes03}
\begin{equation}
	\bm{\mathsf{G}}_{i, \{k_i + j\}_K} = d(u_i - (k_i + j)) \, ,
\end{equation}
where $i$ is the index of the measurement $y_i$, $k_i$ is the closest integer to visibility coordinate $u_i - J/2$ (in units of pixels), and $j = 1 \dots J$ are the possible non-zero entries of the kernel. The modulo-$K$ function is denoted by $\{ \cdot \}_K$, where $K = \alpha \sqrt{N}$ is the dimension of the Fourier grid in 1-D (for notational sake, the 2-D Fourier grid is comprised of $N=\sqrt{N}\times \sqrt{N}$ samples).

The diagonal convolution correction operator $\bm{\mathsf{D}}$ can be calculated in a similar way:
\begin{equation}
	\bm{\mathsf{D}}_{i,i} = s\left(\frac{i}{K} - \frac{1}{2}\right)\, ,
\end{equation}
	where $s(x)$ is the reciprocal of the inverse Fourier transform of $d(u)$. In practice, $\bm{\mathsf{D}}$ can be computed numerically from $\bm{\mathsf{G}}$ or analytically if the inverse Fourier transform of the convolution kernel is tractable.

\subsubsection{Sinc}
	The Sinc convolution kernel is ideal when its infinite support is considered. This convolution kernel can be written as \citep{gre79,sch78a}
	\begin{equation}
		d(u) = \left( \frac{u \pi}{N} \right)^{-1}\sin\left( \frac{u \pi}{N} \right)\, .
	\end{equation}
	The convolution correction is
	\begin{equation}
		s(x) = \begin{cases}
		\frac{1}{N}, &\text{ if } |x| \leq \frac{N}{2}\\
		0, & {\rm otherwise}
	\end{cases}.
	\end{equation}
	The advantage of the Sinc convolution kernel is that it corresponds to multiplication by a Box function in the image domain, which bounds the signal at the edges of the imaged region. Consequently, there is close to no aliasing error.

	\subsubsection{Box}
	The Box function is fast to compute since it is localised in harmonic space, but it does not suppress aliasing error effectively. This kernel has the form \citep{gre79,sch78a}:
	\begin{equation}
		d(u) = \begin{cases}
		\frac{1}{J}, &\text{ if } \vert u \vert \leq \frac{J}{2}\\
		0, & {\rm otherwise}
	\end{cases}.
	\end{equation}
	The Fourier transform of the Box function is the Sinc function, so the convolution correction reads
	\begin{equation}
		s(x) = \left [ \frac{\sin\left( x J \pi \right)}{x J \pi} \right]^{-1} \, .
	\end{equation}
	The Sinc function is not bounded by the edges of the image, and the sidelobes of the Sinc function can cause large aliasing error. This is why the Box function is far from ideal, even if it is fast to compute.

	\subsubsection{Gaussian} 
	The Gaussian kernel is moderately well-localised in both image and Fourier space and takes the form:
	\begin{equation}
		d(u) = {\rm e}^{-\frac{u^2}{2\sigma^2}}\, .
	\end{equation}
	The gridding correction is calculated by the Fourier transform and also takes the form of a Gaussian:
	\begin{equation}
		s(x) = \left[{\frac{\pi}{2\sigma^2}}\right]^{-1/2}{\rm e}^{2x^2\pi^2\sigma^2}\, .
	\end{equation}
	An optimal choice for $\sigma$ as a function of the support size $J$ was found in the work of \citet{fes03}, where it was shown that $\sigma = 0.31 J^{0.52}$ works better than using the typical value $\sigma = 1$. In the early years of radio astronomy, in the 1970's, the Gaussian kernel was used for convolutional gridding \citep{tho08}.

	\subsubsection{Prolate spheroidal wavefunction} Prolate spheroidal wavefunctions (PSWFs) do not have an explicit analytic form but there are several ways of characterising them \citep{str35,sle61,lan61,lan62}. The most useful way to characterise PSWFs is in terms of energy concentration. PSWFs are bandlimited functions that maximise the energy concentration in a given interval, by finding the function $f$ that maximises the ratio
	\begin{equation}
		\frac{\int _{-\tau}^\tau \vert f(t)\vert^2{\rm d}t}{\int _{-\infty}^\infty \vert f(t)\vert^2{\rm d}t}	\, ,
	\end{equation}
	for an interval $[-\tau, \tau]$. For a convolution kernel, this is an ideal property since we want the convolution kernel to have minimal support in the Fourier domain and to have a maximal amount of energy concentrated over the imaged region in the image domain. This allows one to have minimal support in the Fourier domain while maximally suppressing aliasing error in the image domain.

	The standard choice of PSWFs in radio astronomy are a modified version, where more energy is weighted towards the centre of the image, since typically this is the scientific region of interest. The standard choice of weighted PSWFs are described in the work of \citet{sch84,sch80}. The convolution kernel is given by
	\begin{equation}
		d(u) = \vert 1-\eta^2(u)\vert^\kappa\psi_\kappa(\pi J/2, \eta(u))\, ,
	\end{equation}
	where $\eta(u) = 2 u /J$, $\kappa$ is a parameter that varies the weighting, and $\psi_\kappa$ is a zero order PSWF that can be calculated using a rational approximation:
	\begin{equation}
		\psi_\kappa(\pi J/2, \eta) = \frac{\sum_{k=0}^n p_k(\eta^2 - \eta_2^2)^k}{\sum_{k=0}^d q_k(\eta^2 - \eta_2^2)^k}\, ,
	\end{equation}
	where the $p_k$ and $q_k$ polynomial coefficients are specified in \cite{sch80,sch84}. The case of $\kappa = 0$ reduces to an unweighted PSWF. In this work, we use the polynomial coefficients for a support of $J = 6$ and $\kappa = 1$, the standard used in the radio interferometric imaging packages MIRIAD\footnote{\url{http://www.atnf.csiro.au/computing/software/miriad/}} \citep{sau96} and Astronomical Image Processing System (AIPS; \citealt{gre98})\footnote{\url{http://www.aips.nrao.edu/index.shtml}}.
	The correction is provided by \cite{sch84}:
	\begin{equation}
		s(x) \approx \frac{1}{  \psi_0(\pi J/2, 2 x)}\, .
	\end{equation}

	\subsubsection{Kaiser-Bessel} 
	Kaiser-Bessel functions are another useful form of convolution kernel.
	The zeroth order Kaiser-Bessel function can be expressed as
	\begin{equation}
		d(u) = \frac{I_0\left(\beta \sqrt{1 -\left(\frac{2 u}{J}\right)^2}\right)}{I_0(\beta)}\, ,
	\end{equation}
	where $J$ is the support, $I_0$ is the zeroth order modified Bessel function of the first kind, and $\beta$ determines the spread of the Kaiser-Bessel function \citep{jac91,fes03}.
	The gridding correction is calculated from the Fourier transform, yielding \citep{jac91,fes03}:
	\begin{equation}
		s(x) = \left[\frac{\sin\left(\sqrt{\pi^2 x^2 J^2 - \beta^2}\right)}{\sqrt{\pi^2 x^2 J^2 - \beta^2}}\right]^{-1}\, .
	\end{equation}
	An optimal choice for $\beta$ as a function of the support size $J$ was found in the work of \citet{fes03}, where it was shown that for $\beta = 2.34 J$ the Kaiser-Bessel kernel performs similarly to the optimal min-max kernel considered in \citet{fes03}.  In \cite{gre79}, it is suggested that the zeroth order Kaiser-Bessel functions perform similarly to the zeroth-order PSWFs, which is consistent with the results of \cite{jac91}.  Kaiser-Bessel functions, however, have the advantage that they have an analytic expression that can be evaluated easily and accurately. Note that Kaiser-Bessel functions are the standard choice of interpolation kernel in the interferometric imaging package WSCLEAN\footnote{\url{https://sourceforge.net/projects/wsclean/}} \citep{off14}.

\section{Simulations}
\label{sec:simulations}
In the previous section we described how the measurement operator $\bm{\mathsf{\Phi}}$ approximates a direct Fourier transform. If this approximation is inaccurate, it will introduce error when recovering interferometric images. The choice of the interpolation kernel will therefore have an impact on reconstruction quality. In this section we perform simulations to assess the performance of different convolution kernels, using the P-ADMM algorithm \citep{ono16} implemented in the latest release of PURIFY to recover images in the analysis framework, with an additional positivity constraint.

\subsection{Simulations}
To assess the impact that the interpolation kernel has on image reconstruction with PURIFY, we perform quality tests using simulated measurements. We compare the signal to noise ratio (SNR) of the reconstructed image with the ground truth image, reconstructing with different $uv$-coverages and different interpolation kernels. Note that we cannot replicate all of the complexities of the real observational setting with simple simulations. For example our simulated observations do not include contributes from sources outside the field of view. Nevertheless, simulations where the ground truth image is known are useful for a partial assessment of the performance of different convolution kernels.

To ensure the simulated measurements do not limit the reconstruction quality, a high quality `ground truth' measurement operator is applied to test images of HII emission of M31 and of 30 Doradus (30Dor). The Kaiser-Bessel kernel with a support of \mbox{$8\times8$} pixels and an oversampling ratio of $\alpha = 2$ is used for the ground truth measurement operator. The Kaiser-Bessel kernel typically requires only a small support, so choosing a support of \mbox{$8\times8$} provides an accurate measurement model \citep{fes03}.

\begin{figure*}
		\includegraphics[width=0.9\textwidth]{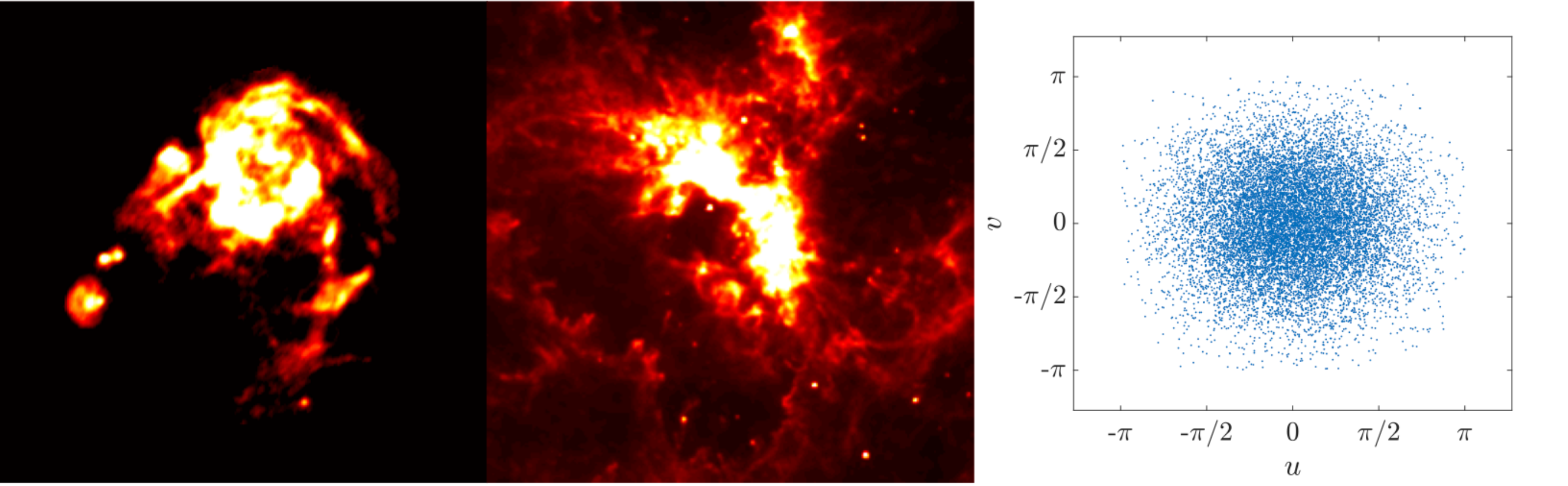}
		\caption{Ground truth images of M31 (left) and 30Dor
(middle) used in simulations (of size 256 $\times$ 256). An example of a variable density visibility coverage in the Fourier plane, normalised to a domain of $\pm\pi$ (right). To generate a simulated observation, the measurement operator was applied to a ground truth image. Each simulation has added thermal noise and a random variable density coverage in the Fourier plane. The reconstruction quality was evaluated as a function of the number of Fourier components measured. The SNR was averaged over ten random coverages, with error bar given by the standard deviation (see Figure \ref{fig:M31_30Dor_SNR_plots}).}
		\label{fig:simulation_images}
\end{figure*}

We calculate the average SNR for reconstructing M31 and 30Dor from $M$ visibilities, in a way that does not depend on a specific $uv$-coverage. The $uv$-coverages are randomly generated to follow a Gaussian variable sampling density with a standard deviation of $\pm\pi/3$ in the $uv$-plane, where the $uv$-plane has been normalised to a maximum height and width of $\pm \pi$. Ten sample $uv$-coverages were generated using $M$ visibilities. The average SNR of a reconstruction from $M$ visibilities was calculated using the ten sample $uv$-coverages. The standard deviation is used to estimate the spread of the SNRs of the reconstructed images.  The test images of M31 and 30Dor and a sample $uv$-coverage are shown in Figure~\ref{fig:simulation_images}.

Gaussian noise was added to the simulated visibilities. The input SNR (ISNR) of the measurements was chosen to be 30 dB. The ISNR can be used to calculate the standard deviation of the Gaussian distribution of noise \citep{car14}:
\begin{equation}
	\sigma_{n} = \frac{\left\| \bm{y}_0 \right\|_{\ell_2}}{\sqrt{M}} \times 10^{-\frac{\rm ISNR}{20}}\, ,
\end{equation}
where $\bm{y}_0$ are the ground truth visibilities, $M$ is the number of visibilities, and ISNR is measured in dB.

The noise is assumed to be Gaussian and independently and identically distributed, which allows the use of the $\chi^2$ distribution to estimate the bound $\epsilon$ for the $\ell_2$-norm \citep{car14}:
\begin{equation}
	\epsilon^2 = (2M + 2\sqrt{4M})\frac{\sigma_{n}^2}{2}\, ,
\end{equation}
where for these tests we set $\epsilon^2$ to two standard deviations above the mean of the $\chi^2$ distribution. 
Following the work of \citet{car14}, we calculate the SNR from the relation 
\begin{equation}
	{\rm SNR} = 20 \log_{10}\left [ \frac{\left\| \bm{x} \right\|_{\ell_2}}{\left\| \bm{x} - \bm{x^\star} \right\|_{\ell_2}}\right]\, ,
\end{equation}
where $\bm{x}$ is the ground truth image and $\bm{x^\star}$ is the reconstructed image.

We solve the $\ell_1$ problem in the analysis setting (Eq.~\ref{eq:l1_analysis}), using P-ADMM.  For the P-ADMM step size $\gamma$, we use the fixed value of
\begin{equation}
	\gamma = \beta \|\bm{\mathsf{\Psi^\dagger}} \bm{\mathsf{\Phi^\dagger}} \bm{y}_0 \|_{\ell_\infty} \, ,
\end{equation}
with $\beta = 10^{-3}$, as recommended in \citet{car14} and \citet{ono16}, where $\| \bm{\mathsf{\Psi^\dagger}} \bm{\mathsf{\Phi^\dagger}} \bm{y}_0 \|_{\ell_\infty}$ returns the maximum coefficient of the measurements in the wavelet representation.
The reconstructions were solved by assuming sparsity in the SARA wavelet dictionary, which includes a Dirac (\textit{i.e.}\ point source) basis and { Daubechies} wavelets 1 to 8 \citep{car12,car12a}.  Note that re-weighting is not considered.
In these simulations, \mbox{P-ADMM} is stopped when the data fidelity constraint is satisfied and the relative difference in the model image between iterations is less than $10^{-3}$. Each reconstruction was run for a maximum of 100 iterations.

\begin{figure}
    \includegraphics[width=1\linewidth]{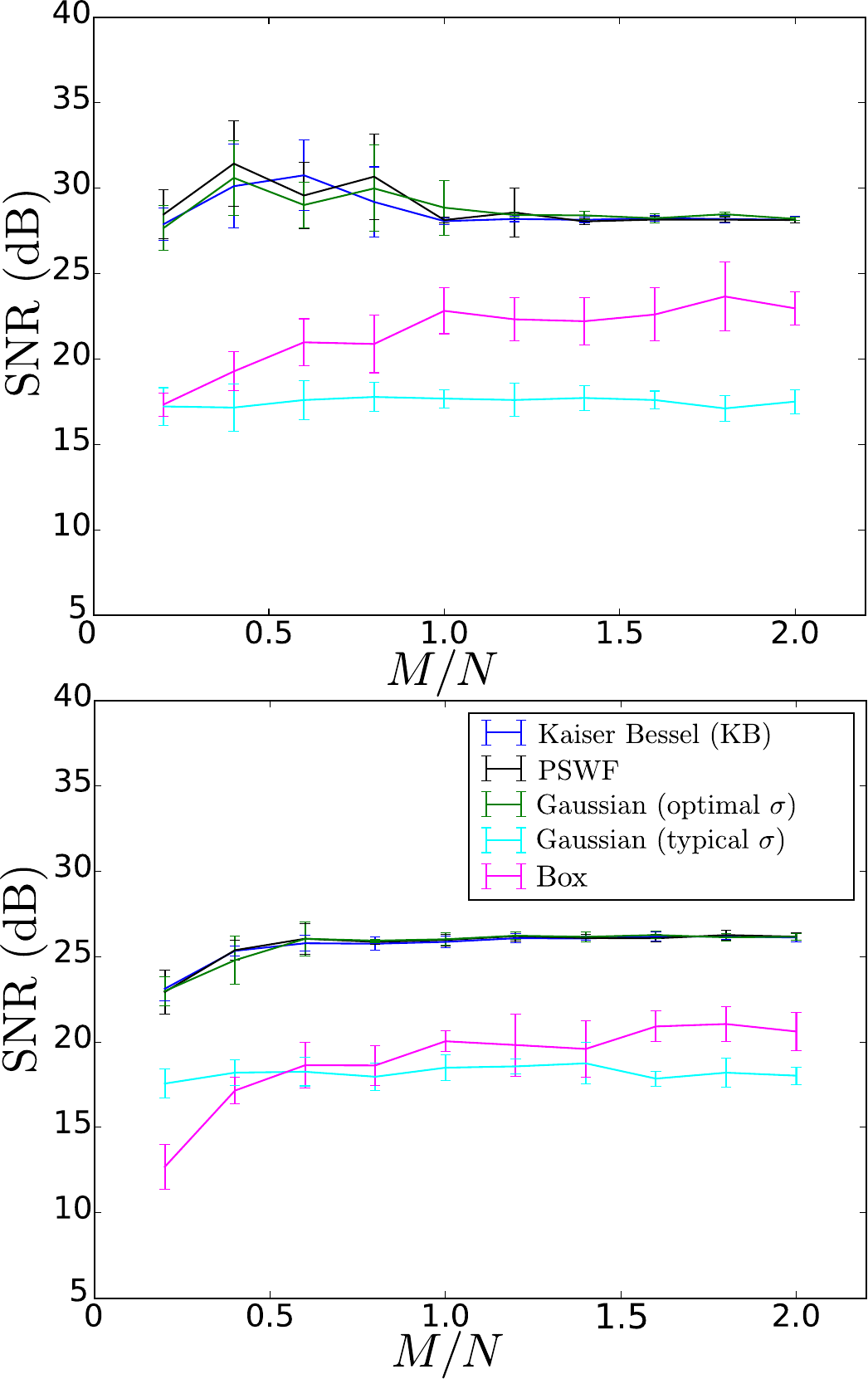}
		\caption{The top and bottom plots of the SNR of the reconstructions of M31 and 30Dor respectively, with an input SNR of 30dB. $M/N$ is the ratio of measurements to pixels. Kaiser-Bessel and optimised Gaussian kernels can perform as well as the PSWF. Furthermore, choosing a bad choice of kernel, like a Box function or a Gaussian kernel with a typical $\sigma$, limits the possible quality of the reconstruction.}
		\label{fig:M31_30Dor_SNR_plots}
\end{figure}

\begin{figure*}
			\begin{minipage}{0.3\textwidth}
    \includegraphics[width=1.15\linewidth]{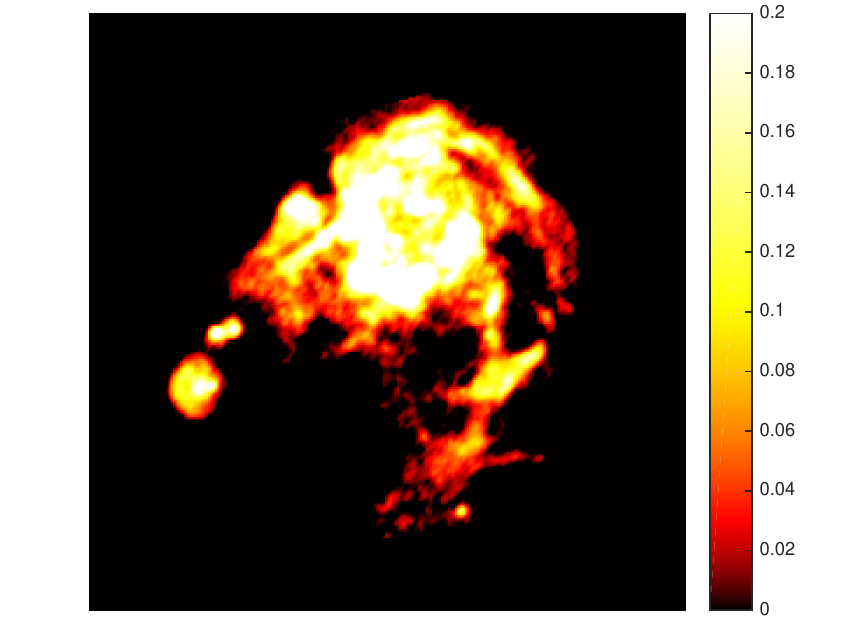}
    \end{minipage}
    \hspace{.5cm} % note: no blank line here
    \begin{minipage}{0.3\textwidth}
    \includegraphics[width=1.15\linewidth]{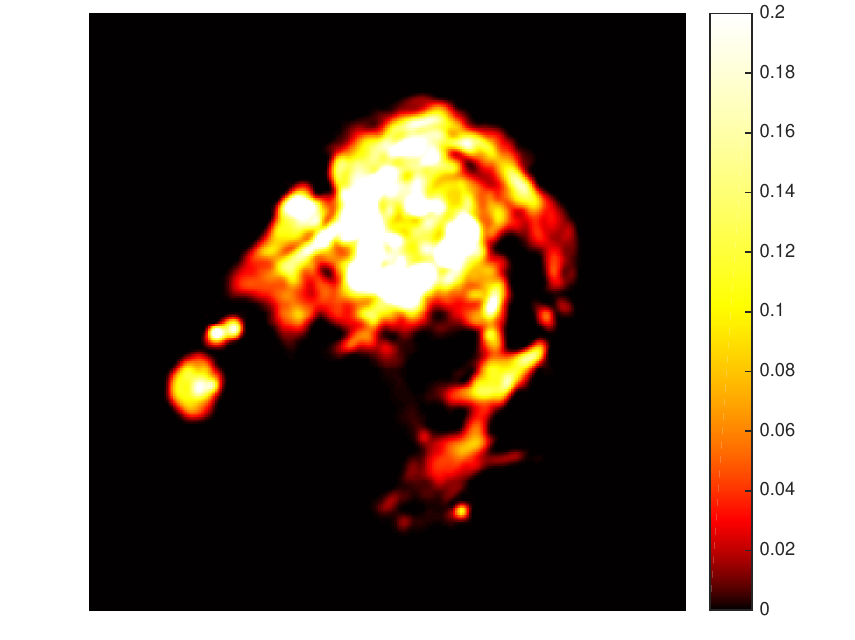}
    \end{minipage}
   \hspace{.5cm} % note: no blank line here
    \begin{minipage}{0.3\textwidth}
    \includegraphics[width=1.15\linewidth]{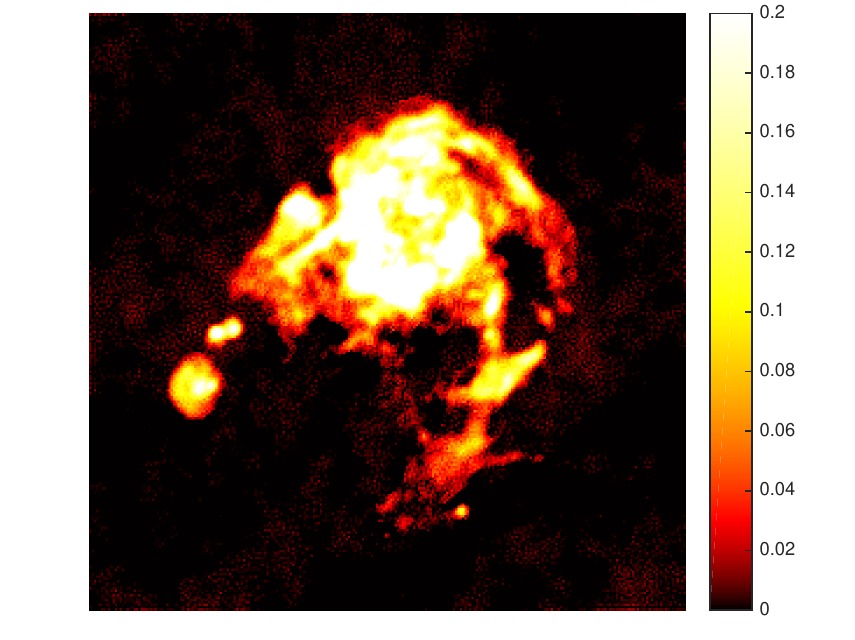}
    \end{minipage}

   \vspace*{0.01cm} % vertical separation
	
	\begin{minipage}{0.3\textwidth}
    \includegraphics[width=1.15\linewidth]{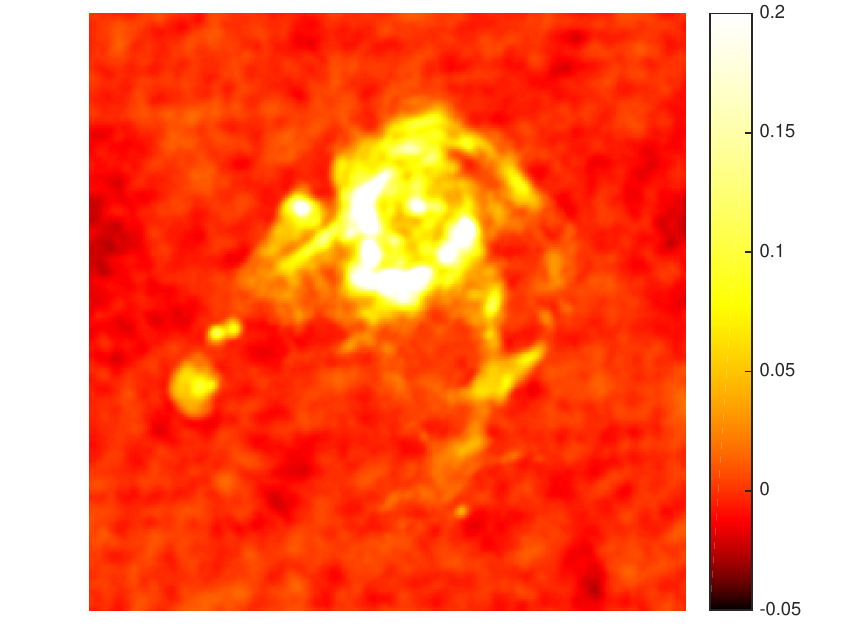}
    \end{minipage}
    \hspace{.5cm} % note: no blank line here
    \begin{minipage}{0.3\textwidth}
    \includegraphics[width=1.15\linewidth]{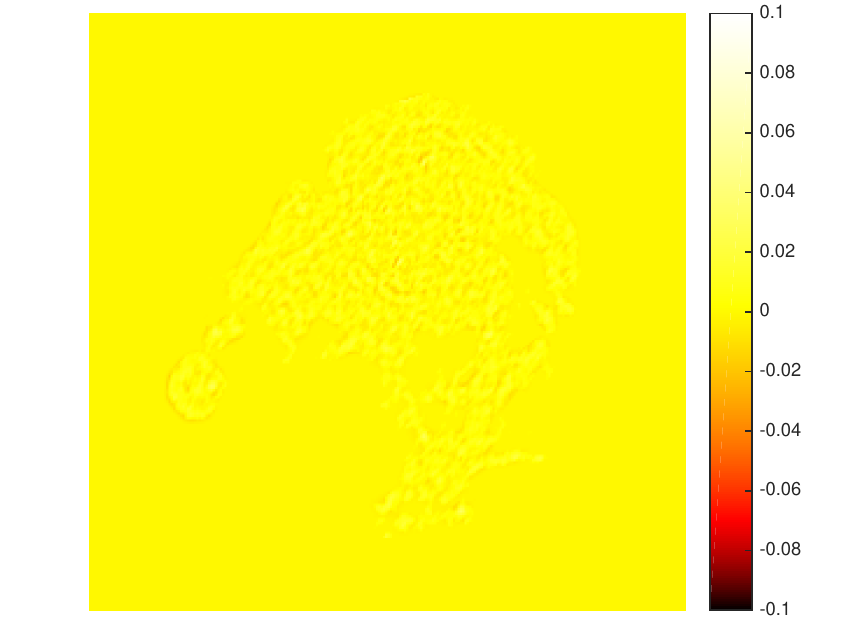}
    \end{minipage}
   \hspace{.5cm} % note: no blank line here
    \begin{minipage}{0.3\textwidth}
    \includegraphics[width=1.15\linewidth]{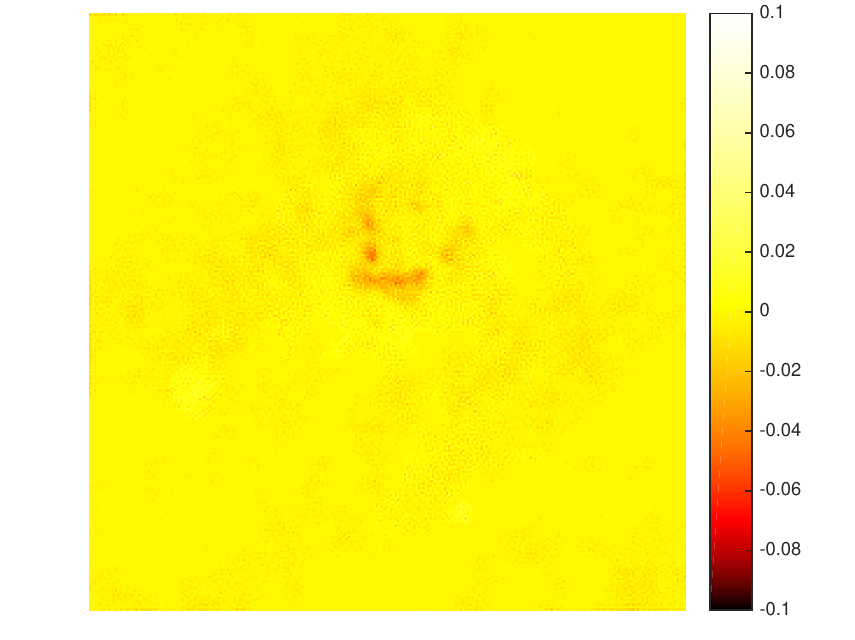}    
    \end{minipage}

		\caption{(M31) Left column shows ground truth (top) and dirty image (bottom). Middle column shows reconstructed image (top) and error (bottom) with Kaiser-Bessel kernel. Right column shows reconstructed image (top) and error (bottom) with Box kernel. For these simulations $M = 2N$ visibilities were used, with an input SNR of 30 dB. The error image shows that the Box kernel reconstruction has artefacts, which explains why the SNR is lower than the Kaiser-Bessel reconstruction. The Box kernel reconstruction did not converge within 100 iterations (based on the convergence criteria described in the text), while the Kaiser-Bessel kernel reconstruction did.}
		\label{fig:M31_simulation_images}
\end{figure*}

\begin{figure*}
			\begin{minipage}{0.3\textwidth}
    \includegraphics[width=1.15\linewidth]{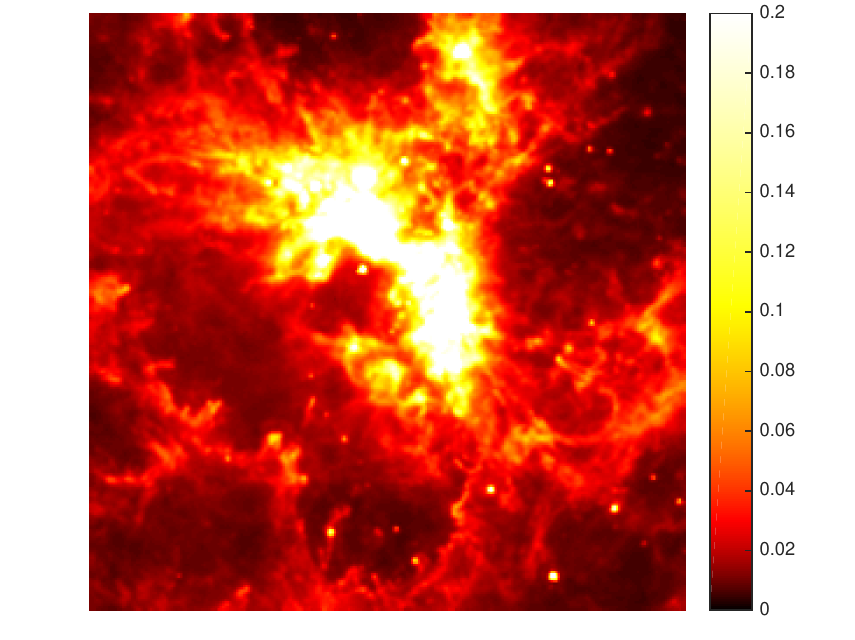}
    \end{minipage}
    \hspace{.5cm} % note: no blank line here
    \begin{minipage}{0.3\textwidth}
    \includegraphics[width=1.15\linewidth]{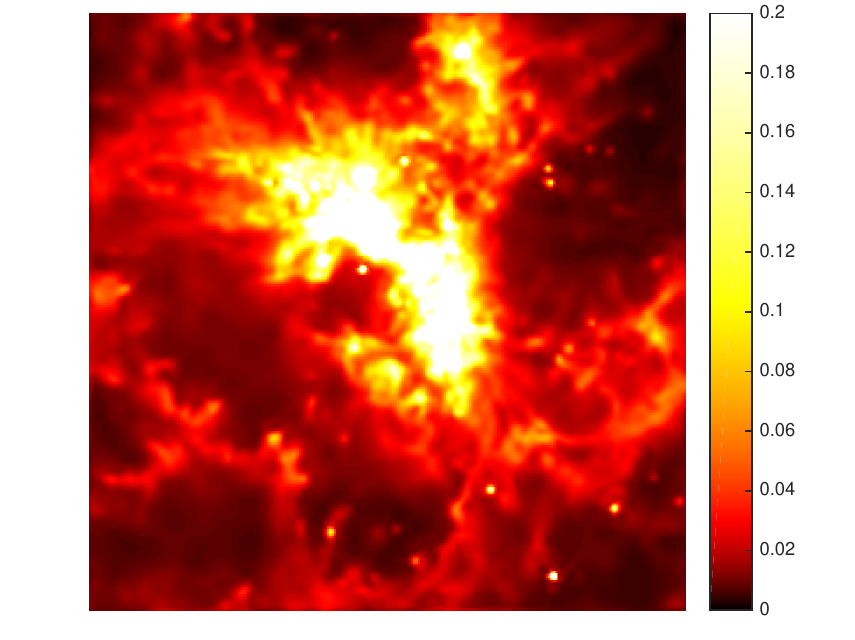}
    \end{minipage}
   \hspace{.5cm} % note: no blank line here
    \begin{minipage}{0.3\textwidth}
    \includegraphics[width=1.15\linewidth]{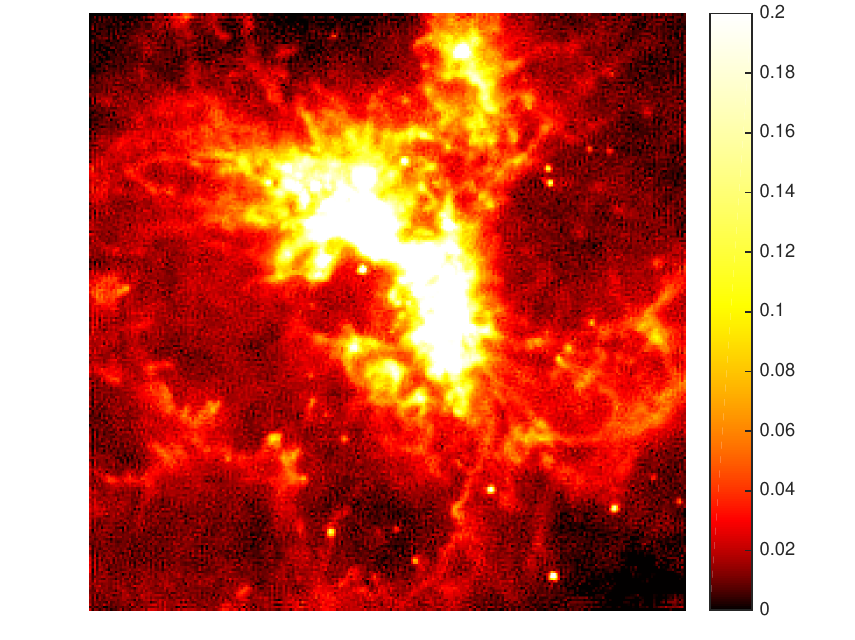}
    \end{minipage}

   \vspace*{0.01cm} % vertical separation
	
	\begin{minipage}{0.3\textwidth}
    \includegraphics[width=1.15\linewidth]{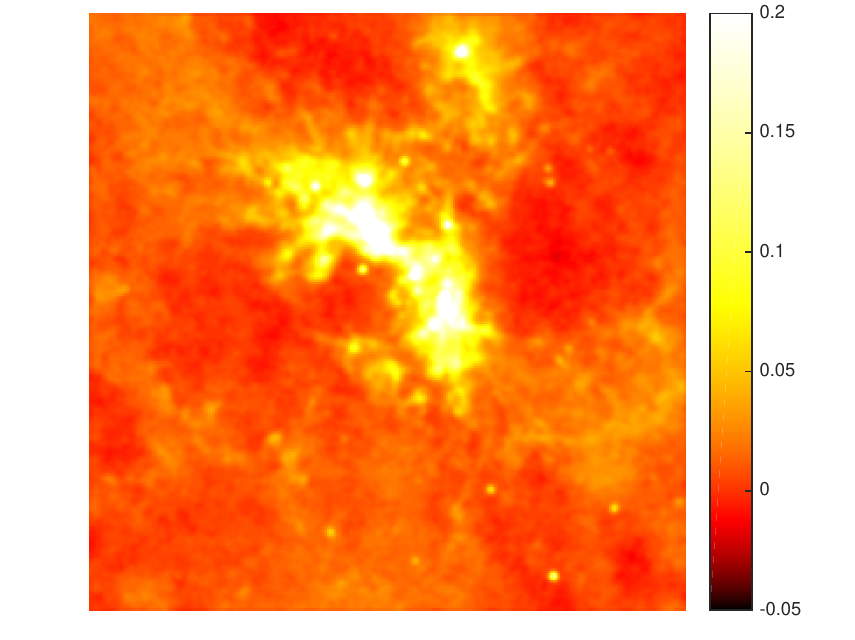}
    \end{minipage}
    \hspace{.5cm} % note: no blank line here
    \begin{minipage}{0.3\textwidth}
    \includegraphics[width=1.15\linewidth]{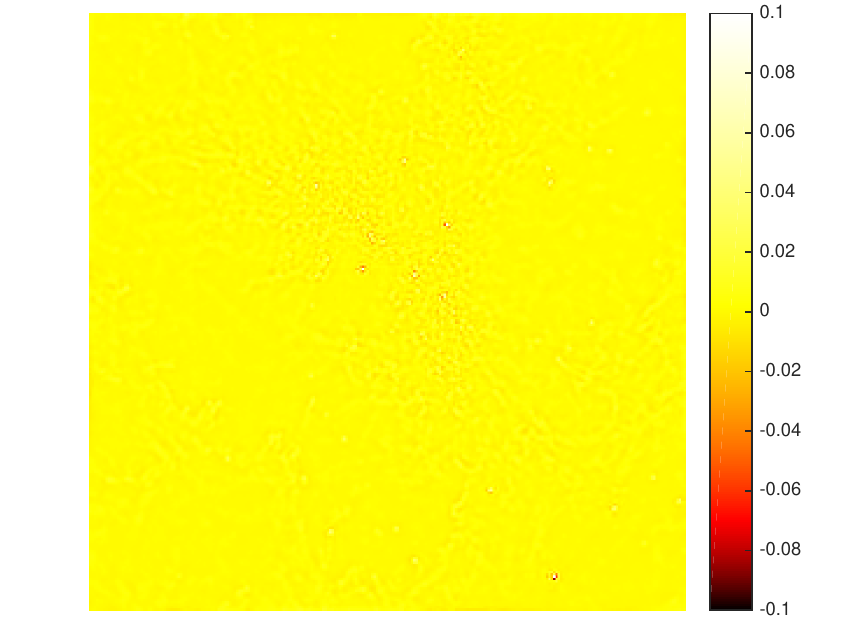}
    \end{minipage}
   \hspace{.5cm} % note: no blank line here
    \begin{minipage}{0.3\textwidth}
    \includegraphics[width=1.15\linewidth]{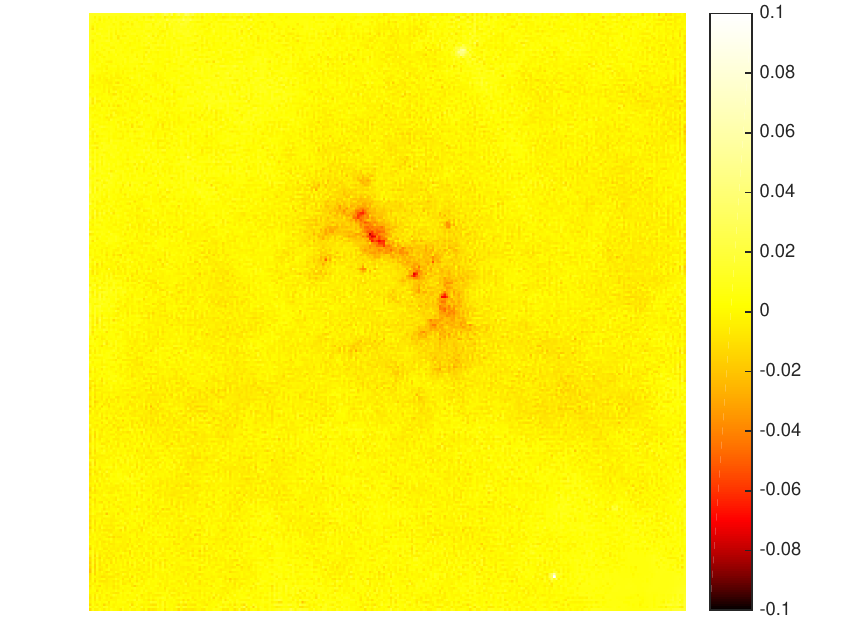}    
    \end{minipage}
		\caption{(30Dor) Left column shows ground truth (top) and dirty image (bottom). Middle column shows reconstructed image (top) and error (bottom) with Kaiser-Bessel kernel. Right column shows reconstructed image (top) and error (bottom) with Box kernel. For these simulations $M = 2N$ visibilities were used, with an input SNR of 30 dB. The error image shows that the Box kernel reconstruction has artefacts, which explains why the SNR is lower than the Kaiser-Bessel reconstruction. The Box kernel reconstruction did not converge within 100 iterations (based on the convergence criteria described in the text), while the Kaiser-Bessel kernel reconstruction did.}
		\label{fig:30dor_simulation_images}
\end{figure*}

\subsection{Results}
The SNR of the reconstructed images as a function of number of visibilities $M/N$ is shown in Figure~\ref{fig:M31_30Dor_SNR_plots} for both M31 and 30Dor. Simulations were performed using five of the different interpolation kernels described in Section~\ref{sec:gridding}, including: Kaiser-Bessel ($J = 4$, $\beta = 2.34J$), PSWF ($J = 6$, $\kappa = 1$), Box function ($J =1$), Gaussian with a typical $\sigma$ ($J=4$, $\sigma = 1$) and optimised $\sigma$ ($J=4$, $\sigma = 0.31 J^{0.52}$).  An oversampling ratio of $\alpha = 2$ was used for all cases.

Similar SNR results were found for reconstructions using the SARA dictionary for both the M31 and 30Dor images. The Kaiser-Bessel, PSWF, and Gaussian kernels with an optimised $\sigma$ were found to provide reconstructions of the same level of quality. The tests for these kernels converged within 100 iterations.

However, the Gaussian kernel with a typical $\sigma$ and the Box function provide reconstructions that have an SNR that is 5 to 10 dB below the other kernels in these tests. Furthermore, for the Box kernel, the reconstructions had often not converged within 100 iterations, while for the Gaussian with a typical $\sigma$ the majority of tests converged.

To illustrate the difference between reconstructions using the Kaiser-Bessel and Box interpolation kernels, Figure~\ref{fig:M31_simulation_images} and Figure~\ref{fig:30dor_simulation_images} show example reconstructions for $M = 2N$.  Error images are also shown, defined as the difference between the reconstructed and ground truth image. The structure in the Kaiser-Bessel kernel error images looks close to Gaussian error. The structure in the Box kernel error images shows artefacts, which spread throughout the reconstructed image, explaining the lower SNR.  

Tests were also performed using only a Dirac basis as the sparsifying dictionary, which provides a proxy for the CLEAN algorithm.  The results obtained were consistent with those found with the SARA wavelet dictionary. This suggests that these results found here are likely to apply also to CLEAN and other similar algorithms.

Additional tests were performed at an ISNR of 10 dB, where it was found that there was minimal difference between the reconstructed SNR with different interpolation kernels. This suggests that the choice of interpolation kernel will limit the reconstruction SNR when the level of artefacts is comparable or greater than the noise level.  
Consequently, for high dynamic range imaging the choice of kernel is important.

\subsection{Discussion}
Many calibration and imaging techniques depend on gridding and degridding methods to approximate the Fourier transform. While it has been understood that gridding methods in radio astronomy can impact image quality \citep{sch78a,gre79,sch80,bri99}, the current study confirms that gridding with poor kernels reduces the quality of images that can be recovered by sparse regularisation approaches, such as those implemented in PURIFY, and also those that can be recovered by CLEAN.  The magnitude of the impact depends on the quality of the measurements.  For high quality measurements with high ISNR, the use of poor interpolation kernels will limit the SNR of the reconstruction.  At low measurement ISNR, noise dominates the limit imposed by the interpolation kernel. 

In particular, we have found that the Gaussian kernel with an optimal $\sigma$ and Kaiser-Bessel kernel perform as well as the PSWF, while using a smaller support.  Moreover, both of the former have analytic forms that can be easily evaluated, which is not the case for the PSWF, where approximations are typically made and look-up-tables used.  This suggests that the Kaiser-Bessel kernel is just as good as the PSWF for sparse image reconstruction, and computationally less expensive with a smaller support. These finding are consistent with previous works, suggesting that the Kaiser-Bessel kernel is on par with optimal kernels \citep{gre79,jac91,fes03}.

\section{Applying PURIFY to observations}
\label{sec:purify_useage}
The application of compressive sensing to radio interferometry is a relatively new development and to date most of the exploration of compressive sensing has been via simulated observations. Simulations are useful for testing the performance of reconstructions because the ground truth and noise level is known, and  appropriate algorithm parameters can be estimated accurately.  However, this is not the case when reconstructing images from real observations.

In the next section (Section~\ref{sec:reconstructions}) we demonstrate that PURIFY can perform high quality image reconstruction on real observations and compare reconstructed images with those recovered by the CLEAN algorithm. However, to compare PURIFY and CLEAN reconstructions, we need to make clear the fundamental differences between the final outputs produced by each approach. In this section we discuss CLEAN in the context of sparse image reconstruction and clarify where the differences lie. In addition we describe how to apply PURIFY to real observations, including how to set the pixel size, weighting, and other parameters of the algorithm.

\subsection{CLEAN comparison}
Variations of CLEAN, such as Clark and Cotton-Schwab CLEAN \citep{cla80,sch84a}, work by iteratively building a model of the sky in major and minor cycles. This can be expressed in terms of iterations \citep{ono16}
\begin{equation}
	\bm{x}^{(t)} = \bm{x}^{(t-1)} + \mathcal{T}\Bigl( \bm{\mathsf{\Phi}}^\dagger\bigl (\bm{y}  - \bm{\mathsf{\Phi}}\bm{x}^{(t-1)}\bigr) \ \Bigr)\, ,
\end{equation}
where $\bm{x}^{(t)}$ represents the solution after $t$ iterations, and $\mathcal{T}$ represents the process of deconvolving the brightest sources in the residuals $\bm{\mathsf{\Phi}}^\dagger\bigl (\bm{y}  - \bm{\mathsf{\Phi}}\bm{x}^{(t-1)}\bigr)$.

CLEAN operates in minor and major cycles, the minor cycles $\mathcal{T}$ are performed after the calculation of a major cycle $\bm{\mathsf{\Phi}}^\dagger\bigl (\bm{y}  - \bm{\mathsf{\Phi}}\bm{x}^{(t-1)}\bigr)$. The minor cycles iteratively subtract the brightest sources from the image using an approximate point-spread function (PSF), which allows the location of the peaks of multiple sources to be found quickly. The major cycle performs an accurate subtraction of sources located in the minor cycle to generate the residuals for the next round of minor cycles.

CLEAN is essentially a matching pursuit algorithm \citep{mar87}, with a threshold constraint as suggested by \cite{hog74}, where the algorithm stops when the peak pixel of the residual image is below $\epsilon_{\rm threshold}$, \mbox{$\|\bm{\mathsf{\Phi}}^\dagger \left(\bm{y} - \bm{\mathsf{\Phi}}\bm{x} \right)\|_{\ell_\infty} \leq \epsilon_{\rm threshold}$}. Most variations of CLEAN impose the prior that the sky is sparse in a Dirac representation (CLEAN components/point sources), while multi-scale and { adaptive scale pixel decomposition} (ASP) CLEAN consider atoms with wider support to better model a sky containing extended sources \citep{bha04,cor08,zha16}. The solution obtained by the CLEAN algorithm $\bm{x}$ is typically called a CLEAN \emph{component} image. 

\subsubsection{CLEAN restoration}
\label{sec:purify_useage:clean_restoration}
In the case that the CLEAN components $\bm{x}$ could accurately model the entire sky, there would be nothing but noise remaining in the residuals. However, often it is not possible for CLEAN components to model diffuse structures that cannot be represented efficiently by point sources. For this reason, a final \emph{restored} image is constructed to include structures not deconvolved by CLEAN. The final restored image is found by convolving the CLEAN components with a Gaussian and then adding the residual image:
\begin{equation}
	\bm{x}^{\rm restored} = \bm{\mathsf{P}} \bm{x} + \bm{\mathsf{\Phi}}^\dagger \left(\bm{y} - \bm{\mathsf{\Phi}}\bm{x} \right)\, ,
\end{equation}
where $\bm{\mathsf{P}}$ is a post-processing operator that convolves $\bm{x}$ with a Gaussian of the same full width at half maximum as the dirty beam. The final restored image is expressed in units of Jy/Beam. These modifications mean the process of constructing a final restored image is not consistent with finding a solution that best fits the data for a given prior, even if the motivations are pragmatic. 

The CLEAN residuals are therefore not a true representation of how well the restored image models the true sky. Rather, the residuals $\bm{\mathsf{\Phi}}^\dagger \left(\bm{y} - \bm{\mathsf{\Phi}}\bm{x} \right)$ of a reconstructed CLEAN image are due to the CLEAN components $\bm{x}$, not the final restored image $\bm{x}^{\rm restored}$.

An additional systematic that can occur with the CLEAN method is that the dirty beam may not be well approximated by a Gaussian, which is assumed in constructing the restored image \citep{obe03}. This could impact studies that require accurate characterisation of point sources, such as weak lensing \citep{pat15}. Additionally, in low frequency imaging the ionospheric distortion on short timescales can produce a non-Gaussian dirty beam. For low frequency radio astronomy this is a serious issue, as discussed in \cite{Hur16}.

\subsection{PURIFY}
\label{sec:purify_useage:purify}
PURIFY adopts the prior that the sky has a sparse representation. This can include a representation as a collection of point sources and/or single or multiple wavelet dictionaries. This allows more flexibility when modelling both point sources and extended sources simultaneously, providing more accurate deconvolution of complex structure. As a result diffuse structures are not expected in the residual image, hence, there is no need to combine the model with the residuals as is done with the CLEAN algorithm. PURIFY provides a final image that is completely deconvolved, eliminating the need to convolve the model with a Gaussian beam. 

PURIFY therefore provides several advantages over CLEAN (in addition to improved image quality and the ability to scale to big data). First, it means the residuals correspond to the final image used for scientific analysis, such that the final image is the model that minimizes the error (this is not true for the CLEAN restored image). Second, the final model image recovered has units of Jy/Pixel, rather than Jy/Beam. This provides an advantage when computing statistics on an image and for general scientific interpretation, because there is no need to include Gaussian and dirty beam dependence.

\subsection{Choice of pixel size}
The final image recovered by PURIFY is sampled at discrete pixel values, hence there is a choice in the size of a pixel of the discrete image representing the sky brightness. The size and number of pixels can be determined by the resolution and field of view of the telescope. The size of the pixel can be estimated from the resolving power of the longest baseline and number of pixels determined by the field of view imaged (by the Nyquist relation).

However, radio astronomy packages such as Common Astronomy Software Applications (CASA) or MIRIAD typically assume between 3 and 5 pixels across the FWHM of the synthesised beam, found by least squares fitting a Gaussian to the main lobe of the synthesised beam \citep{mcm07,sau96,off14}.

Ideally, the size of the image should include all of the bright sources within the telescope's field of view.
When bright sources are outside the imaged field of view they cannot be modelled but may be aliased into the imaged region, which can limit image fidelity. 

PURIFY is flexible with regard to the pixel sampling rate and size and these parameters can be set by the user.  However, the default approach to setting the pixel size is to adopt Nyquist sampling since the resolution of the model is fundamentally limited by the $uv$-sampling pattern.

\subsection{Weighting}
In radio interferometry it is standard practice to weight the measurements according to natural, uniform, or robust weighting schemes, which are described in detail in \cite{bri95}.
The visibilities are weighted by the natural weighting scheme to optimize the sensitivity of an observation. However, for observations containing extended emission, the sidelobes in the image domain due to natural weighting can dominate the synthesised beam. In this case, CLEAN can perform badly, so the visibilities are uniformly weighted to minimize sidelobes.  We concisely review different weighting schemes, including the standard natural, uniform and robust weighting schemes used in radio interferometry.  \mbox{PURIFY} supports all of these schemes.

\subsubsection{Natural}
Natural weighting maximises the sensitivity of the observation, with weights set to $\bm{\mathsf{W}}^{\rm natural}_{i,i} = \sigma_i^{-1}$, where $\sigma_i$ is the standard deviation of the error for visibility $\bm{y}_i$. Note that here we consider the weighting operator as a component of the measurement operator following Eq.~\ref{eq:measurement_operator}, hence its entries are given by $\sigma_i^{-1}$, rather than a scaling of the visibilities only, in which case the weights are given by $\sigma_i^{-2}$.
Natural weighting is also known as whitening: each measurement has the same (unit) variance after weighting \citep{car14}. Whitening is a standard weighting approach in statistical data analysis and image processing.  Using natural weighting for interferometric imaging allows one to use a $\chi^2$ distribution when comparing how well the model visibilities fit the data, which can be used for a statistical interpretation of the bound on the $\ell_2$-norm. 

\subsubsection{Uniform}
Uniform weighting minimises the amplitude of sidelobes over a given field of view, which is achieved by calculating an average weighting from the nearest neighbours of a visibility. Explicitly, an average weight is calculated by
% \begin{equation}
% 	\bm{\mathsf{W}}^{\rm Gridded}_{i,i} = \sum_{k}\Theta\left(\frac{1}{\rm FoV} -\sqrt{(u(i) - u(k))^2 + (v(i) - v(k))^2}\right)\bm{\mathsf{W}}^{\rm Natural}_{kk}\,
% 	\label{eq:gridded_weights}
% \end{equation}
%where $\Theta$ is the indicator function, and FoV averaging region determined by the field of view. 
\begin{equation}
	\bm{\mathsf{W}}^{\rm gridded}_{i,i} = \sqrt{\frac{1}{{\vert \mathcal{S}_i \vert}}{\sum_{k \in \mathcal{S}_i}\left(\bm{\mathsf{W}}^{\rm natural}_{k,k}\right)^2}}\, ,
	\label{eq:gridded_weights}
\end{equation}
where $\mathcal{S}_i$ denotes the set of visibility indices that are included in the grid cell corresponding to visibility $i$, and $\vert \mathcal{S}_i \vert$ denotes the number of elements in $\mathcal{S}_i$.
% If the natural weights are assumed constant, this corresponds to the number of nearby neighbours. 
The uniform weights are then calculated by normalising the natural weights:
\begin{equation}
	\bm{\mathsf{W}}^{\rm uniform}_{i,i} = \frac{\bm{\mathsf{W}}^{\rm natural}_{i,i}}{\bm{\mathsf{W}}^{\rm gridded}_{i,i}}\, .
\end{equation}

It is possible to control the field of view at which the synthesised beam sidelobe suppression due to weighting occurs by changing the resolution of the grid cells. As the grid resolution increases, the field of view for dirty beam sidelobe suppression increases, although the suppression level is reduced. As the field of view for suppression increases, the weighting tends to natural weighting.

\subsubsection{Robust}
Robust weighting allows one to vary a robustness parameter $R$ to continuously move between natural and uniform weighting:
\begin{equation}
	\bm{\mathsf{W}}^{\rm robust}_{i,i} = \frac{\bm{\mathsf{W}}^{\rm natural}_{i,i}}{\sqrt{1 + \rho\left(\bm{\mathsf{W}}^{\rm gridded}_{i,i}\right)^2 } }\,
\end{equation}
where 
\begin{equation}
	\rho = \frac{\sum_{k} \left(\bm{\mathsf{W}}^{\rm natural}_{k,k}\right)^2}{\sum_{k} \left (\bm{\mathsf{W}}^{\rm gridded}_{k,k}\right)^4} \times 10^{-2R + {\rm log}_{10}\left( 25\right)}\, .
\end{equation}

\subsection{Parameter choice}
The parameters of PURIFY are set automatically, following the recommendations of \citet{car14} and \citet{ono16}. We also consider some minor modifications of these schemes that can be useful when analysing real observations, where, for example, the errors on the visibilities that are provided (\textit{i.e.}\ weights) may not be accurate.  Two parameters that need to be set carefully are the bound on the data fidelity error bound $\epsilon$ and the step size of the algorithm $\gamma$. We suggest a method to estimate $\epsilon$ using the Stokes V visibilities and to adaptively estimate the step size $\gamma$ during the first steps of the algorithm.

\subsubsection{Choosing the error bound $\epsilon$}
The parameter $\epsilon$ determines the error on how closely the model visibilities are required to match the measured visibilities. If $\epsilon$ is too small the model will start to fit to noise and if $\epsilon$ is too large the model will not model structures accurately. 

In the case of natural weighting, $\epsilon$ can be estimated by \citep{car14}
\begin{equation}
	\epsilon^2 = (2M + q\sqrt{4M})\frac{\sigma_{n}^2}{2}\,,
\end{equation}
where $\epsilon^2$ is set to $q$ standard deviations above the mean of the $\chi^2$ distribution. However, for typical observations $2M \gg \sqrt{4M}$, so this interpretation is less useful (due to the concentration of measure in high dimensions).  For real observations with large $M$ we simply estimate $\epsilon$ from the mean of the $\chi^2$ distribution and allow a scaling:
\begin{equation}
	\epsilon_\eta = \eta\sqrt{M}\sigma_{n}\, ,
\end{equation}
where $\eta$ allows one to vary $\epsilon$ to include non-thermal noise contributions, such as instrumental errors and radio frequency interference (RFI).  When using this latter approach to set $\epsilon$ we explicitly denote the $\eta$ dependence by $\epsilon_\eta$.

{ In principle, standard calibration and self-calibration methods can be applied with PURIFY but to date these have not yet been tested. Such an approach may be considered by choosing a high error bound for $\epsilon$ to generate a sky model of the brightest sources, applying a calibration algorithm to recover calibration parameters, before iterating.}

In the case that the source of noise in the visibilities is thermal, the weights should be accurate. However, if the weights are not accurate it is possible to use Stokes V to estimate the noise level and thus $\epsilon$. This is because Stokes V rarely contains astrophysical sources and so is dominated by thermal noise. To estimate the noise on a measurement, we use the median absolute deviation (MAD) method \citep{rou93,hoa00}
\begin{equation}
	\sigma_n = \sqrt{\left[\frac{{\rm Median}({\rm Real}(\bm{\mathsf{W}y}_{\rm V}))}{0.67449}\right]^2 + \left[\frac{{\rm Median}({\rm Imag}(\bm{\mathsf{W}y}_{\rm V}))}{0.67449}\right]^2}\, ,
\end{equation}
where $\bm{\mathsf{W}y}_{\rm V}$ is the weighted Stokes V visibilities. The MAD method provides a robust way to estimate $\sigma_n$ given Gaussian noise, and should be reliable when Stokes V is dominated by thermal noise.  

Furthermore, if the weights are only proportional to the standard deviation of noise, they will be incorrect by a scaling factor. The MAD method can be used to determine the standard deviation of the noise from a sample distribution. While using the MAD method to estimate $\sigma_n$ is intended to work for thermal noise contributions, it might not be accurate when there are polarimetric, amplitude, and phase calibration errors or RFI.

\subsubsection{Adapting the step size $\gamma$}
\label{sec:adaptive_step_size}
In \cite{car14}, it is suggested that the algorithm step size $\gamma$ can be set by
\begin{equation}
	{\gamma} = \beta \|\bm{\mathsf{\Psi^\dagger}} \bm{x}^{(0)} \|_{\ell_\infty}\, ,
\end{equation}
$\bm{x}^{(0)}$ is an initial estimate of the image. Typically, the initial estimate is chosen as $\bm{x}^{(0)} = \bm{\mathsf{\Phi^\dagger}}\bm{y}$ (\emph{i.e.} the dirty image). While the choice of $\gamma$ should not affect the final result of the algorithm, it does affect the rate of convergence.

We adapt this approach and allow $\gamma$ to be re-estimated as the algorithm progresses, before settling on a fixed value of $\gamma$ to guarantee convergence.  In this case, a candidate adaptive step size for the $i$-{\rm th} iteration can be calculated $\tilde{\gamma}_i = \beta \|\bm{\mathsf{\Psi^\dagger}} \bm{x}^{(i)} \|_{\ell_\infty}$.  If the current candidate for the step size changes by a small amount only, there is no need to change the step size used. In this case, a general rule for adapting the step size can be set:
\begin{equation}
  \gamma_i = \begin{cases}
    \tilde{\gamma}_i, &\text{ if } \frac{\tilde{\gamma}_i - \gamma_{i-1}}{\gamma_{i-1}} > \delta_{\rm adapt}\\
    \gamma_{i-1}, &\text{ if } \quad\frac{\tilde{\gamma}_i - \gamma_{i-1}}{\gamma_{i-1}} \leq \delta_{\rm adapt}\\
    \gamma_{i-1}, &\text{ if } \quad i \geq i_{\rm adapt}
  \end{cases},
\end{equation}
where $\delta_{\rm adapt}$ is the minimum relative difference needed for adapting the step size and $i_{\rm adapt}$ is the number of iterations after which the step size will not be adapted and will remain fixed.

\subsection{Input parameters of PURIFY}

As described already, the parameters of PURIFY are set automatically and so PURIFY can be run simply be providing the filename of an input measurement set and the output filename of the image to be recovered.  The user does not need to set any parameters.  However, the default settings can be overridden.

The main parameters of interest that a user may want to overwrite are specified in Table~\ref{tab:params}.  These include the $\eta$ value in setting $\epsilon_\eta$, the $\beta$ parameter in setting $\gamma$, the $\delta_{\rm adapt}$ and $i_{\rm adapt}$ parameters that control adapting $\gamma$, the relative variation of the solution criteria $\delta$, the residual norm convergence criteria $\xi$, and the maximum number of iterations $i_{\rm max}$. 

In analysing the observations considered in the next section, the value of $\eta$ varies from 1.4 to 7, and depends on the quality of the data set, such as how free it is from calibration error and RFI.  The $i_{\rm adapt}$ parameter is set to a fraction of the maximum number of iterations. It is important to set $i_{\rm adapt}$ such that the step size $\gamma$ stops adapting before convergence.  The relative variation criteria of the objective function was chosen to be $\delta = 5 \times 10^{-3}$. The choice of residual norm convergence criteria $\xi$ also depends on the quality of the data set. 
\begin{table*}
\caption{Description of main user parameters for using PURIFY to reconstruct an observation. All parameters are set automatically but can be overwritten.}
\label{tab:params}
\begin{tabular}{c p{3.5cm} p{6.5cm} p{4.5cm}}
\toprule
Parameter & PURIFY option & Description & Value \\
\toprule
$\eta$ & \texttt{--l2\_bound} & Parameterisation of the fidelity constraint: \mbox{$\epsilon_\eta = \eta\sqrt{M}\sigma_{n}$}. & $\eta = 1.4$ (default); $\eta \in [1, 10]$ (typical).  \\
\midrule
$\beta$ & \texttt{--beta} & Parameterisation of the step size of the algorithm: $\tilde{\gamma}_i = \beta \|\bm{\mathsf{\Psi^\dagger}} \bm{x}^{(i)} \|_{\ell_\infty}$ (default). One can also fix \mbox{$\gamma = \beta \|\bm{\mathsf{\Psi^\dagger}} \bm{x}^{(0)} \|_{\ell_\infty}$}.  & $\beta=10^{-3}$ (default) \\
\midrule
$\delta_{\rm adapt}$ & \texttt{--relative\_gamma\_adapt} & Relative difference criteria for adapting $\gamma_i$. & $\delta_{\rm adapt} = 0.01$ (default).\\
\midrule
$i_{\rm adapt}$ & \texttt{--adapt\_iter} & Number of iterations to consider adapting the step size $\gamma_i$ (should be before convergence). & $i_{\rm adapt} = 100$ (default). \\
\midrule
$\delta$ & \texttt{--relative\_variation} & Relative difference convergence criteria on the $\ell_2$-norm of the solution: $\frac{\|\bm{x}^{(i)} - \bm{x}^{(i-1)} \|_{\ell_2}}{\| \bm{x}^{(i)}\|_{\ell_2}} \leq \delta$. & $\delta = 5 \times 10^{-3}$ (default).\\
\midrule
$\xi$ & \texttt{--residual\_convergence} & Convergence criteria on the $\ell_2$ residual norm: \mbox{$\|\bm{y} - \bm{\mathsf{\Phi}\bm{x}} \|_{\ell_2} \leq \xi\epsilon_\eta$} & $\xi = 1$ (default); require $\xi \geq 1$. \\
\midrule
$i_{\rm max}$ & \texttt{--niters} & Maximum number of iterations. & $i_{\rm max}=\infty$ (default). \\ 
\bottomrule
\end{tabular}
\end{table*}

\section{PURIFY reconstruction of observations}
\label{sec:reconstructions}
In this section we compare the use of PURIFY and Cotton-Schwab CLEAN for reconstructing total intensity (Stokes I) observations made by the Very Large Array (VLA) and the Australia Telescope Compact Array (ATCA). 
In particular, we consider observations of the radio galaxies 3C129, Cygnus A, PKS J0334-39, and PKS J0116-473. To perform the Cotton-Schwab CLEAN algorithm, we use  WSCLEAN \citep{off14}. WSCLEAN is a standard choice for imaging in several MWA \citep{tin13} science pipelines including continuum, transients, EoR and polarisation modes \citep{way15,jac16,len16,mur15,off16}.  For PURIFY, we present results using the P-ADMM algorithm \citep{ono16}, in the analysis setting, with a positivity constraint and the SARA wavelet dictionary \citep{car12}, without reweighting.  Results with alternative algorithms that are being implemented in PURIFY (\textit{e.g.}\ the primal dual algorithm; \citealt{ono16}) will be presented in future work.

\subsection{Observations}
\label{sec:reconstructions:observations}
\begin{figure*}
		\begin{minipage}{0.44\textwidth}
    \includegraphics[width=1\linewidth]{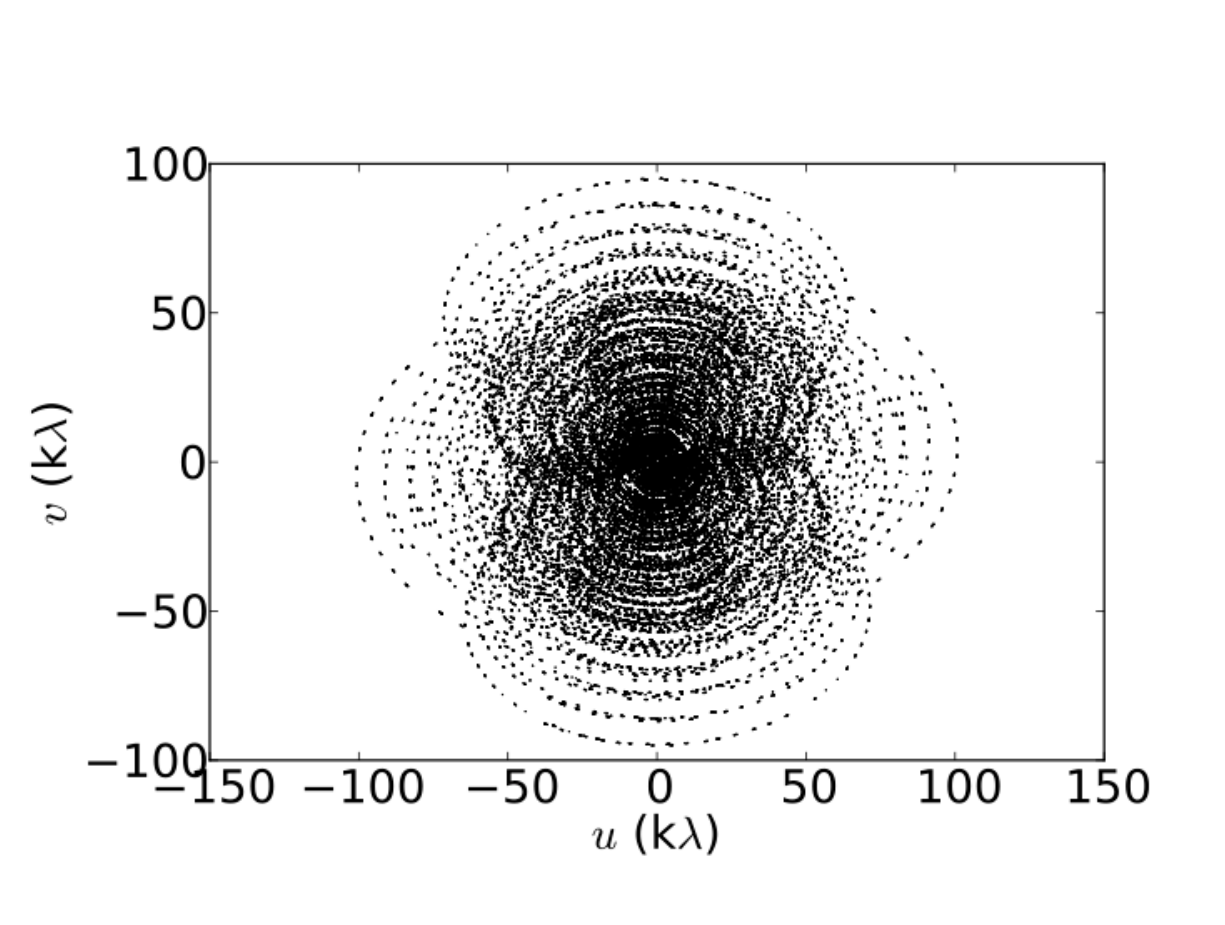}
    \end{minipage}
    \hspace{.5cm} % note: no blank line here
    \begin{minipage}{0.44\textwidth}
    \includegraphics[width=1\linewidth]{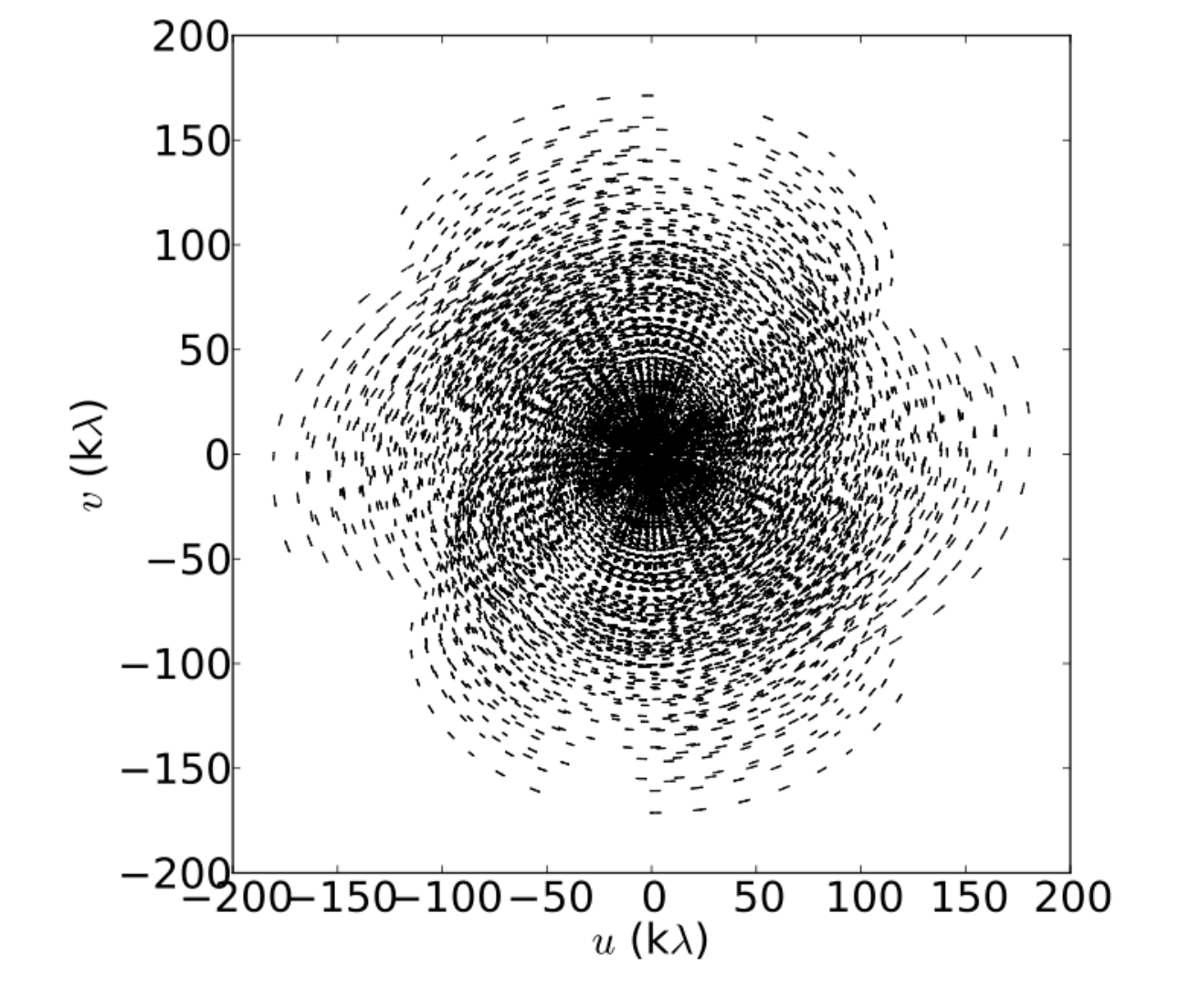}
    \end{minipage}
		\vspace*{0.01cm} % vertical separation
		\begin{minipage}{0.44\textwidth}
    \includegraphics[width=1\linewidth]{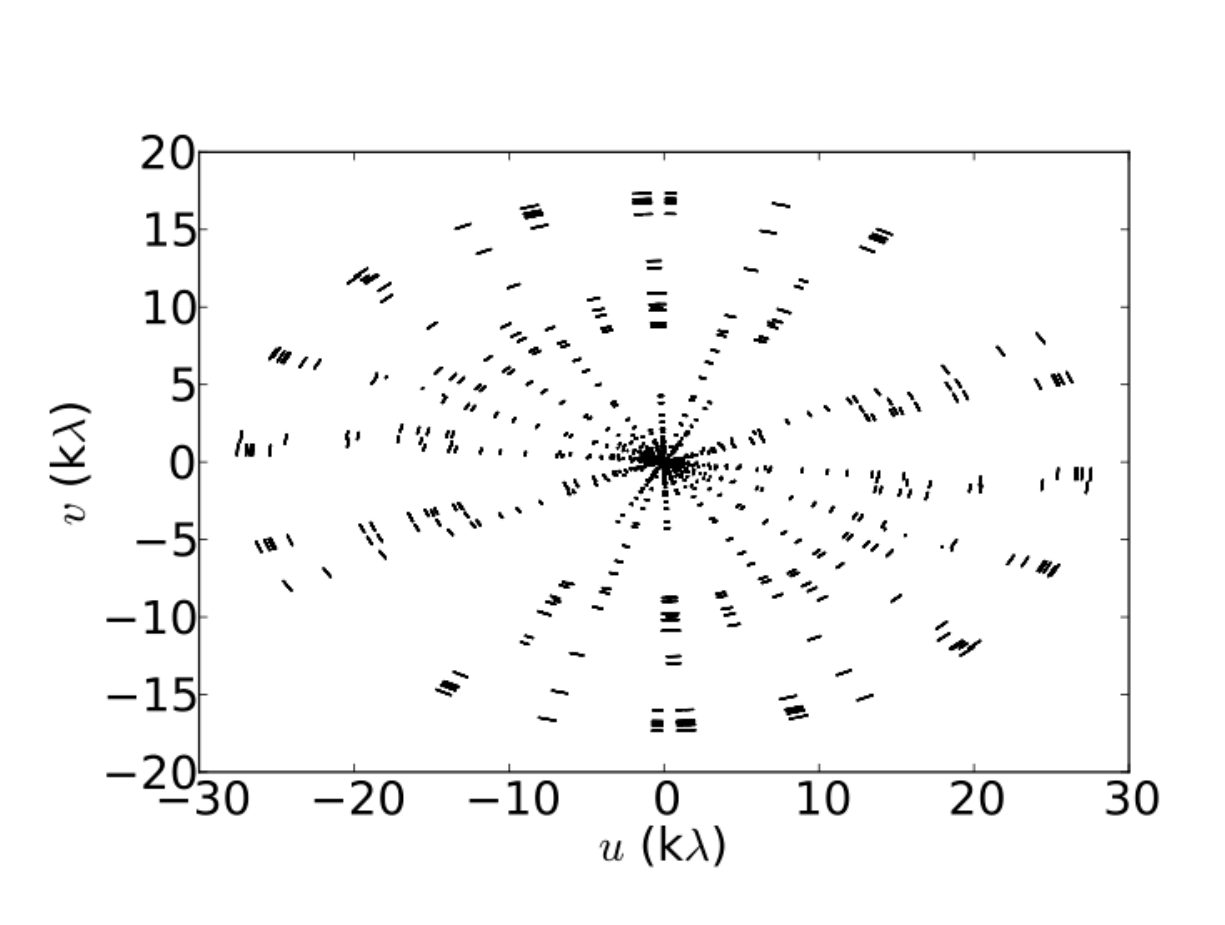}
    \end{minipage}
    \hspace{.5cm} % note: no blank line here
    \begin{minipage}{0.44\textwidth}
    \includegraphics[width=1\linewidth]{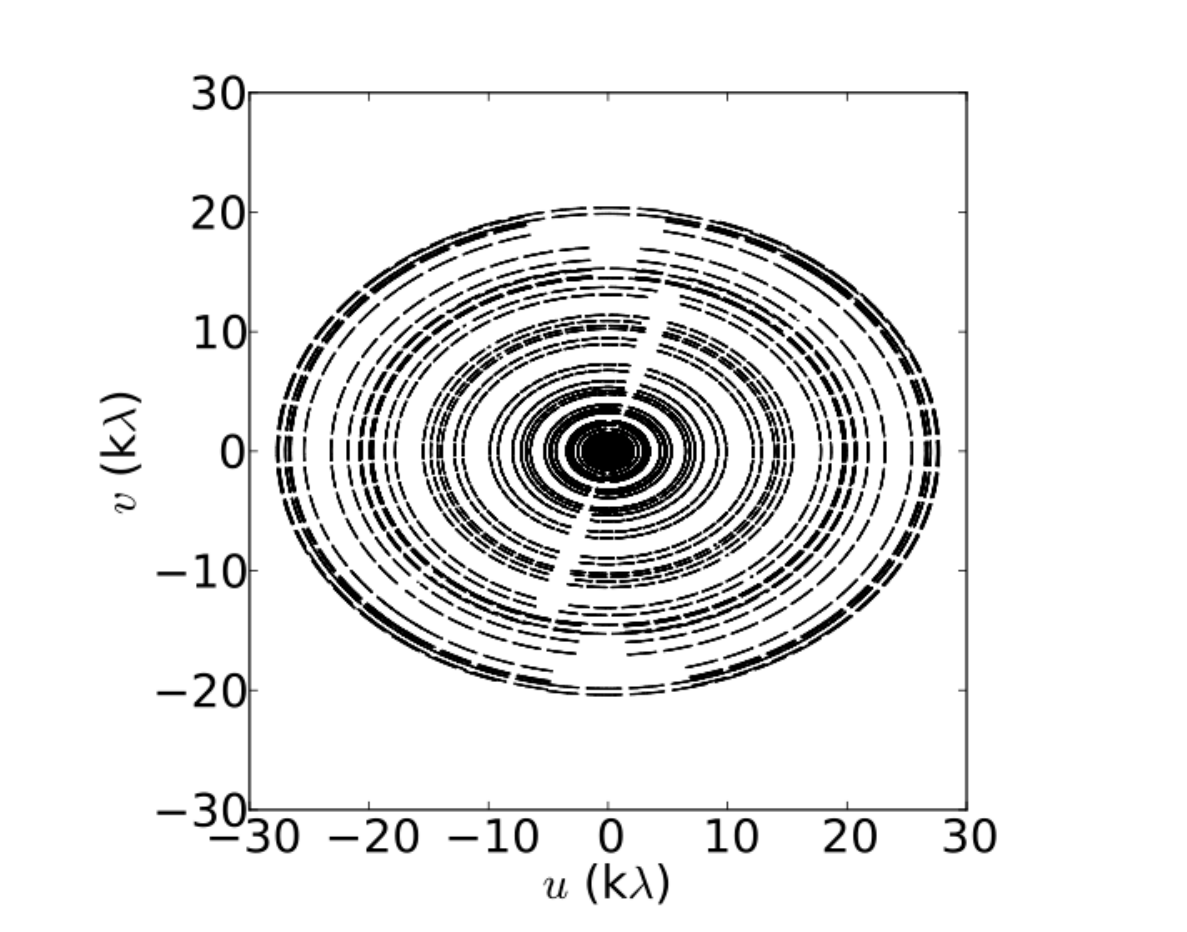}
    \end{minipage}
    \caption{Plots showing the $uv$-coverage of the observations of 3C129 (top left), Cygnus A (top right), PKS J0334-39 (bottom left), and PKS J0116-473 (bottom right). Units of $u$ and $v$ are kilo-wavelengths (kilo-$\lambda$).}
    \label{fig:uv_coverage}
\end{figure*}

In this section we discuss the details of the observations considered. The sampling patterns in the $uv$-plane for each observation are shown in Figure \ref{fig:uv_coverage}.

\subsubsection{3C129} The observation of the bent tailed radio galaxy 3C129 has a phase center of RA = 04h 45m 31.695s, DEC = +44$^\circ$ 55$^\prime$ 19.95$^{\prime\prime}$ (J2000), and was obtained from the NRAO archive. It was performed using the VLA with the project code AT0166, with two 50 MHz channels centered at 4.59 and 4.89 GHz. The observations were performed on the 25$^{\rm th}$ of July 1994 in configuration B and 3$^{\rm rd}$ of November 1994 in configuration C respectively. The total integration time was 79.7 minutes in configuration B and 15.8 minutes in configuration C. The calibration and flagging of radio frequency interference was performed using CASA, following the standard procedure found in the CASA manual. The gains were calibrated using sources 0420+417, 0518+165, and 0134+329, to solve for the instrumental and source polarisation. Source 0420+417 was observed alternately to solve the polarimetric calibration solutions with paralactic angle coverage. 

\subsubsection{Cygnus A} The VLA observation and reduction of Cygnus A in the X band (central frequency of 8.953 GHz, and 92 MHz bandwidth) was performed by Rick Perley\footnote{Private communication.} (PI:Perley, project code 14B-336 (legacy: AP658)). Cygnus A was observed in 2014 between the $3^{\rm rd}$ of November (18:39:44.0 UTC) to $4^{\rm th}$ November (04:28:12.0 UTC), using configuration C. The pointing centre was located at RA = 19h 59m 28.356s, DEC = +40$^\circ$ 44$^\prime$ 02.075$^{\prime\prime}$ (J2000). The data was reduced and calibrated using AIPS, following standard procedure that can be found in the AIPS Cookbook\footnote{\url{http://www.aips.nrao.edu/cook.html}}.

\subsubsection{PKS J0334-39} The observation of PKS J0334-39 was first presented in the work of \citet{pra13}, where the tailed radio galaxy's polametric structure was used to probe the environment of the galaxy cluster Abell 3135. The observation was also reprocessed using self calibration in \citet{pra16}, where it was used as an example of applying Generalised Complex CLEAN \citep{pra16} to a observation. The observation was performed using the ATCA (with the pre-CABB correlator) in 2001 is centered on RA = 03h 34m 07.18s DEC = -39$^\circ$00$^\prime$03.19$^{\prime\prime}$ (J2000), at a central frequency of 1.384 GHz. There are 12 channels, each with a width of 8 MHz. The observation was performed in configuration 6A for 59 minutes, 1.5A for 76 minutes, 750A for 79.7 minutes, 375 for 75.4 minutes. A detailed description of the calibration procedure, performed using MIRIAD, can be found in \cite{pra13}.

\subsubsection{PKS J0116-473} The observation of PKS J0116-473 used in this work was first presented in \cite{sar02}. The total intensity, polametric structure, and morphology of PKS J0116-473 have been studied in detail at 12 and 22 cm emission. The ATCA observations of PKS J0116-473 used in this work were extracted from the archive (PI:Shankar, project code C770), then calibrated and flagged following a standard ATCA data reduction procedure found in the MIRIAD manual\footnote{\url{http://www.atnf.csiro.au/computing/software/miriad/userguide/userhtml.html}}. The phase center is located at RA = 14h 59m, 15.75s DEC =-36$^\circ$ 55$^\prime$ 47.87$^{\prime\prime}$ (J2000), and the central observation frequency is 1.384 GHz. After flagging and removing channels due to cross-channel interference, there are 12 channels each with 8 MHz channel width. The observations were performed in 1999, on the 10$^{\rm th}$ and 12$^{\rm th}$ of January (configuration 375, 1115 minutes integration), on the 7$^{\rm th}$ (750C, 1088.3 minutes) and 20$^{\rm th}$ (6C, 1109.3 minutes) of February, and on the 24$^{\rm th}$ and 25$^{\rm th}$ of April (1.5C, 1112 minutes). Sources PKS B1934-638 and PKS B0823-500 were used to set the flux density scale at 1.384 GHz. The time variations in complex antenna gains and bandpass were calibrated using alternating observations of the unresolved source PKS B0153-410. 

\subsection{Reconstructions}
In this section we present the reconstructions from real observations. We show the reconstructed model image, alongside the residuals. For the CLEAN reconstructions we show the post-processed restored image (see Section~\ref{sec:purify_useage:clean_restoration}), while for PURIFY there is no need for post-processing so there is no restored image but only a reconstructed model image (see Section~\ref{sec:purify_useage:clean_restoration}). For PURIFY reconstructions we use natural weighting, and for CLEAN we use both natural and uniform weightings. \footnote{Rather than using measurement sets for the ATCA data sets, the tables were read with PURIFY from uvfits files. In all other cases, the observations were read from measurement sets.}

The CLEAN thresholds and FWHM of the restoring beams can be found in Table \ref{tab:clean}. The CLEAN components are restricted to be positive valued. CLEAN has not been restricted to regions around the source. CLEAN was run until the residual peak reached the cutoff flux value. 
We are careful to make the distinction between the \emph{restored} image and the \emph{reconstructed} image for CLEAN (see Section~\ref{sec:purify_useage:clean_restoration}), since the restored image is not used to generate the residuals.  When we refer to the reconstructed image, we are referring to the CLEAN component image.

For PURIFY, the error constraint in the model is set using $\epsilon_\eta$. The \mbox{P-ADMM} step size was set adaptively as described in Section~\ref{sec:adaptive_step_size}. PURIFY images have a resolution set by the longest baseline in the observation.  

\begin{figure*}
	
	\begin{minipage}{0.28\textwidth}
    \includegraphics[width=1.15\linewidth]{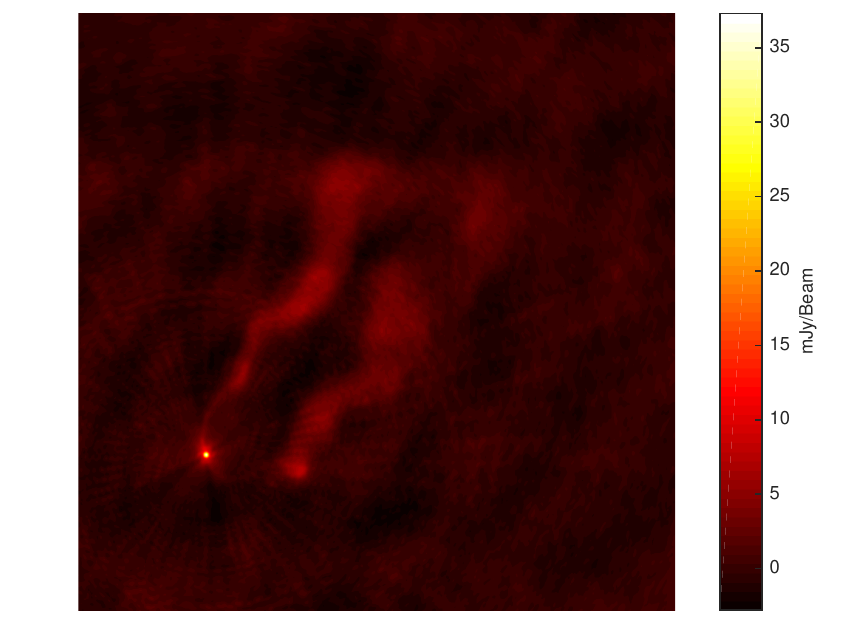}
    \end{minipage}
    \hspace{.5cm} % note: no blank line here
    \begin{minipage}{0.28\textwidth}
    \includegraphics[width=1.15\linewidth]{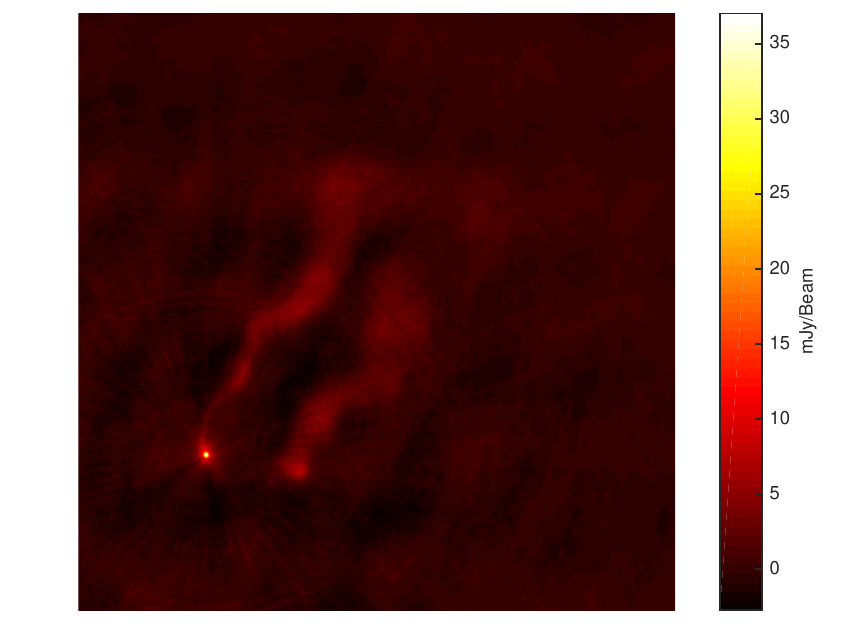}
    \end{minipage}
   \hspace{.5cm} % note: no blank line here
    \begin{minipage}{0.28\textwidth}
    \includegraphics[width=1.15\linewidth]{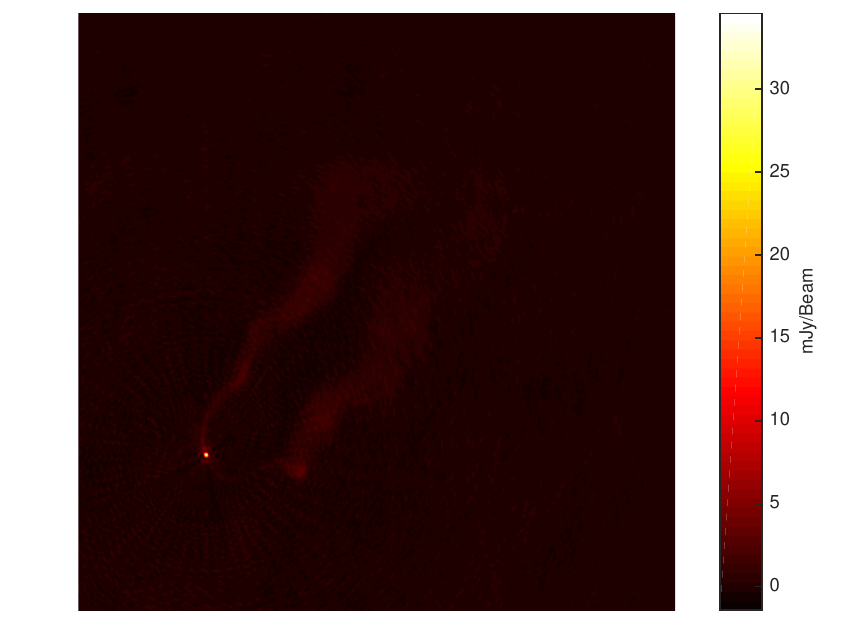}
    \end{minipage}

    \vspace*{0.01cm} % vertical separation
	
	\begin{minipage}{0.28\textwidth}
    \includegraphics[width=1.15\linewidth]{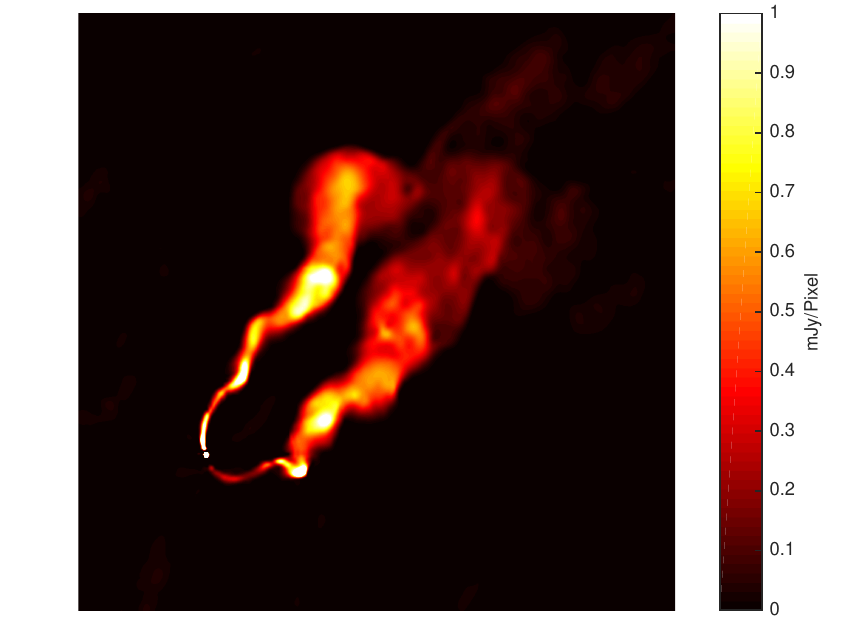}
    \end{minipage}
    \hspace{.5cm} % note: no blank line here
    \begin{minipage}{0.28\textwidth}
    \includegraphics[width=1.15\linewidth]{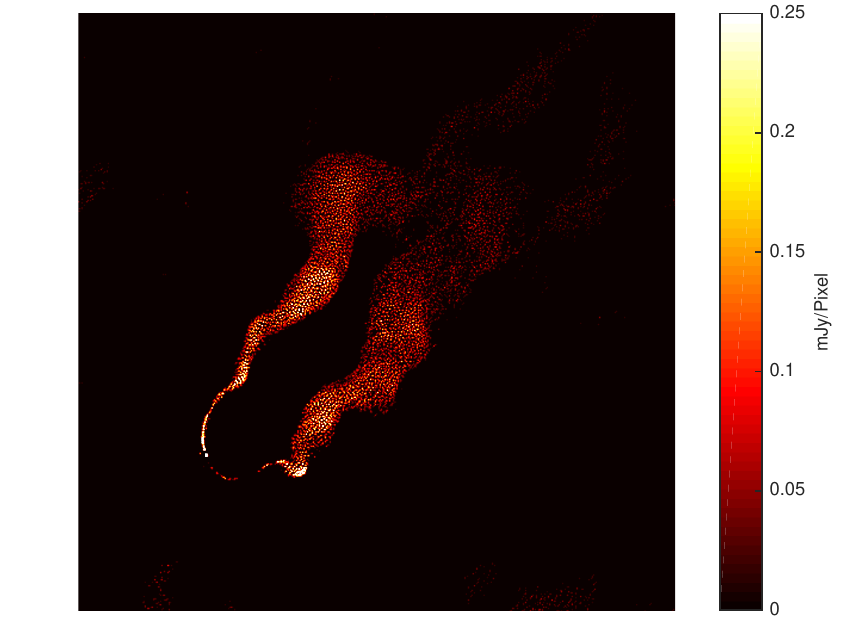}
    \end{minipage}
   \hspace{.5cm} % note: no blank line here
    \begin{minipage}{0.28\textwidth}
    \includegraphics[width=1.15\linewidth]{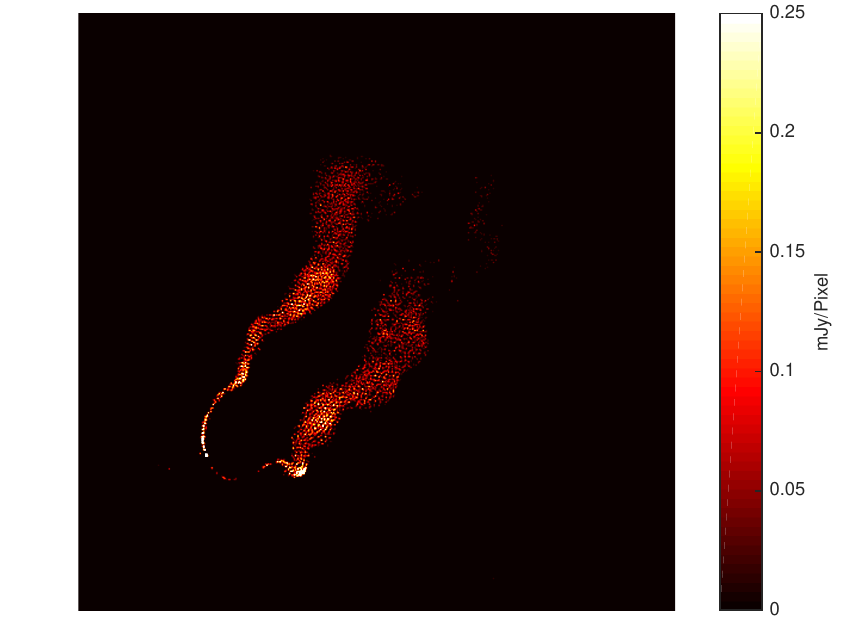}
   \end{minipage}
   
    \vspace*{0.01cm} % vertical separation
	
	\rule{5.6cm}{0cm}
    \begin{minipage}{0.28\textwidth}
    \includegraphics[width=1.15\linewidth]{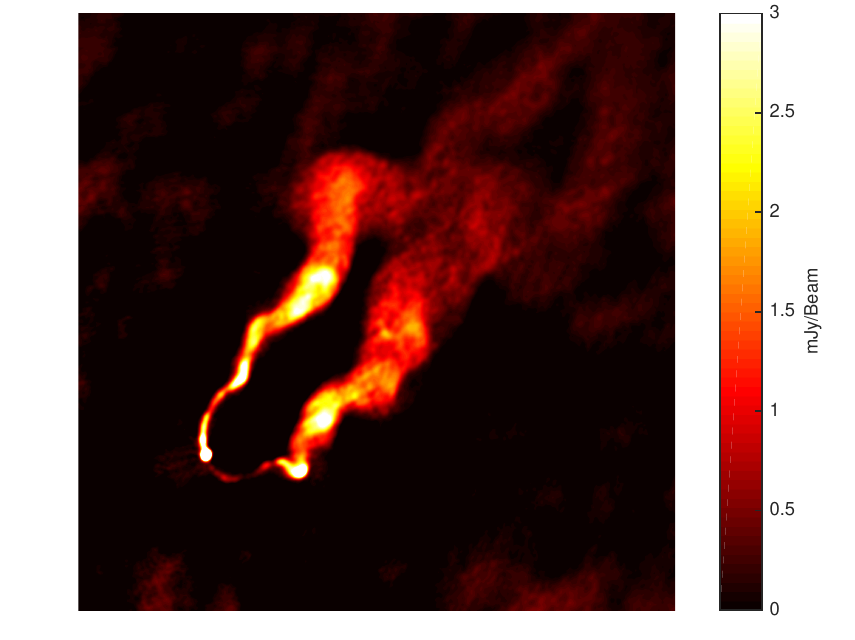}
    \end{minipage}
   \hspace{.5cm} % note: no blank line here
    \begin{minipage}{0.28\textwidth}
    \includegraphics[width=1.15\linewidth]{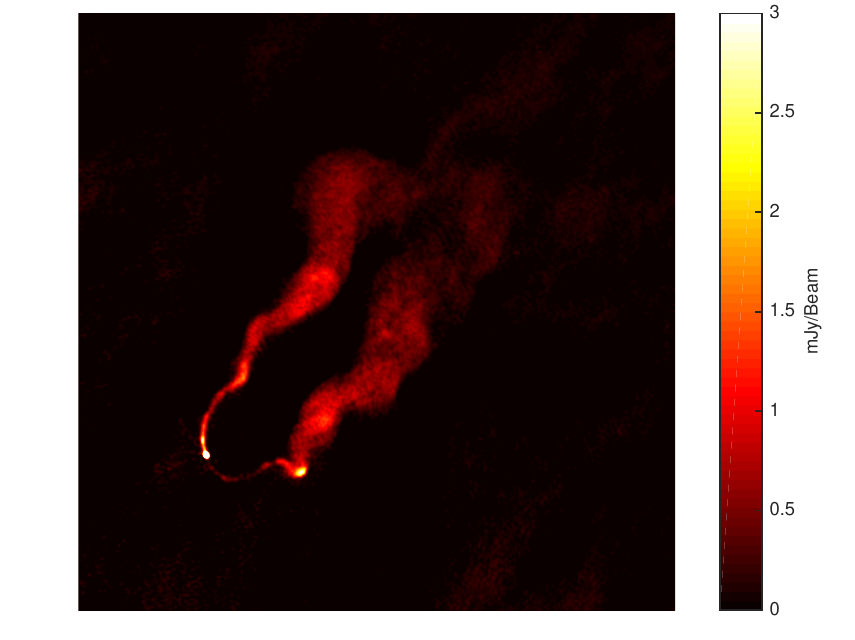}
    \end{minipage}
\vspace*{0.01cm} % vertical separation
	
	\begin{minipage}{0.28\textwidth}
    \includegraphics[width=1.15\linewidth]{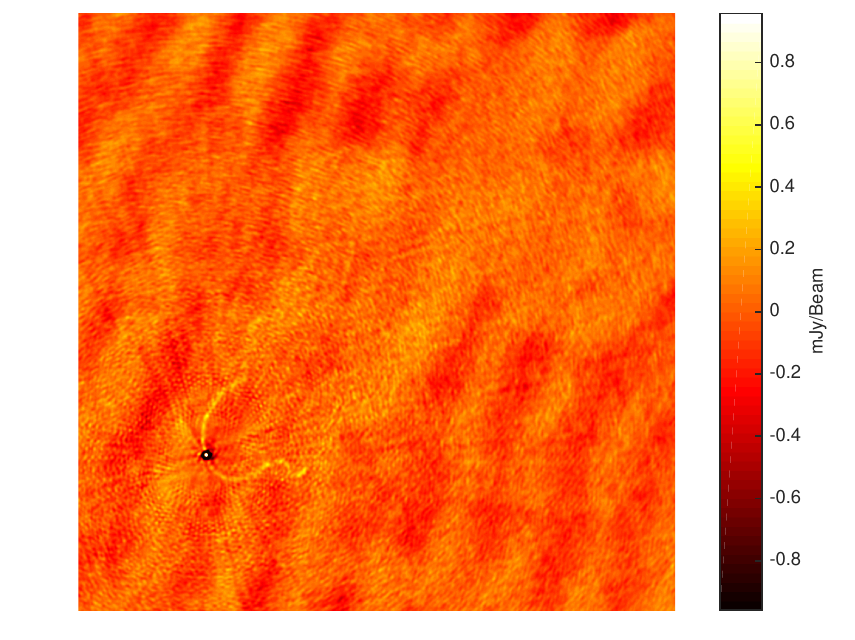}
    \end{minipage}
    \hspace{.5cm} % note: no blank line here
    \begin{minipage}{0.28\textwidth}
    \includegraphics[width=1.15\linewidth]{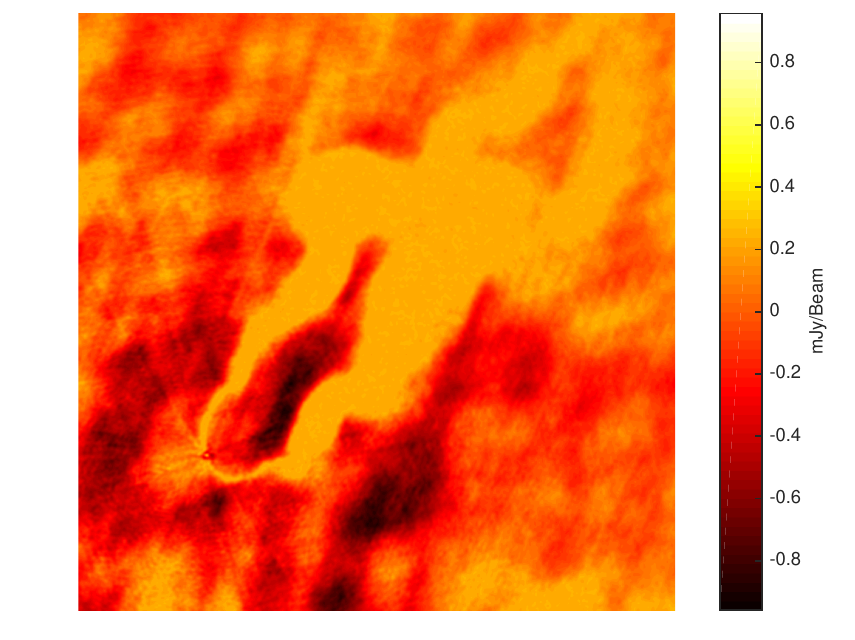}
    \end{minipage}
   \hspace{.5cm} % note: no blank line here
    \begin{minipage}{0.28\textwidth}
    \includegraphics[width=1.15\linewidth]{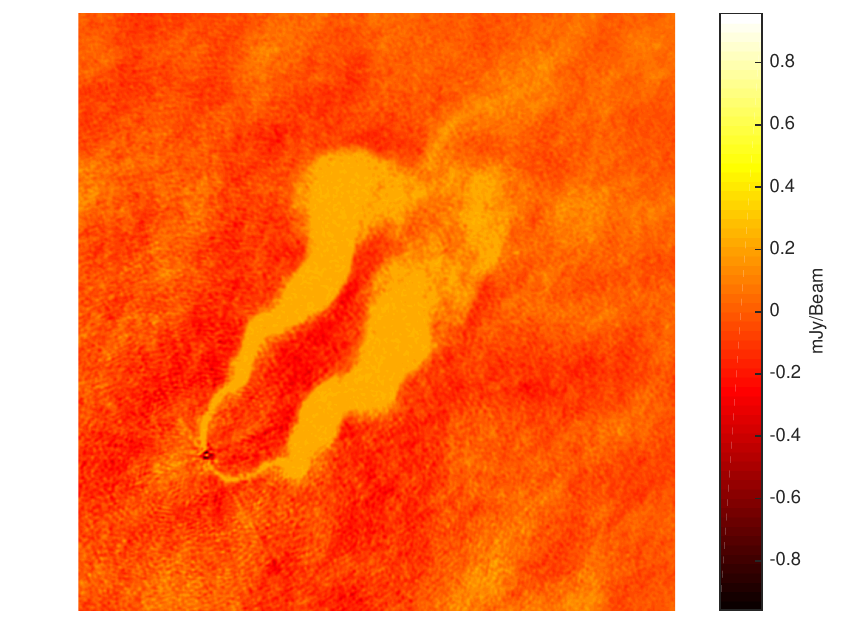}
	\end{minipage}
    
    \vspace*{0.01cm} % vertical separation
	
	\begin{minipage}{0.28\textwidth}
    \includegraphics[width=1.15\linewidth]{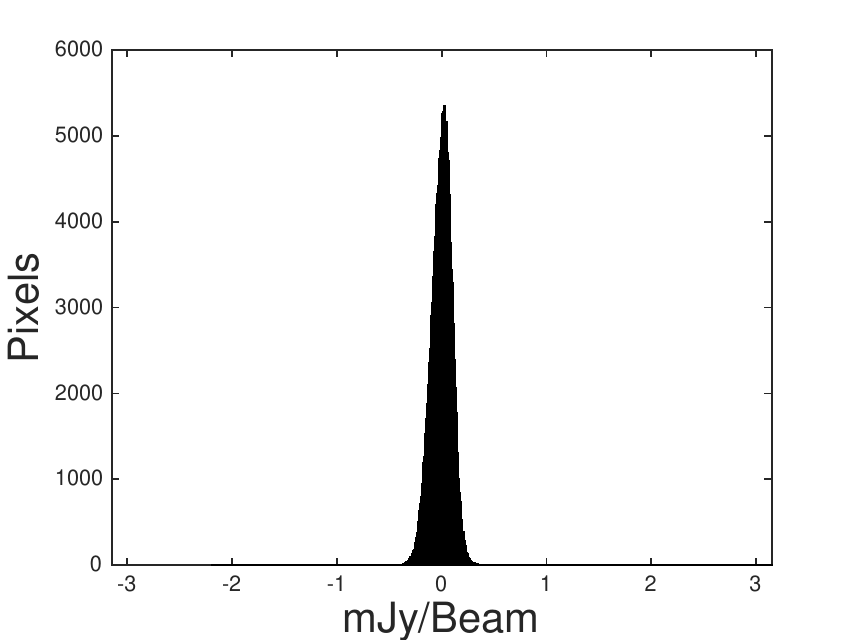}
    \end{minipage}
    \hspace{.5cm} % note: no blank line here
    \begin{minipage}{0.28\textwidth}
    \includegraphics[width=1.15\linewidth]{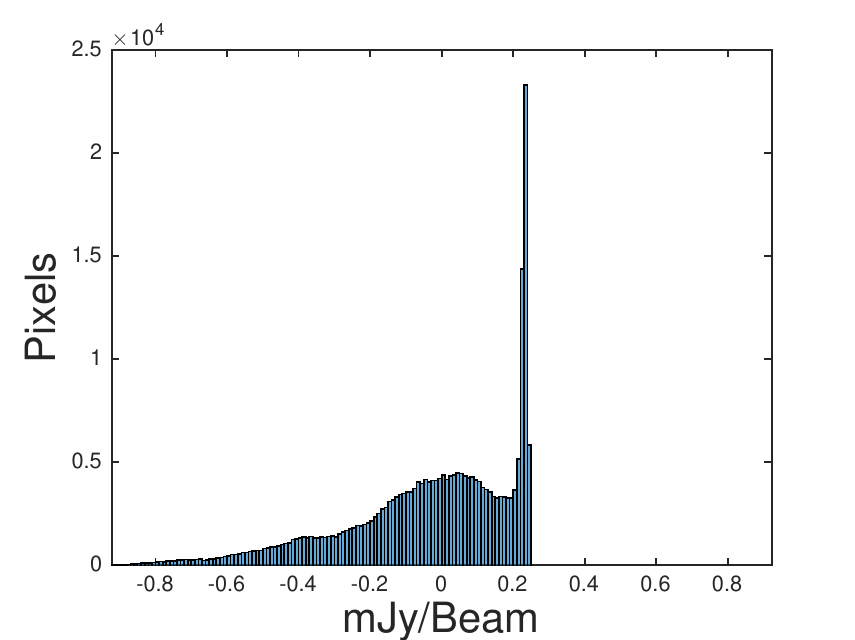}
    \end{minipage}
   \hspace{.5cm} % note: no blank line here
    \begin{minipage}{0.28\textwidth}
    \includegraphics[width=1.15\linewidth]{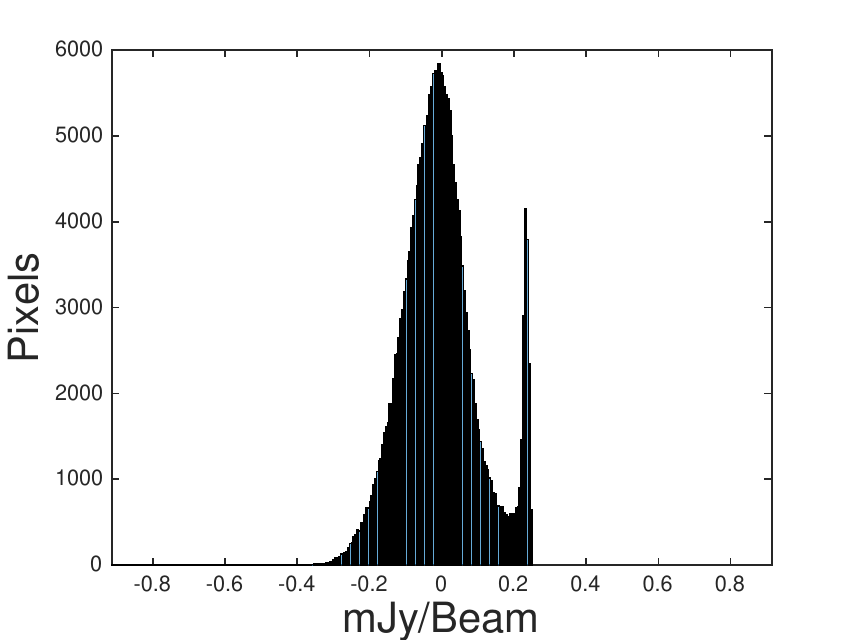}
    \end{minipage}
		\caption{PURIFY and CLEAN reconstructions of 3C129. Each pixel is 0.4 arcseconds, and the images are $1024 \times 1024$ pixels. { The pixels within $\left[ 400, 900\right] \times \left[400, 900 \right]$ are shown in the images and histogram of this figure.} Left column shows a PURIFY reconstruction with natural weighting. Middle and right columns show CLEAN reconstructions with natural and uniform weightings, respectively.  From the top to bottom row: synthesised (\emph{i.e.} dirty) image, model image, restored image, residual image, and a histogram of residual image. PURIFY does not require any post-processing and so does not produce a restored image.}
		\label{fig:3C129}
\end{figure*}

\begin{figure*}
	
	\begin{minipage}{0.28\textwidth}
    \includegraphics[width=1.15\linewidth]{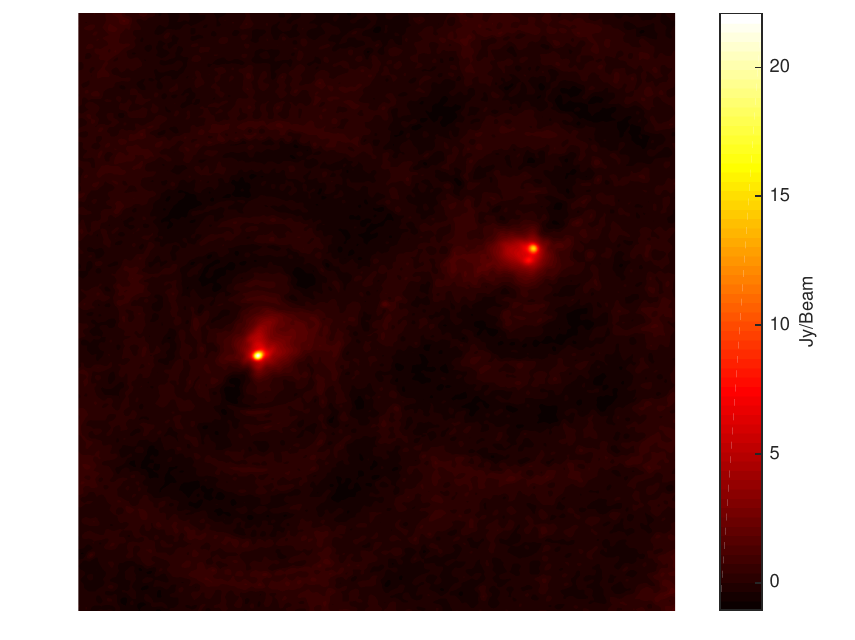}
    \end{minipage}
    \hspace{.5cm} % note: no blank line here
    \begin{minipage}{0.28\textwidth}
    \includegraphics[width=1.15\linewidth]{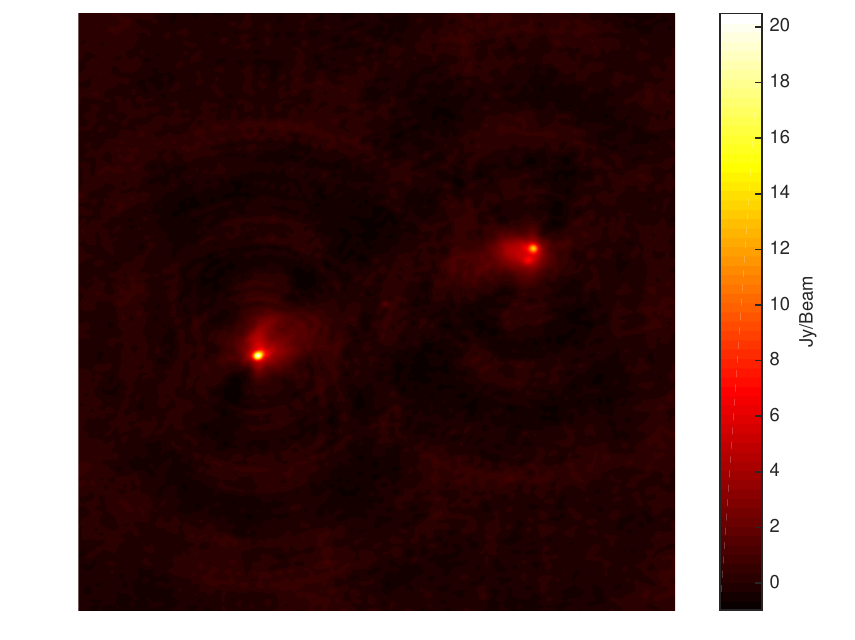}
    \end{minipage}
   \hspace{.5cm} % note: no blank line here
    \begin{minipage}{0.28\textwidth}
    \includegraphics[width=1.15\linewidth]{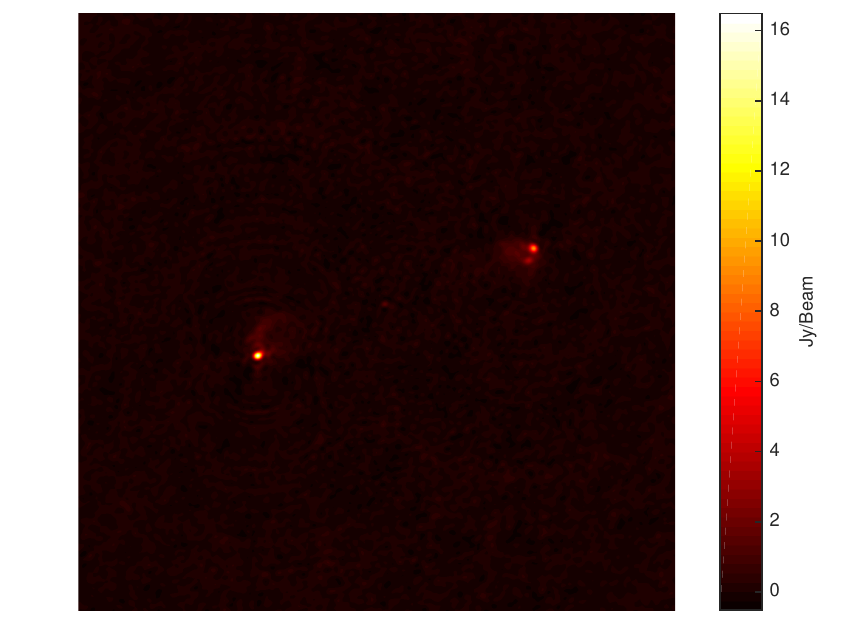}
    \end{minipage}

    \vspace*{0.01cm} % vertical separation
	
	\begin{minipage}{0.28\textwidth}
    \includegraphics[width=1.15\linewidth]{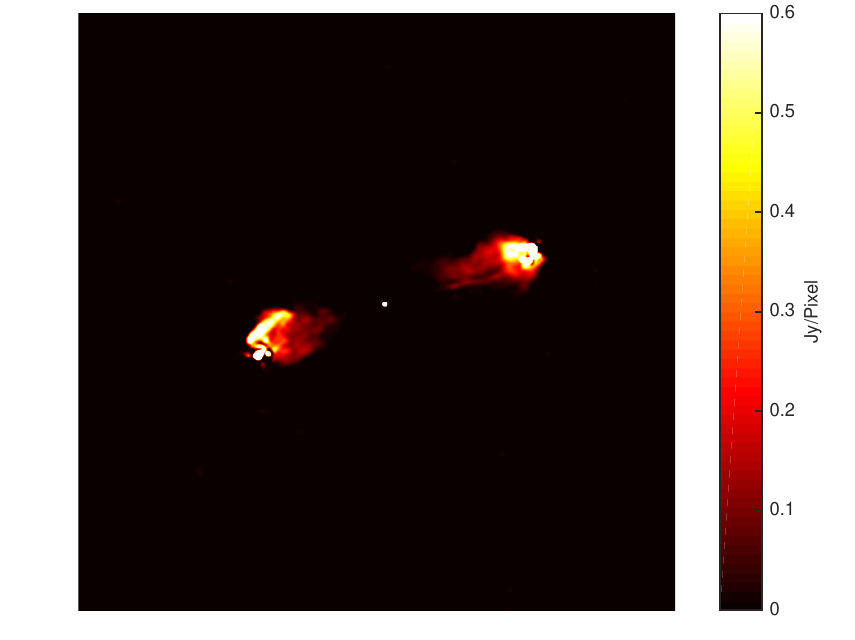}
    \end{minipage}
    \hspace{.5cm} % note: no blank line here
    \begin{minipage}{0.28\textwidth}
    \includegraphics[width=1.15\linewidth]{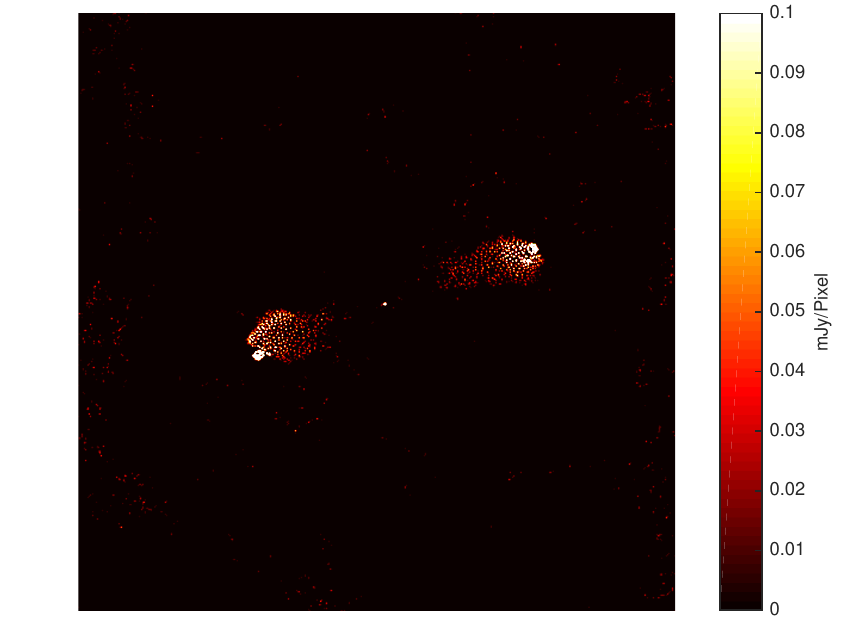}
    \end{minipage}
   \hspace{.5cm} % note: no blank line here
    \begin{minipage}{0.28\textwidth}
    \includegraphics[width=1.15\linewidth]{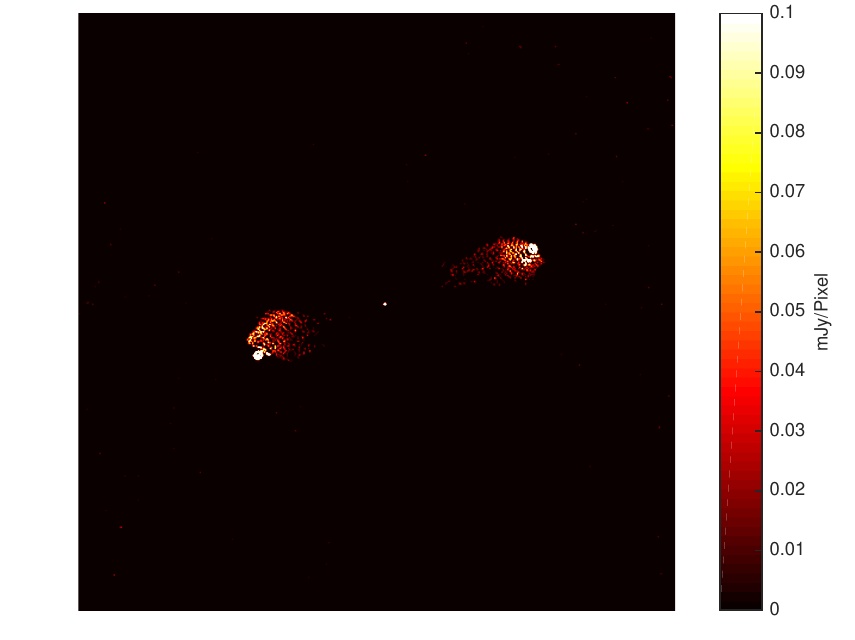}
   \end{minipage}
   
    \vspace*{0.01cm} % vertical separation
	
\rule{5.6cm}{0cm}
    \begin{minipage}{0.28\textwidth}
    \includegraphics[width=1.15\linewidth]{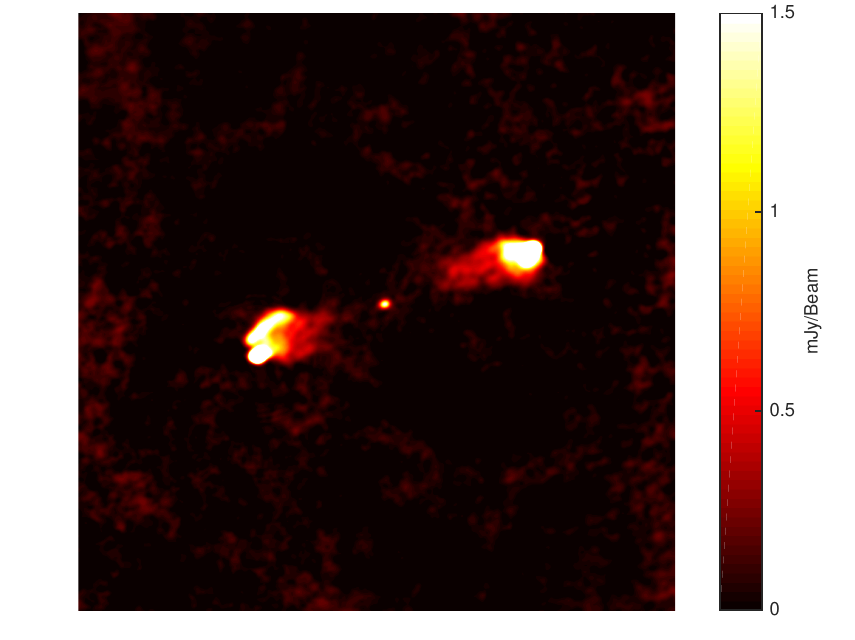}
    \end{minipage}
   \hspace{.5cm} % note: no blank line here
    \begin{minipage}{0.28\textwidth}
    \includegraphics[width=1.15\linewidth]{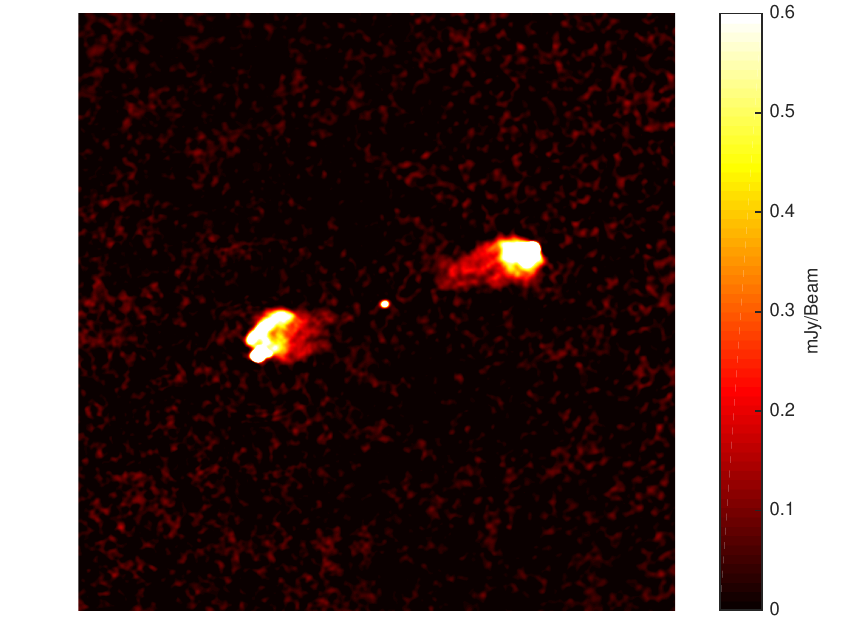}
    \end{minipage}
    \vspace*{0.01cm} % vertical separation
	
	\begin{minipage}{0.265\textwidth}
    \includegraphics[width=1.15\linewidth]{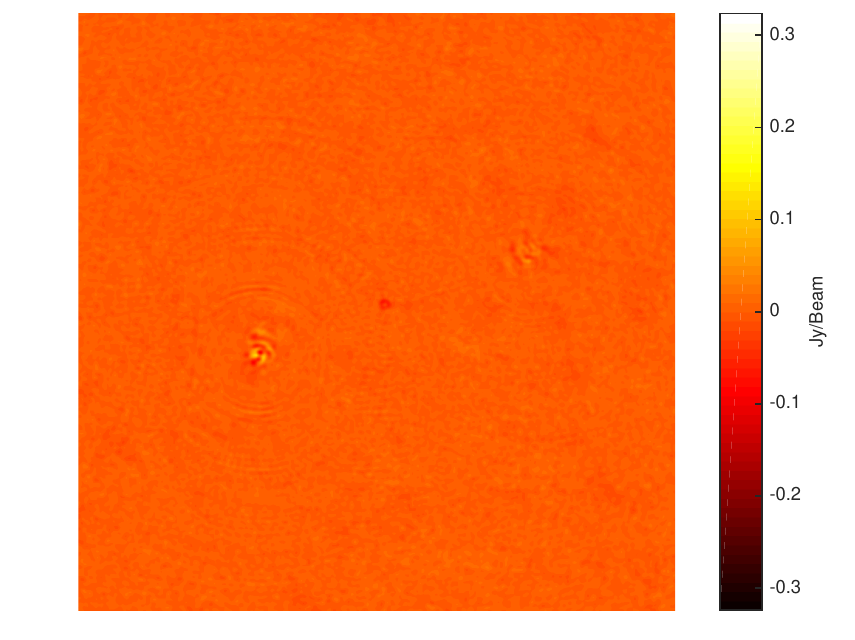}
    \end{minipage}
    \hspace{.5cm} % note: no blank line here
    \begin{minipage}{0.265\textwidth}
    \includegraphics[width=1.15\linewidth]{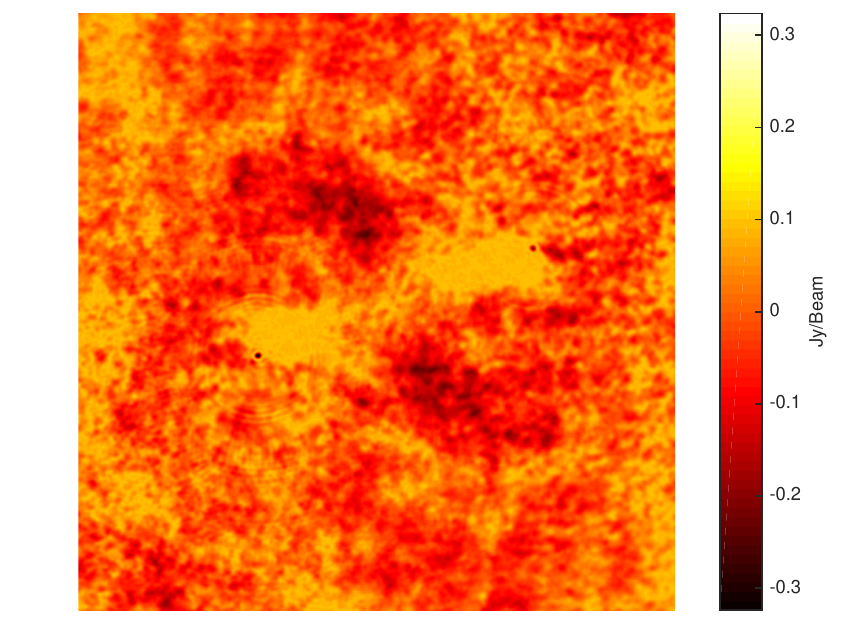}
    \end{minipage}
   \hspace{.5cm} % note: no blank line here
    \begin{minipage}{0.265\textwidth}
    \includegraphics[width=1.15\linewidth]{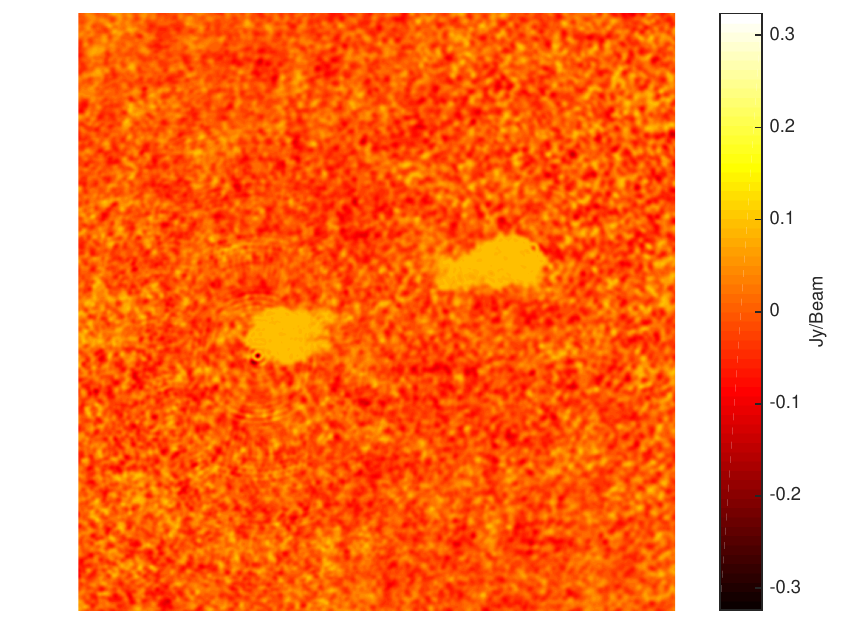}
	\end{minipage}
    
    \vspace*{0.01cm} % vertical separation
	
	\begin{minipage}{0.28\textwidth}
    \includegraphics[width=1.15\linewidth]{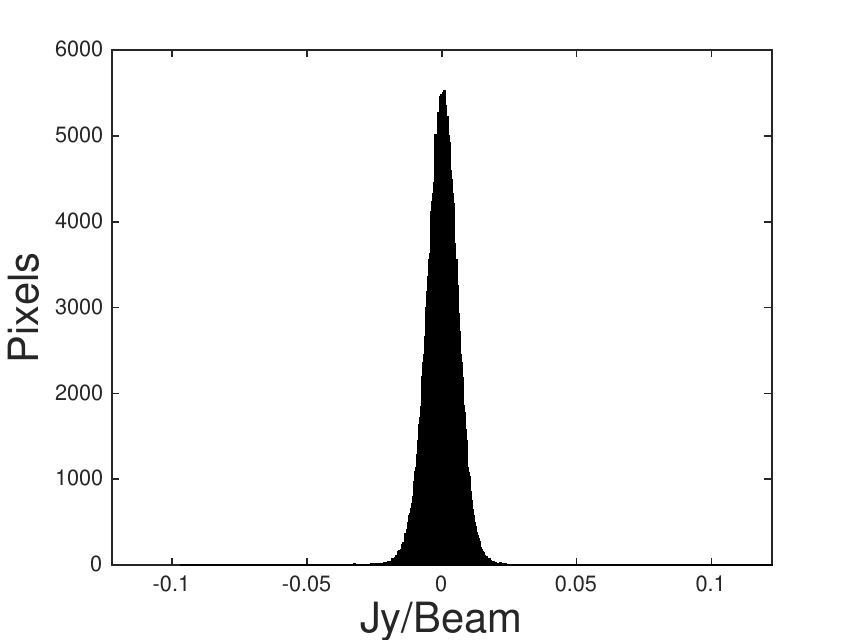}
    \end{minipage}
    \hspace{.5cm} % note: no blank line here
    \begin{minipage}{0.28\textwidth}
    \includegraphics[width=1.15\linewidth]{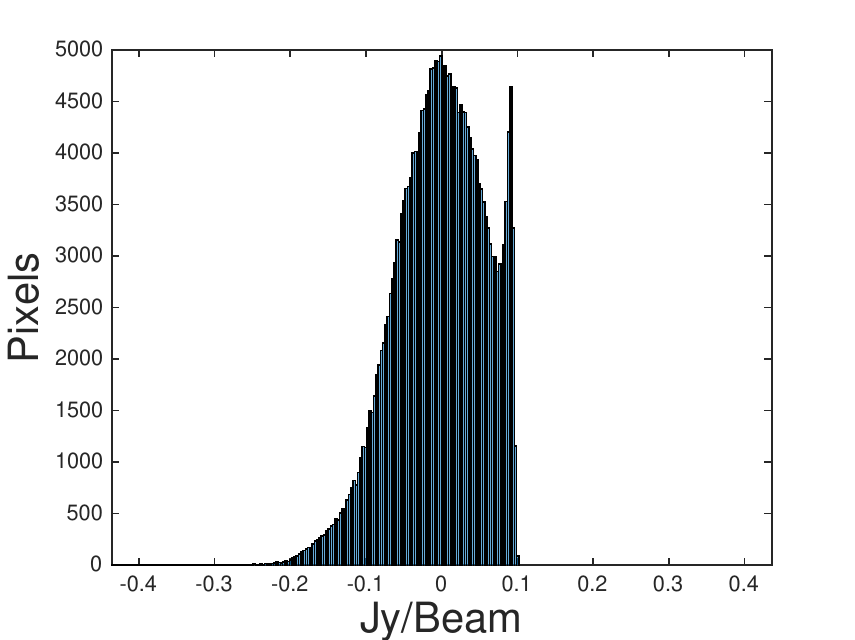}
    \end{minipage}
   \hspace{.5cm} % note: no blank line here
    \begin{minipage}{0.28\textwidth}
    \includegraphics[width=1.15\linewidth]{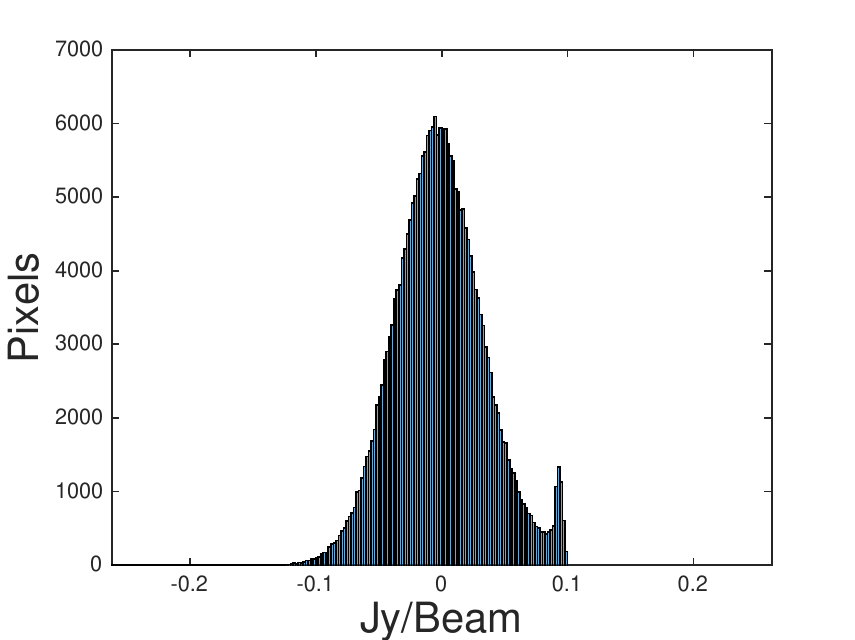}
    \end{minipage}
    \caption{PURIFY and CLEAN reconstructions of Cygnus A. Each pixel is 0.5 arcseconds, and the images are $1024 \times 1024$ pixels. { The pixels within $\left[ 256, 756\right] \times \left[256, 756 \right]$ are shown in the images and histogram of this figure.} Left column shows a PURIFY reconstruction with natural weighting. Middle and right columns show CLEAN reconstructions with natural and uniform weightings, respectively.  From the top to bottom row: synthesised (\emph{i.e.} dirty) image, model image, restored image, residual image, and a histogram of residual image. PURIFY does not require any post-processing and so does not produce a restored image.}
		\label{fig:CygA}
\end{figure*}

\begin{figure*}
	
	\begin{minipage}{0.28\textwidth}
    \includegraphics[width=1.15\linewidth]{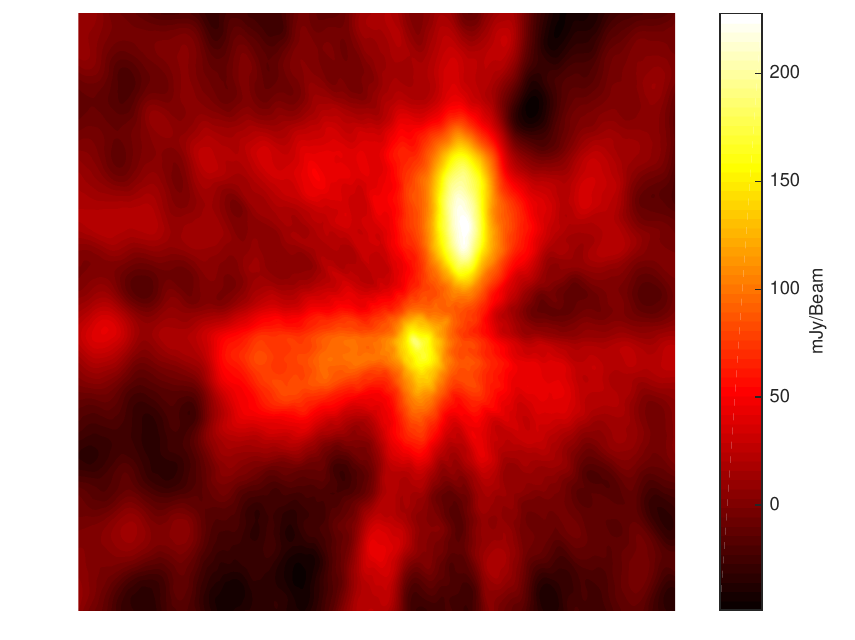}
    \end{minipage}
    \hspace{.5cm} % note: no blank line here
    \begin{minipage}{0.28\textwidth}
    \includegraphics[width=1.15\linewidth]{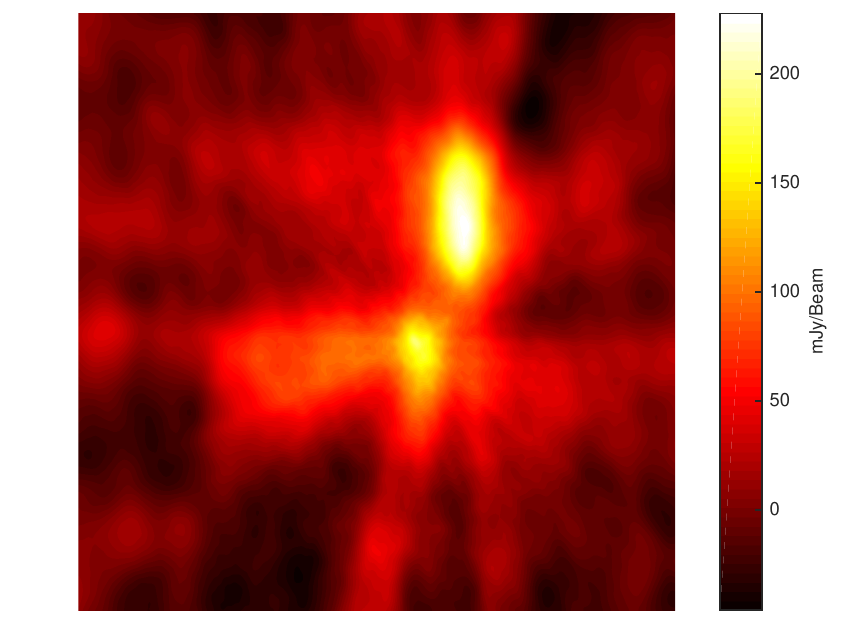}
    \end{minipage}
   \hspace{.5cm} % note: no blank line here
    \begin{minipage}{0.28\textwidth}
    \includegraphics[width=1.15\linewidth]{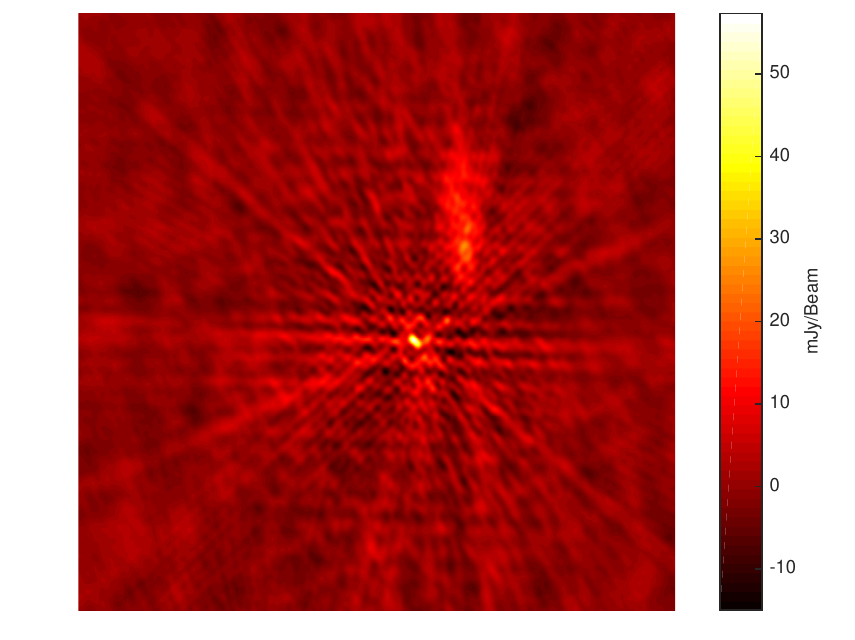}
    \end{minipage}

    \vspace*{0.01cm} % vertical separation
	
	\begin{minipage}{0.28\textwidth}
    \includegraphics[width=1.15\linewidth]{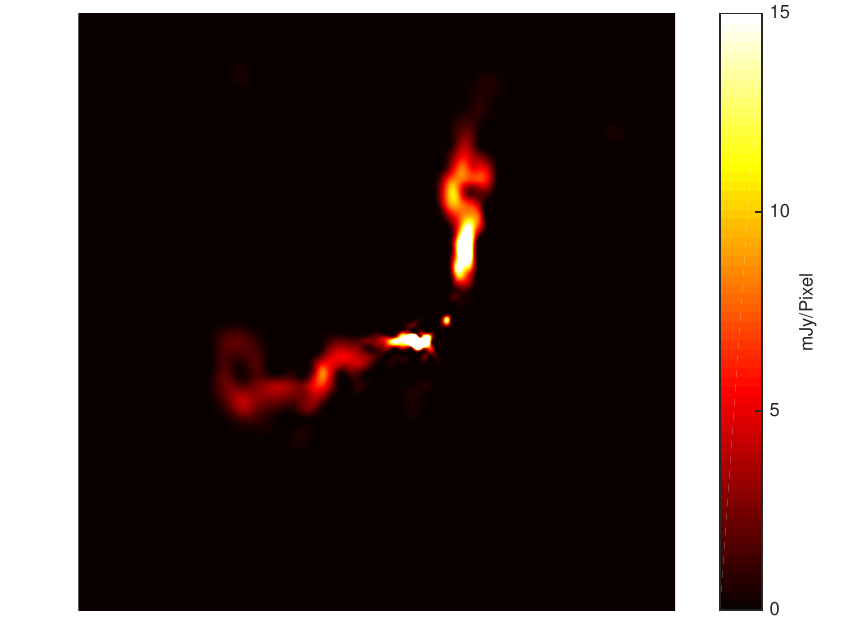}
    \end{minipage}
    \hspace{.5cm} % note: no blank line here
    \begin{minipage}{0.28\textwidth}
    \includegraphics[width=1.15\linewidth]{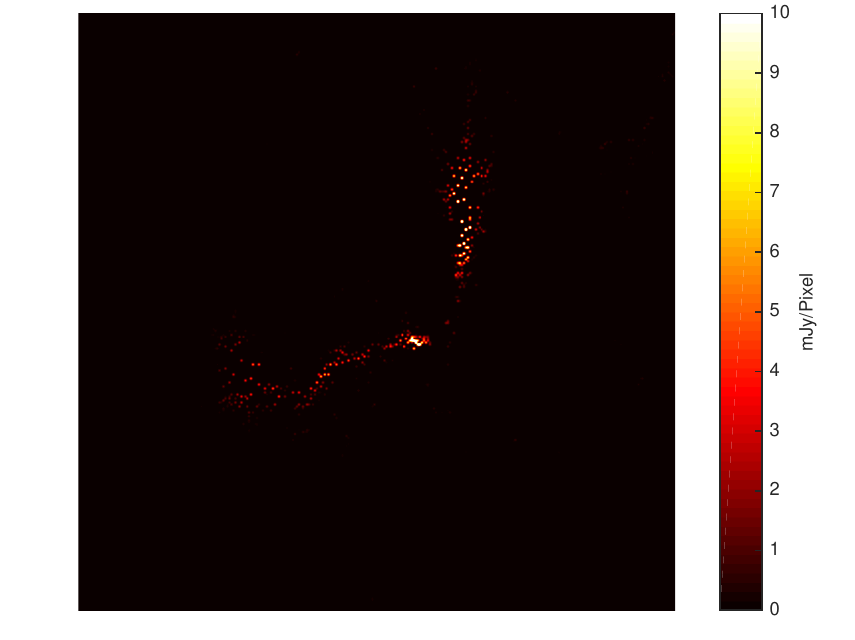}
    \end{minipage}
   \hspace{.5cm} % note: no blank line here
    \begin{minipage}{0.28\textwidth}
    \includegraphics[width=1.15\linewidth]{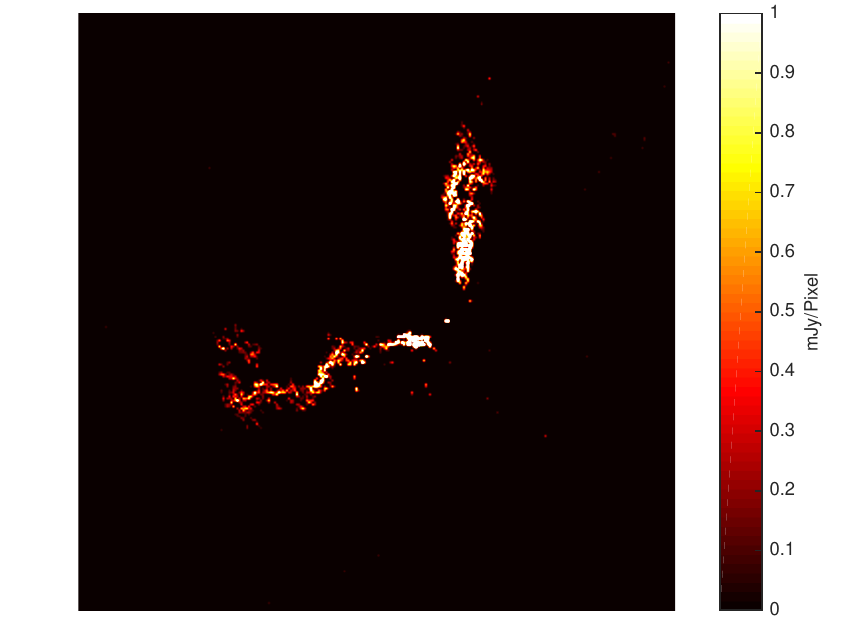}
   \end{minipage}
   
\rule{5.6cm}{0cm}
    \begin{minipage}{0.28\textwidth}
    \includegraphics[width=1.15\linewidth]{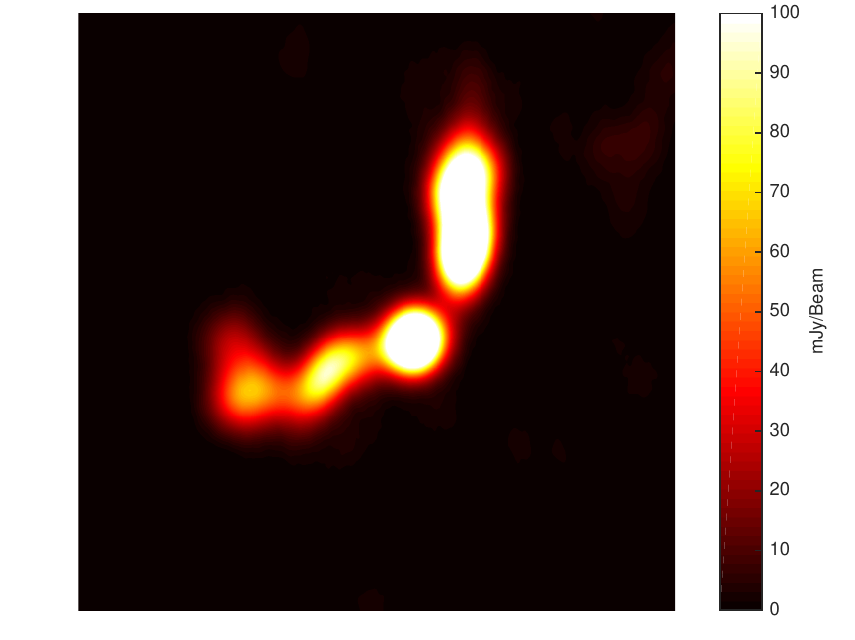}
    \end{minipage}
   \hspace{.5cm} % note: no blank line here
    \begin{minipage}{0.28\textwidth}
    \includegraphics[width=1.15\linewidth]{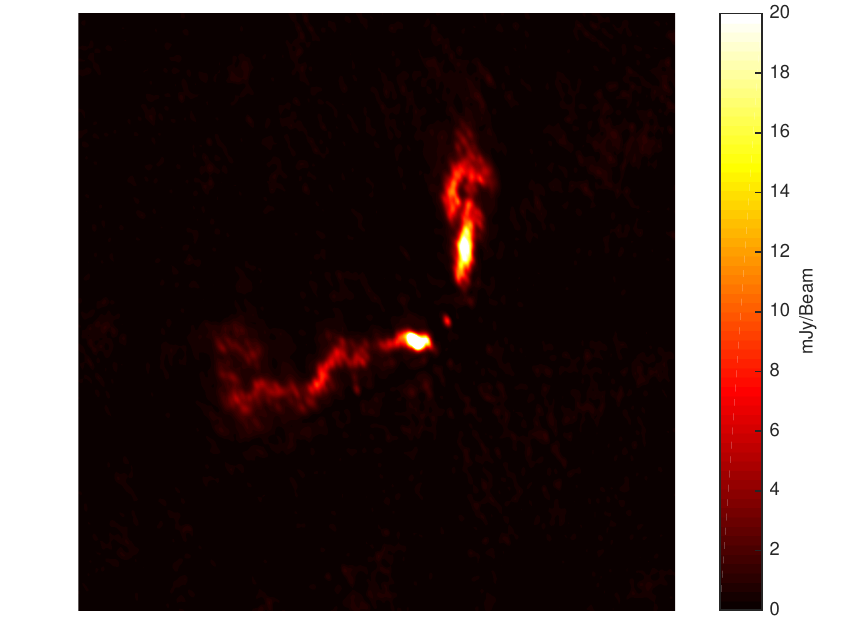}
    \end{minipage}
   \vspace*{0.01cm} % vertical separation
	
	\begin{minipage}{0.28\textwidth}
    \includegraphics[width=1.15\linewidth]{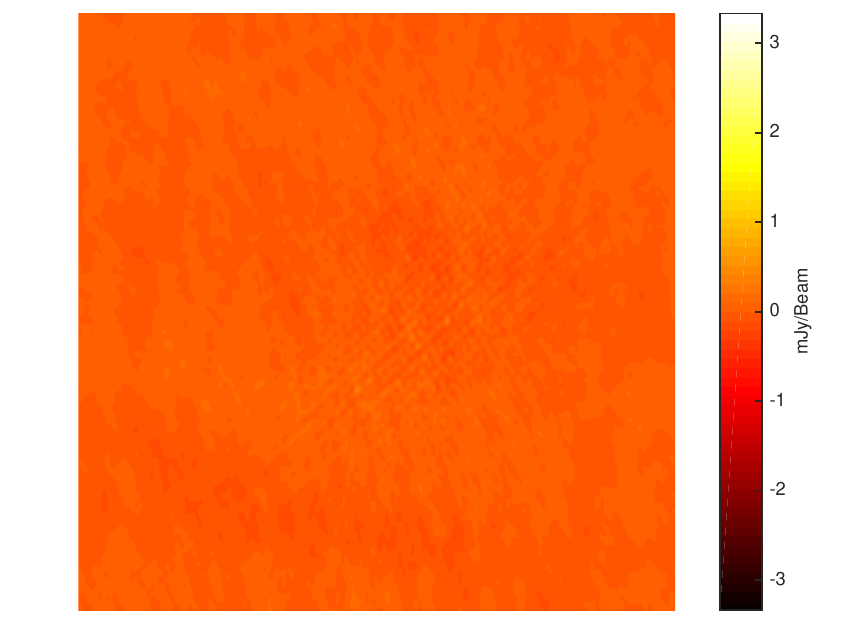}
    \end{minipage}
    \hspace{.5cm} % note: no blank line here
    \begin{minipage}{0.28\textwidth}
    \includegraphics[width=1.15\linewidth]{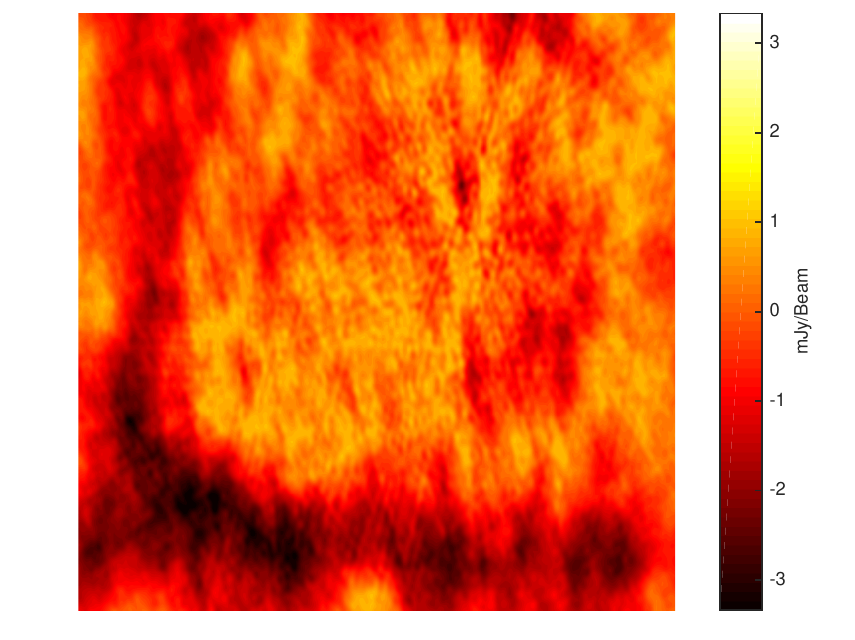}
    \end{minipage}
   \hspace{.5cm} % note: no blank line here
    \begin{minipage}{0.28\textwidth}
    \includegraphics[width=1.15\linewidth]{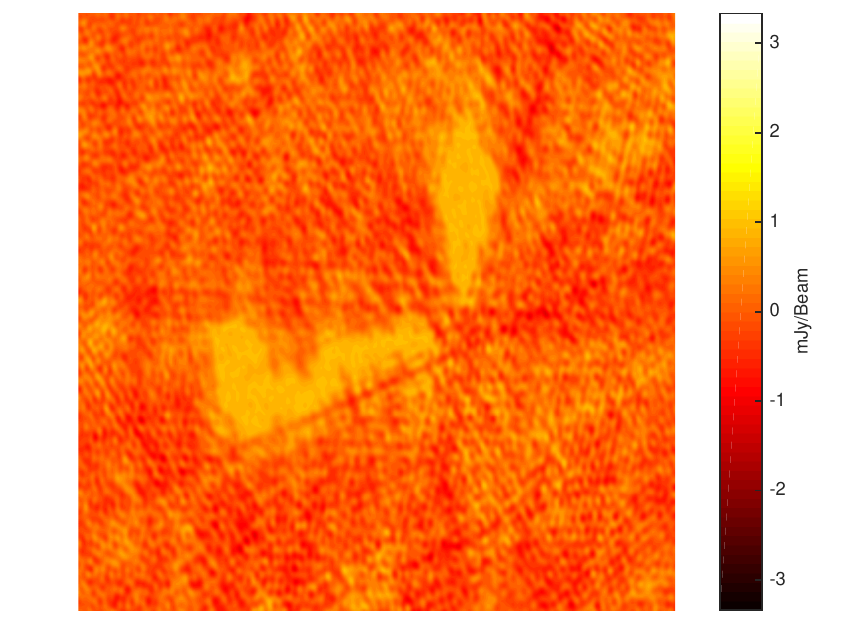}
	\end{minipage}
    
    \vspace*{0.01cm} % vertical separation
	
	\begin{minipage}{0.28\textwidth}
    \includegraphics[width=1.15\linewidth]{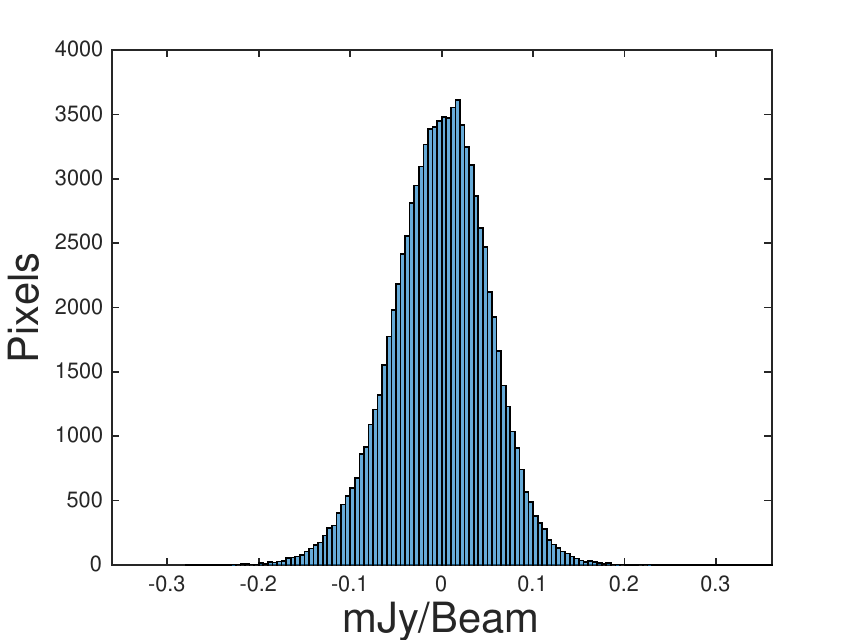}
    \end{minipage}
    \hspace{.5cm} % note: no blank line here
    \begin{minipage}{0.28\textwidth}
    \includegraphics[width=1.15\linewidth]{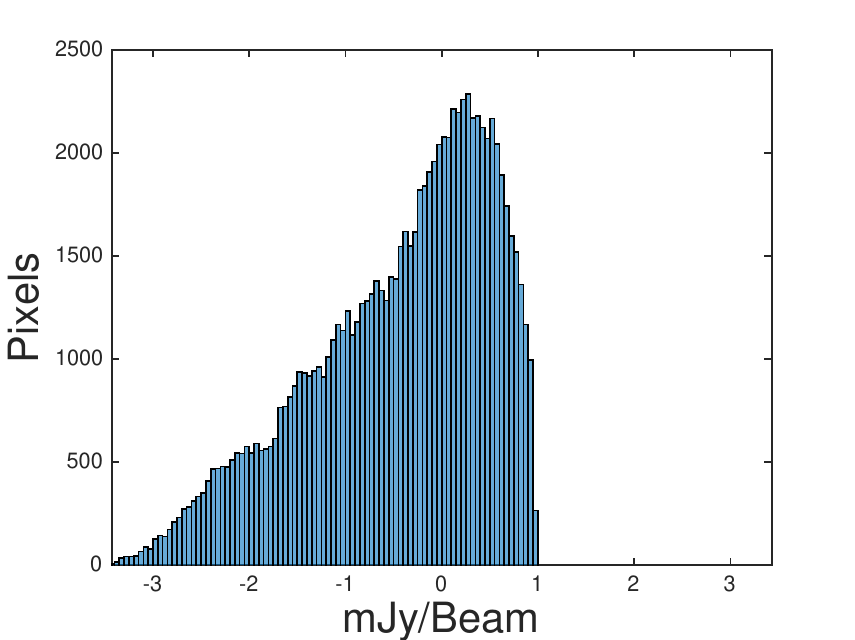}
    \end{minipage}
   \hspace{.5cm} % note: no blank line here
    \begin{minipage}{0.28\textwidth}
    \includegraphics[width=1.15\linewidth]{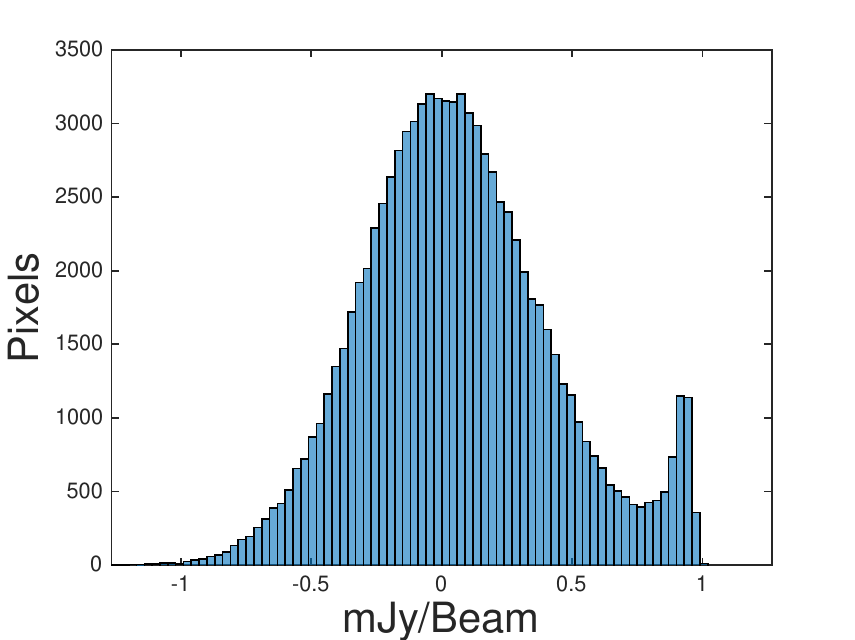}
    \end{minipage}
    \caption{PURIFY and CLEAN reconstructions of PKS J0334-39. Each pixel is 2 arcseconds, and the images are $2048 \times 2048$ pixels. { The pixels within $\left[ 862, 1162\right] \times \left[862, 1162 \right]$ are shown in the images and histogram of this figure.} Left column shows a PURIFY reconstruction with natural weighting. Middle and right columns show CLEAN reconstructions with natural and uniform weightings, respectively.  From the top to bottom row: synthesised (\emph{i.e.} dirty) image, model image, restored image, residual image, and a histogram of residual image. PURIFY does not require any post-processing and so does not produce a restored image.}
		\label{fig:0332-391}
\end{figure*}

\begin{figure*}
	
	\begin{minipage}{0.28\textwidth}
    \includegraphics[width=1.15\linewidth]{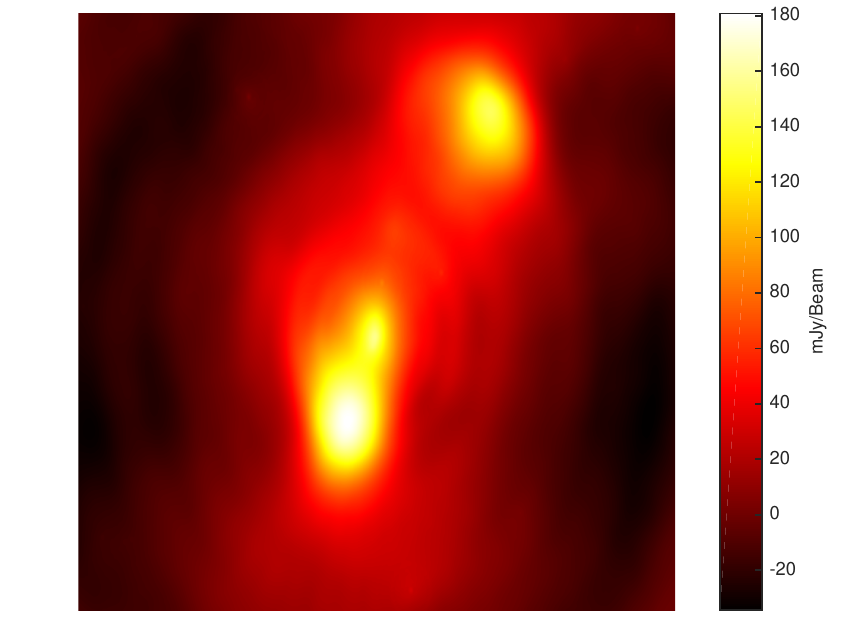}
    \end{minipage}
    \hspace{.5cm} % note: no blank line here
    \begin{minipage}{0.28\textwidth}
    \includegraphics[width=1.15\linewidth]{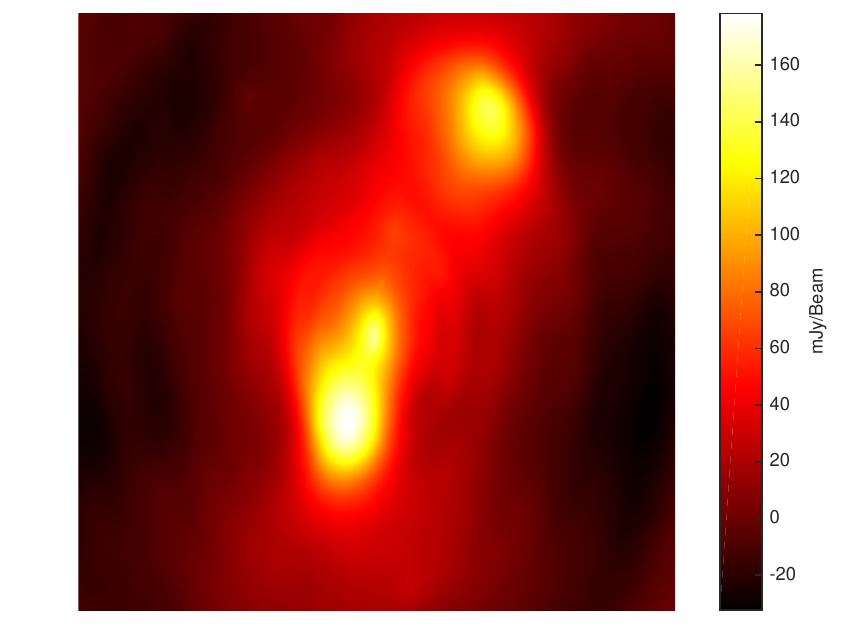}
    \end{minipage}
   \hspace{.5cm} % note: no blank line here
    \begin{minipage}{0.28\textwidth}
    \includegraphics[width=1.15\linewidth]{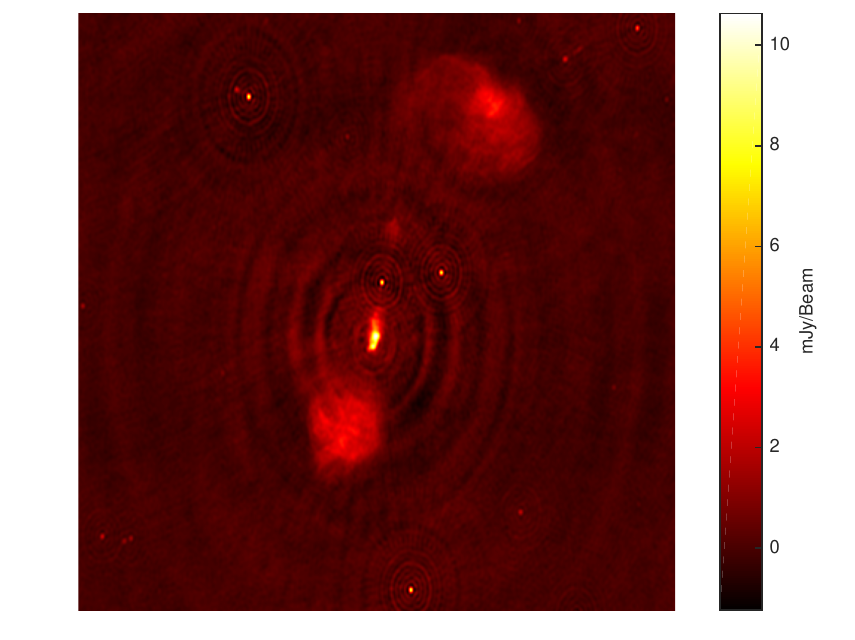}
    \end{minipage}
    
    \vspace*{0.01cm} % vertical separation
	
	\begin{minipage}{0.28\textwidth}
    \includegraphics[width=1.15\linewidth]{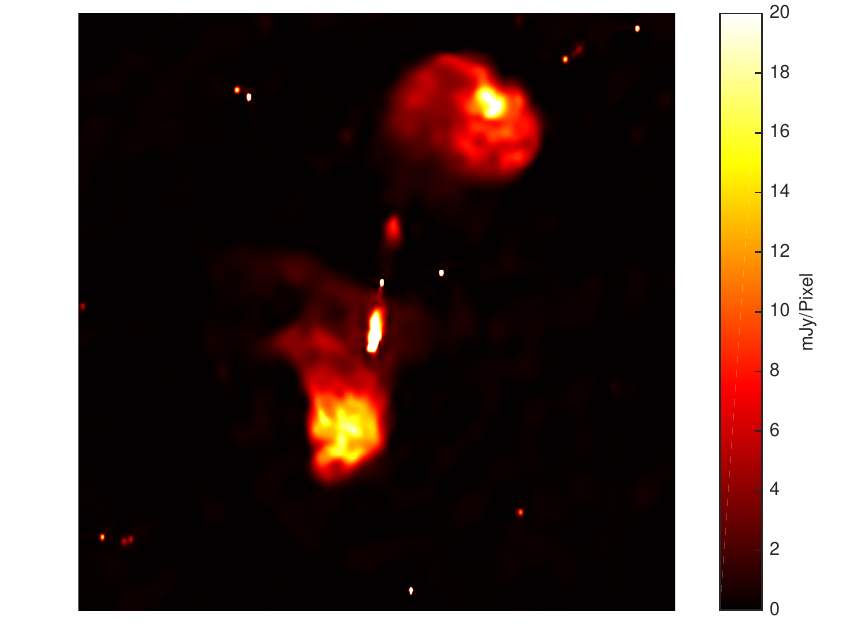}
    \end{minipage}
    \hspace{.5cm} % note: no blank line here
    \begin{minipage}{0.28\textwidth}
    \includegraphics[width=1.15\linewidth]{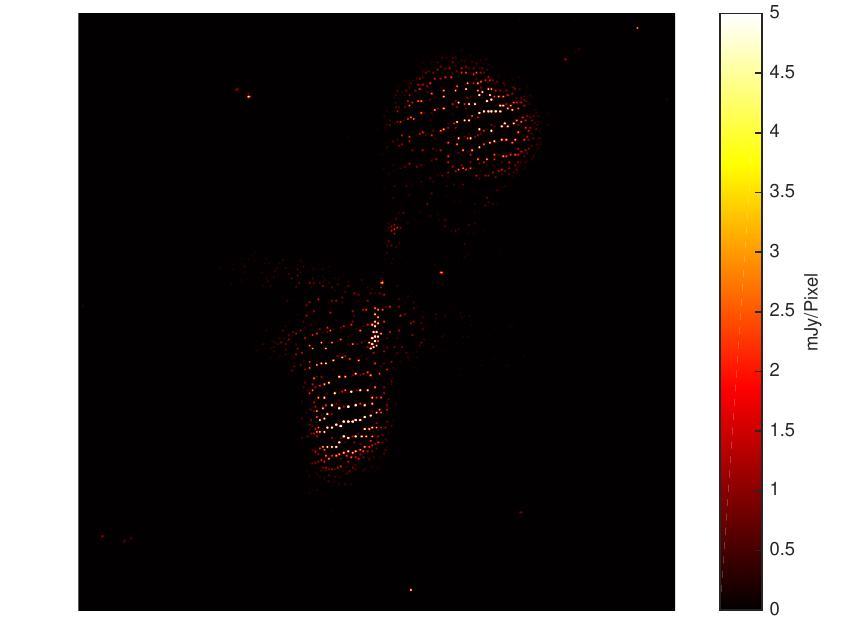}
    \end{minipage}
   \hspace{.5cm} % note: no blank line here
    \begin{minipage}{0.28\textwidth}
    \includegraphics[width=1.15\linewidth]{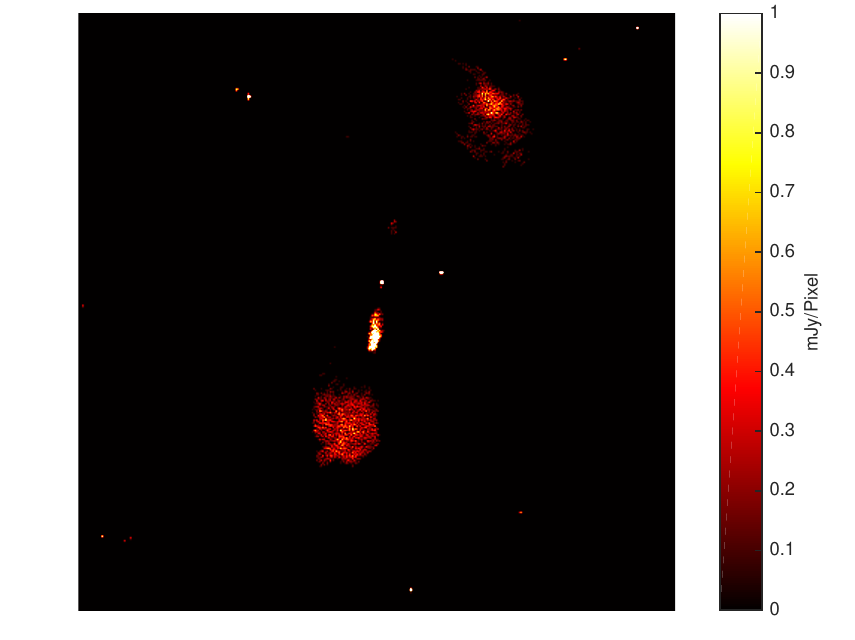}
   \end{minipage}
   
  \rule{5.6cm}{0cm}
    \begin{minipage}{0.28\textwidth}
    \includegraphics[width=1.15\linewidth]{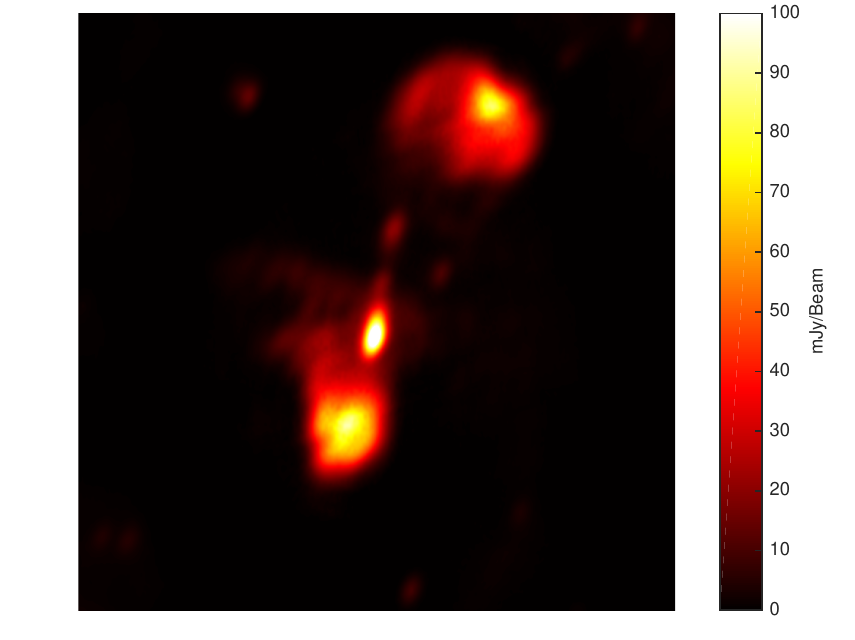}
    \end{minipage}
   \hspace{.5cm} % note: no blank line here
    \begin{minipage}{0.28\textwidth}
    \includegraphics[width=1.15\linewidth]{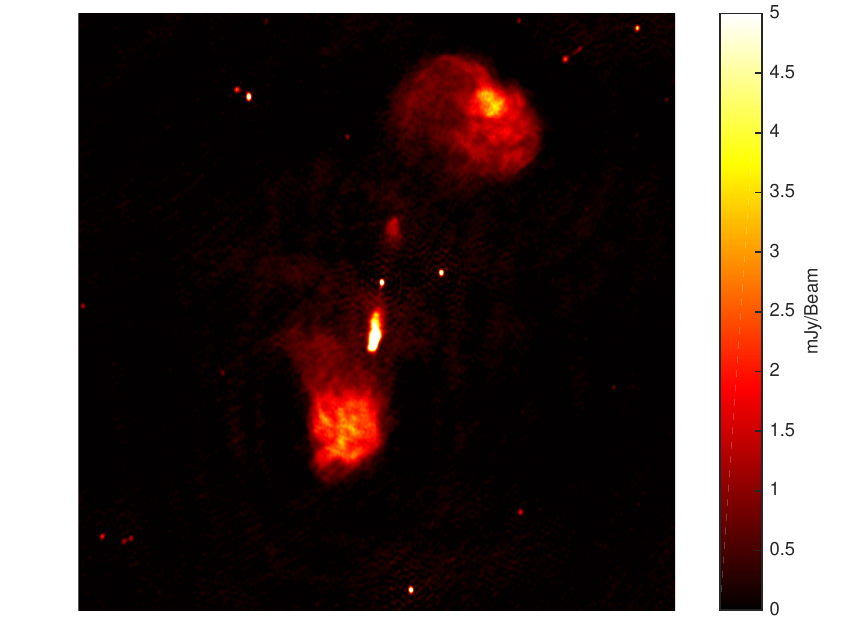}
    \end{minipage}
    \vspace*{0.01cm} % vertical separation
	
	\begin{minipage}{0.28\textwidth}
    \includegraphics[width=1.15\linewidth]{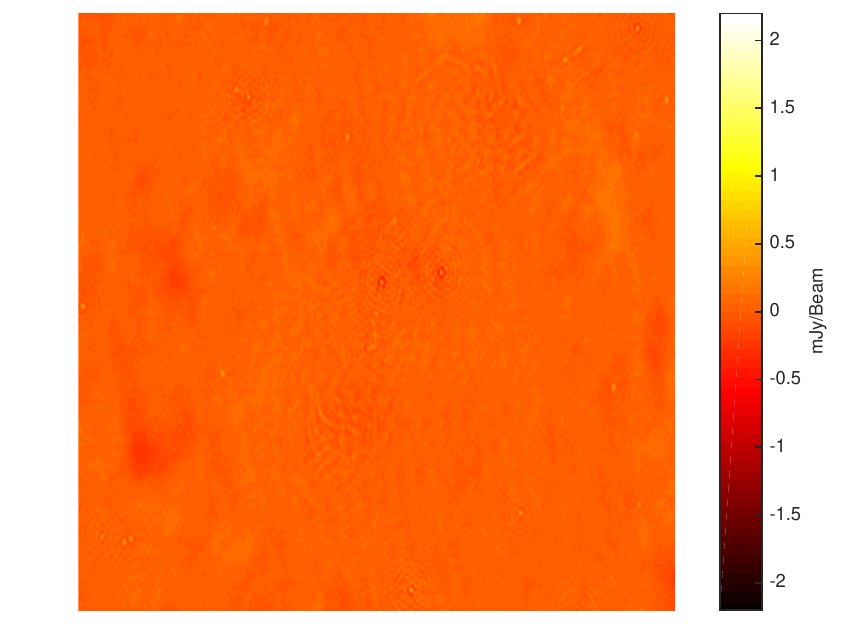}
    \end{minipage}
    \hspace{.5cm} % note: no blank line here
    \begin{minipage}{0.28\textwidth}
    \includegraphics[width=1.15\linewidth]{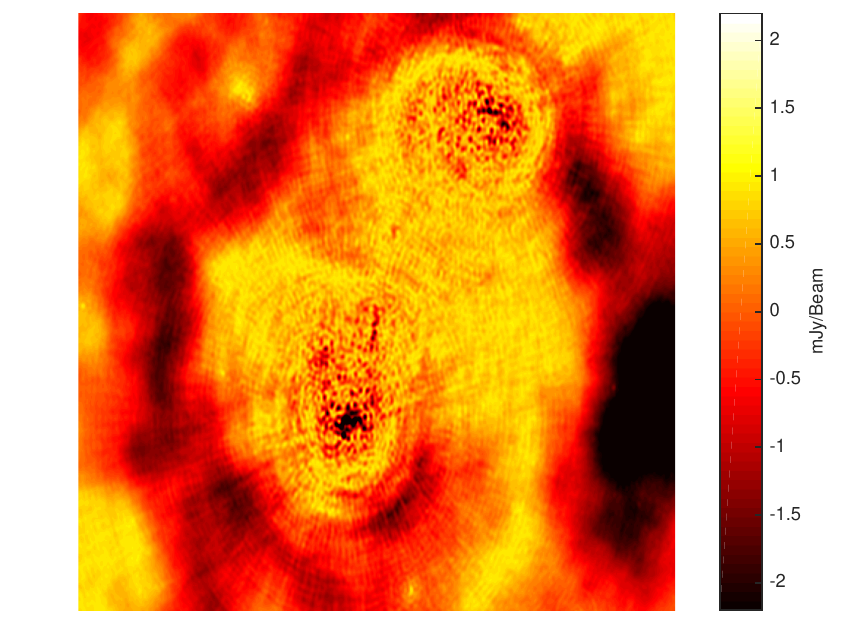}
    \end{minipage}
   \hspace{.5cm} % note: no blank line here
    \begin{minipage}{0.28\textwidth}
    \includegraphics[width=1.15\linewidth]{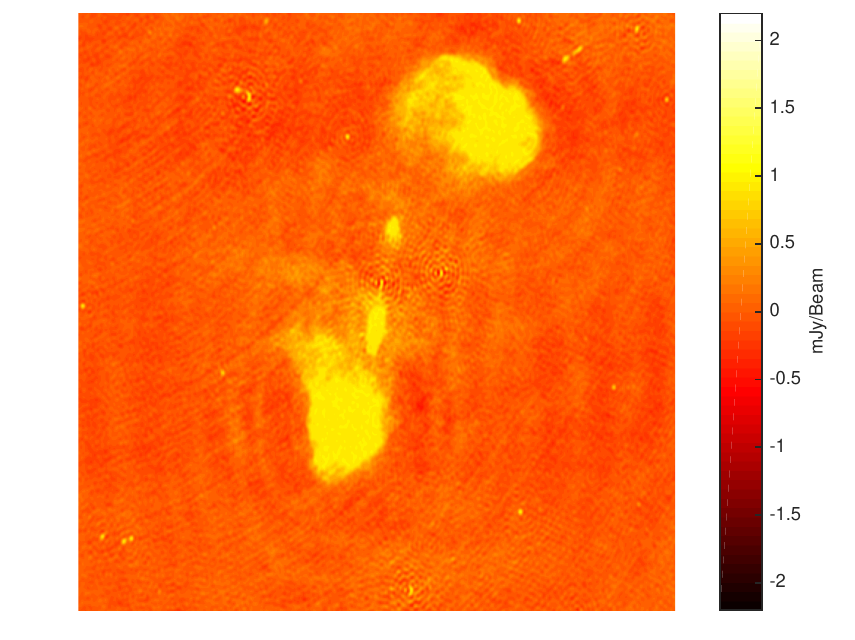}
	\end{minipage}
    
    \vspace*{0.01cm} % vertical separation
	
	\begin{minipage}{0.28\textwidth}
    \includegraphics[width=1.15\linewidth]{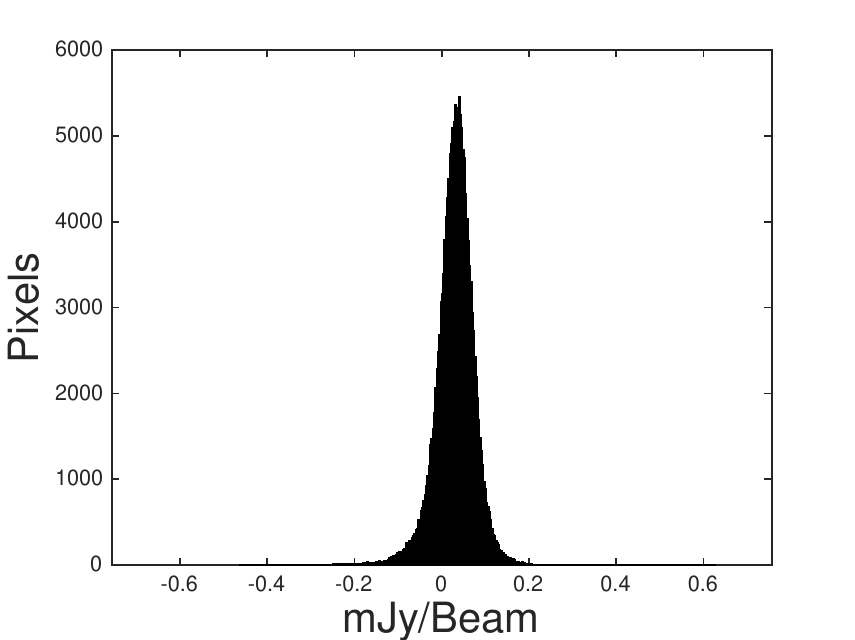}
    \end{minipage}
    \hspace{.5cm} % note: no blank line here
    \begin{minipage}{0.28\textwidth}
    \includegraphics[width=1.15\linewidth]{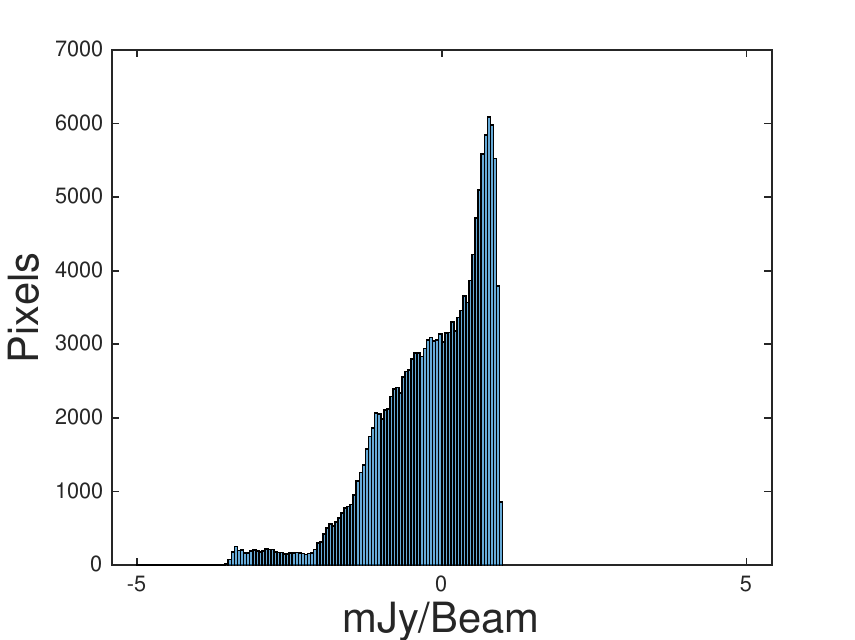}
    \end{minipage}
   \hspace{.5cm} % note: no blank line here
    \begin{minipage}{0.28\textwidth}
    \includegraphics[width=1.15\linewidth]{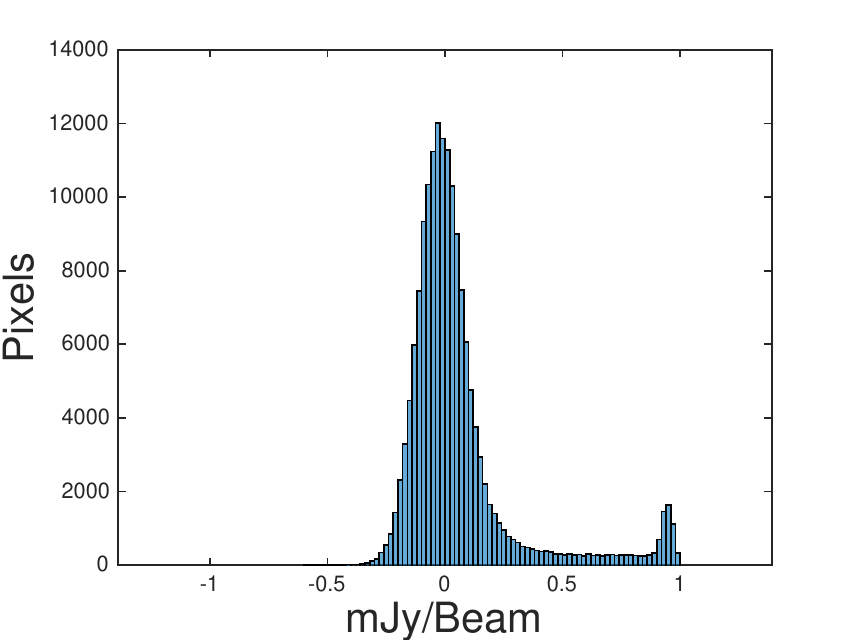}
    \end{minipage}
    \caption{PURIFY and CLEAN reconstructions of PKS J0116-473. Each pixel is 2.4 arcseconds, and the images are $2048\times 2048$ pixels. { The pixels within $\left[ 800, 1200\right] \times \left[800, 1200 \right]$ are shown in the images and histogram of this figure.} Left column shows a PURIFY reconstruction with natural weighting. Middle and right columns show CLEAN reconstructions with natural and uniform weightings, respectively.  From the top to bottom row: synthesised (\emph{i.e.} dirty) image, model image, restored image, residual image, and a histogram of residual image. PURIFY does not require any post-processing and so does not produce a restored image.}
		\label{fig:0114-476}
\end{figure*}

Images recovered by CLEAN and PURIFY, and auxiliary plots, are shown in Figures~\ref{fig:3C129}, \ref{fig:CygA}, \ref{fig:0332-391}, and \ref{fig:0114-476}.
Reconstructions of the source 3C129 are shown in Figure~\ref{fig:3C129}, for a pixel width and height of 0.4 arcseconds. The PURIFY reconstruction was performed using a value of $\eta = \bm{0.9}$ and $\xi = 1$, and ran for { 75} iterations. The step size was adapted for the first $i_{\rm adapt} = 20$ iterations.
Figure~\ref{fig:CygA} contains the reconstructions of Cygnus A, for a pixel width and height of 0.5 arcseconds. The PURIFY reconstruction was performed using $\eta = \bm{2.14}$ and $\xi = 7.07$, and ran for { 183} iterations. The step size was adapted for the first $i_{\rm adapt} = 100$ iterations.
Reconstructions of the source PKS J0334-39 are shown in Figure~\ref{fig:0332-391}, for a pixel width and height of 2 arcseconds. The PURIFY reconstruction was performed using $\eta = 1$ and $\xi = 2$, and ran for 372 iterations. The step size was adapted for the first $i_{\rm adapt} = 200$ iterations.
Reconstructions of the source PKS J0116-473 are shown in Figure~\ref{fig:0114-476}, for a pixel width and height of 2.4 arcseconds. The PURIFY reconstruction was performed using $\eta = 1$ and $\xi = 2.3$, and ran for 707 iterations. The step size was adapted for the first $i_{\rm adapt} = 500$ iterations.

The run times for these reconstructions range from an hour to several hours using a high-performance desktop computer, to produces images of sizes $1024\times 1024$ and $2048\times 2048$ pixels. { Currently, a large factor in the computational cost and run time for PURIFY is computing wavelet transforms for a number of dictionaries. In the case that only a Dirac basis is used and no wavelet transforms are performed, the run time is reduced considerably for large image sizes. However, this greatly reduces the quality of the reconstructed image, because a Dirac basis is not an efficient representation of extended structures.} As discussed, highly distributed and parallelised algorithms will be implemented in future work to reduce the run-time significantly \citep{ono16}. { While CLEAN methods appear computationally efficient, this comes at a significant cost to reconstruction quality and with additional restrictions on the ability for distribution.}

In all cases PURIFY provides more complete reconstructions than CLEAN.  When comparing with the CLEAN component images, the CLEAN component images are not smooth and do not reconstruct the diffuse emission well (due to the point source model of CLEAN), while the PURIFY recovered images model diffuse emission.  After post-processing the CLEAN component image to yield the CLEAN restored image and comparing with \mbox{PURIFY}, it is also clear that PURIFY provides higher quality reconstructions.

{ The dirty and residual images of PURIFY are shown in Jy/Beam for comparison. To convert from units of Jy/Pixel to Jy/Beam, the image is divided by the peak of the point spread function $\bm{\mathsf{\Phi}}^\dagger\bm{\mathsf{W}}\bm{1}$, where $\bm{1}$ denotes a vector of ones. This allows direct comparisons of the residual images between CLEAN and PURIFY, since they will have the same units without arbitrary scaling. To compare the residuals the scale of the colour axis has been set to a common scale, using 3 times the median root-mean-squared (RMS) values between the residual images in Table \ref{tab:residual_rms}. The histograms show the full range of pixel values in the residuals, determined by the peak of the absolute residuals, to allow one to inspect outliers.}

For all observations, PURIFY can model faint extended structure while also modeling the bright compact sources. Additionally, the PURIFY model has left little structure in the residuals. This is also clear from the histogram of the residual pixel brightness, which shows the residuals are dominated by Gaussian noise. The CLEAN reconstruction leaves visible diffuse structure in the residuals. The histogram of the residual images show large peaks below the clean cutoff.

The primary difference that natural and uniform weightings have on CLEAN is that uniform weighting suppresses the synthesised beam sidelobes. While this lowers the sensitivity of the observation, CLEAN then performs better at modelling fine structure with CLEAN components, with diffuse structure left in the residuals, which are then added back in the CLEAN restored image.

The dynamic range is used to assess the quality of reconstructions quantitatively and is calculated by
\begin{equation}
	{\rm DR} = \frac{\sqrt{N}\| \bm{\mathsf{\Phi}}\|^2}{\| \bm{\mathsf{\Phi}}^\dagger\left(\bm{y} - \bm{\mathsf{\Phi}} \bm{x}\right)\|_{\ell_2}} \max \{\bm{x}_k \}\, ,
\end{equation}
\textit{i.e.}\ the ratio of the peak of the recovered image to the root-mean-square (RMS) of the residuals (for a normalised measurement operator). { The weights are assumed to be in the measurement operator. The norm of the measurement operator is included so that the dynamic range does not scale arbitrarily under the choice of the normalisation of the measurement operator.} For CLEAN, we follow the standard approach and use the peak of the \emph{restored} image, divided by the RMS of the residual image. The dynamic ranges of the images recovered by CLEAN and PURIFY can be found in Table~\ref{tab:performance}, where PURIFY consistently recovers images with higher dynamic range.
{ The RMS of the residuals around the scientific region of interest (see Table \ref{tab:residual_rms}) show that PURIFY consistently fits the measurements better than CLEAN.}

{ Table \ref{tab:residual_rms} compares the RMS of the residual images with in the regions shown in Figures \ref{fig:3C129}, \ref{fig:CygA}, \ref{fig:0332-391}, \ref{fig:0114-476}. Other than 3C129, PURIFY shows a consistent order of magnitude improvement in the RMS of the residuals.}

\subsection{Discussion}

From a scientific standpoint, the PURIFY models show more structure than those recovered by CLEAN. This is clear when looking at the surface brightness variation of the jets of 3C129 and Cygnus A. For 3C129 and Cygnus A, unlike the CLEAN restored images, the surface brightness structure is well resolved in the images recovered by PURIFY.

The CLEAN restored images of PKS J0334-39 and PKS J0116-473 with uniform weighting show an improvement over natural weighting for deconvolving the fine structure, as well as containing diffuse structure. However, uniform weighting is known to suppress large scale structure, and lowers the sensitivity of the observation (as discussed in \citealt{hin14}). However, PURIFY has the ability to reconstruct the fine details of PKS J0334-39 and PKS J0116-473 without uniform weighting. This demonstrates that PURIFY has the potential to reconstruct observations that can be used to perform a more detailed analysis of morphology and structure of diffuse sources. The reconstruction of Cygnus A shows that it is possible to accurately reconstruct diffuse bright structures in the presence of compact bright sources.

Modeling extended structure accurately is particularly important for understanding the underlying physics of radio sources and their environment. Bent tailed radio galaxies, such as 3C129, are a example of where this is important \citep{mil72}. The morphology of bent tailed radio galaxies can be used as a probe of their local cluster environment\linebreak \citep{gun72,fre08,dou11,pfr11,pra13,pra15}.

Additionally, an important class of diffuse, low surface brightness radio sources are cluster relics and halos (\textit{e.g.} \citealt{bru08,hin14,sha16,mar16}),  which are believed to be caused by shocks and turbulence in the outskirts of galaxy clusters \citep{cas13,cas15}. Radio halos and relics are not well understood, and they are prime examples of sources with diffuse low surface brightness structure that relates to the physics within the intra-cluster medium and merging galaxy clusters. However, galaxy clusters often contain bright compact sources, providing a challenge in deconvolving low surface brightness sources. PURIFY's ability to accurately model extended structure and reconstruct images with high dynamic range has the potential to improve scientific interpretations of many radio interferometric observations.

\begin{table}
\caption{Table listing details of settings used to recover CLEAN images.}
\label{tab:clean}
\centering
\begin{tabular}{c c c c}
\toprule
Observation & Weighting & Beam Size & Cutoff \\
\midrule
3C129 & Natural&	2.07$^{\prime\prime}$ $\times$ 1.88$^{\prime\prime}$, 158$^\circ$ & 0.0025 Jy\\
 & Uniform&	1.30$^{\prime\prime}$ $\times$ 1.06$^{\prime\prime}$, 33$^\circ$ \\
Cygnus A & Natural&	3.48$^{\prime\prime}$ $\times$ 2.81$^{\prime\prime}$, 105$^\circ$ & 0.1 Jy\\
 & Uniform&	2.25$^{\prime\prime}$ $\times$ 1.99$^{\prime\prime}$, 97.4$^\circ$ \\
PKS J0334-39 & Natural&	45.6$^{\prime\prime}$ $\times$ 36.8$^{\prime\prime}$, 171$^\circ$ & 0.001 Jy\\
 & Uniform&	8.6$^{\prime\prime}$ $\times$ 4.3$^{\prime\prime}$, 17$^\circ$ \\
PKS J0116-473 & Natural&	40.0$^{\prime\prime}$ $\times$ 24.6$^{\prime\prime}$, 158$^\circ$ & 0.001 Jy\\
 & Uniform&	6.33$^{\prime\prime}$ $\times$ 4.72$^{\prime\prime}$, 3$^\circ$ \\
\bottomrule
\end{tabular}
\end{table}

\begin{table}
\caption{Table listing the dynamic range of each reconstruction. {{When computing the dynamic range for PURIFY reconstructions the calculation includes the norm of the measurement operator, so the dynamic range does not scale arbitrarily under the choice of the norm of the measurement operator. For CLEAN, we follow the standard approach and use the peak of the \emph{restored} image, divided by the RMS of the residual image.}}}
\label{tab:performance}
\centering
\begin{tabular}{c c c c c}
\toprule
Observation & PURIFY & CLEAN & CLEAN \\
 &  &  (natural) &  (uniform) \\
\midrule
3C129 & { 72\:444} & 220 & 495 \\
Cygnus A & { 312\:928} & 372 & 472\\
PKS J0334-39 & { 1\:701\:050} & 208 & 263 \\
PKS J0116-473 & { 1\:185\:700} & 153 & 361 \\
\bottomrule
\end{tabular}
\end{table}

\begin{table}
\caption{ Table listing the root-mean-squared of each reconstruction (units are in mJy/Beam).}
\label{tab:residual_rms}
\centering
\begin{tabular}{c c c c c}
\toprule
Observation & PURIFY & CLEAN & CLEAN \\
 &  &  (natural) &  (uniform) \\
\midrule
3C129 &  0.10 &  0.23 & 0.11 \\
Cygnus A &  6.1 &  59 & 36 \\
PKS J0334-39  &  0.052 &  1.00 &  0.37 \\
PKS J0116-473 & 0.054 &  0.88 & 0.24 \\
\bottomrule
\end{tabular}
\end{table}

% Temporarily turn off to help with loading pdf.

\section{Conclusions}
\label{sec:concusions}

In this work we have further developed the PURIFY software package so that it can be easily applied to observational data from radio interferometric telescopes.  PURIFY has been completely redesigned and reimplemented in C++ and now supports the \mbox{P-ADMM} algorithm developed recently by \citet{ono16}.  Furthermore, the capabilities of convolutional degridding in the measurement operator have been expanded.  

Using simulations we studied the impact of a number of different interpolation kernels on the quality of images recovered by sparse reconstruction approaches to interferometric imaging.  The Kaiser-Bessel kernel was found to perform very well---as well as other optimal kernels---while requiring a smaller support size, thereby reducing computation cost, and having an analytic expression that can be evaluated easily and efficiently.

PURIFY was applied to observational data from the VLA and ATCA telescopes, recovering high-quality interferometric images superior to those recovered by CLEAN.  Firstly, the PURIFY residuals contain less extended structure and are more Gaussian { with a lower RMS}. Secondly, the model images recovered by PURIFY are of sufficient quality that there is no need to perform any post-processing as is done for CLEAN (such as restoring the image).  Thirdly, all images recovered by PURIFY show an increase in dynamic range when compared to those recovered by CLEAN, in some cases in excess of an order of magnitude.  On visual inspection, the images recovered by PURIFY reveal extended structure in greater detail.  For example, in reconstructed images of 3C129 the internal structure of the radio jets is much more apparent (Figure~\ref{fig:3C129}).  Such an improvement in reconstruction quality can be important in facilitating a better scientific understanding of astrophysical processes.

While the current version of PURIFY can be readily used to recover high-fidelity images from observations made by radio interferometric telescopes, numerous extensions and improvements are planning for future releases. In future we will implement the primal dual algorithm of \cite{ono16}, highly distribute and parallelise the algorithms supported following the strategies outlined in \cite{car14} and \cite{ono16}, and add support for direction-dependent effects following the approach outlined in \cite{wol13}, for example.

% unnumbered section
\section*{Acknowledgments}

LP thanks Melanie Johnston-Hollitt for discussions on the practical constraints of current interferometric imaging and useful comments on the manuscript, Xiaohao Cai for discussions on optimisation and useful comments on the manuscript, and Arwa Dabbech for general discussions. We thank Rick Perley and Oleg Smirnov for making the VLA observation of Cygnus A available and thank Andr\'e Offringa for assistance with WSCLEAN.  This work was supported by the UK Engineering and Physical Sciences Research Council (EPSRC, grants EP/M011089/1 and EP/M008843/1) and the UK Science and Technology Facilities Council (STFC, grant ST/M00113X/1).

\bibliographystyle{mymnras_eprint}
\bibliography{refs}

%%%%%%%%%%%%%%%%% APPENDICES %%%%%%%%%%%%%%%%%%%%%

% \appendix

%%%%%%%%%%%%%%%%%%%%%%%%%%%%%%%%%%%%%%%%%%%%%%%%%%
\label{lastpage}
\end{document}